\documentclass[a4paper,10pt]{article}

\usepackage{mathrsfs,amstext,amsmath,amssymb,amsfonts,cite,epsf,color,epsfig,epstopdf,graphics,graphicx,mathtools,physics,verbatim,setspace,subfig,soul}
\usepackage{slashed}
\usepackage[left=2cm,right=2cm,top=2.5cm,bottom=2.5cm]{geometry}
\usepackage[font=small]{caption}
\usepackage{bbm}

\definecolor{ultramarine}{rgb}{0.07, 0.04, 0.56}
\definecolor{cadmiumgreen}{rgb}{0.0, 0.42, 0.24}
\definecolor{indigo(dye)}{rgb}{0.0, 0.25, 0.42}
\usepackage[linktocpage=true,breaklinks]{hyperref}
\hypersetup{
	colorlinks=true,
	citecolor=ultramarine,
	linkcolor=cadmiumgreen,
	urlcolor=indigo(dye),
}

\usepackage[capitalise]{cleveref}
\usepackage{upgreek}
\usepackage[export]{adjustbox}
\usepackage{tikz}
\usetikzlibrary{decorations.pathmorphing,decorations.pathreplacing,calligraphy,calc,snakes,cd,external,decorations.markings}
\usetikzlibrary{hobby}
\usepackage[compat=1.1.0]{tikz-feynman}
\numberwithin{equation}{section}
\graphicspath{{figures/}}

\newcommand{\p}{{\partial}}

\newcommand{\eq}[1]{\begin{equation}#1\end{equation}}
\newcommand{\eqa}[1]{\begin{align}#1\end{align}}
\newcommand{\spl}[1]{\begin{split} #1 \end{split}}
\newcommand{\fg}[1]{\begin{figure}[tbp]\centering #1 \end{figure}}

\newcommand{\vp}{\varphi}
\newcommand{\bsk}{\boldsymbol{k}}
\newcommand{\vbx}{\boldsymbol{x}}
\newcommand{\vbk}{\boldsymbol{k}}
\newcommand{\vbq}{\boldsymbol{q}}
\newcommand{\vbxi}{\boldsymbol{\xi}}
\newcommand{\oo}{\mathcal{O}}
\newcommand{\D}{\mathcal{D}}
\newcommand{\PP}{\mathcal{P}}

\newcommand{\tikzxmark}{%
	\tikz[scale=0.12] {
		\draw[line width=0.7,line cap=round] (0,0) to [bend left=6] (1,1);
		\draw[line width=0.7,line cap=round] (0.2,0.95) to [bend right=3] (0.8,0.05);
}}

\usepackage[framemethod=TikZ]{mdframed}
\mdfsetup{roundcorner=3pt}   
\newmdenv[
skipabove=5pt,
skipbelow=7pt,
rightline=false,
leftline=false,
topline=false,
bottomline=false,
backgroundcolor=gray!15,
innerleftmargin=5pt,
innerrightmargin=5pt,
innertopmargin=7pt,
innerbottommargin=7pt,
leftmargin=0cm,
rightmargin=0cm,
linewidth=4pt]{eBox}

\begin{document}
%%%%%%%%%%%%%%%%%%%%%%%%%%%%%%%%%%%%%%%%%%%%%%%%%%%%%%%%
\begin{titlepage}
\setcounter{page}{1} \baselineskip=15.5pt \thispagestyle{empty}
\begin{flushright} {}  \end{flushright}
\vspace{0.5cm}
\def\thefootnote{\fnsymbol{footnote}}
\bigskip

\begin{center}
{\LARGE  \bf Positivity bounds on electromagnetic properties of media}
\\[0.5cm]
\end{center}
\vspace{0.2cm}

\begin{center}
\large{Paolo Creminelli$^{1,2}\footnote{creminel@ictp.it}$, Oliver Janssen$^{3}\footnote{oliver.janssen@epfl.ch}$, Borna Salehian$^{1,2}\footnote{bsalehia@ictp.it}$ and Leonardo Senatore$^{4}\footnote{lsenatore@phys.ethz.ch}$}
\\[.7cm]

{\small \textit{$^{1}$ICTP, International Centre for Theoretical Physics, 34151 Trieste, Italy}} \\ 

\vspace{0.2cm}
{\small \textit{$^{2}$IFPU, Institute for Fundamental Physics of the Universe, 34014 Trieste, Italy}} \\

\vspace{0.2cm}
{\small \textit{$^{3}$Laboratory for Theoretical Fundamental Physics, EPFL, 1015 Lausanne, Switzerland}} \\ 

\vspace{0.2cm}
{\small \textit{$^{4}$Institut f\"ur Theoretische Physik, ETH Z\"urich, 8093 Z\"urich, Switzerland}} \\

\end{center}

\vspace{1cm}
%%%%%%%%%%%%%%%%%%%%%%%%%%%%%%%%%%%%%%%%%%%%%%%%%%%%%

\hrule \vspace{0.5cm}
{\small  \noindent \textbf{Abstract} \\[0.2cm]
\noindent
We study the constraints imposed on the electromagnetic response of general media by microcausality (commutators of local fields vanish outside the light cone) and positivity of the imaginary parts (the medium can only absorb energy from the external field). The equations of motion for the average electromagnetic field in a medium -- the macroscopic Maxwell equations -- can be derived from the in-in effective action and the effect of the medium is encoded in the electric and magnetic permeabilities $\varepsilon(\omega,|\bsk|)$  and $\mu(\omega,|\bsk|)$. Microcausality implies analyticity of the retarded Green's functions when the imaginary part of the $4$-vector $(\omega,\bsk)$ lies in forward light cone. With appropriate assumptions about the behavior of the medium at high frequencies one derives dispersion relations, originally studied by Leontovich. In the case of dielectrics these relations, combined with the positivity of the imaginary parts, imply bounds on the low-energy values of the response, $\varepsilon(0,0)$ and $\mu(0,0)$. In particular the quantities $\varepsilon(0,0)-1$ and $\varepsilon(0,0) - 1/\mu(0,0)$ are constrained to be positive and equal to integrals over the imaginary parts of the response. We discuss various improvements of these bounds in the case of non-relativistic media and with additional assumptions about the UV behavior.
\vspace{0.5cm} \hrule}

\end{titlepage}

\setcounter{footnote}{0} 

%%%%%%%%%%%%%%%%%%%%%%%%%%%%%%%%%%%%%%%%%%%%%%%%%%%%%
\tableofcontents
%%%%%%%%%%%%%%%%%%%%%%%%%%%%%%%%%%%%%%%%%%%%%%%%%%%%%
\vspace{1cm}
\hrule height 0.15mm 
\vspace{1cm}

%%%%%%%%%%%%%%%%%%%%%%%%%%%%%%%%%%%%%%%%%%%%%%%%%%%%
\section{Introduction}
%%%%%%%%%%%%%%%%%%%%%%%%%%%%%%%%%%%%%%%%%%%%%%%%%%%%
The coefficients of operators in an effective field theory (EFT) are dictated by symmetries and by the requirement of a healthy low-energy theory. More interestingly, other constraints, not readily discernible from the low-energy regime, can be obtained by making general assumptions about the theory and its UV behavior. Under the mild assumptions that the UV completion is Lorentz-invariant, local and unitary, one can derive various inequalities that the low-energy coefficients must satisfy. The crucial link between low-energy properties and the UV completion lies in the analyticity of the $S$-matrix, which ultimately stems from microcausality -- the property that local operators commute outside the light cone. Cauchy's theorem allows one to relate low-energy and high-energy contours in the complex plane. This, together with the positivity of the imaginary parts of the amplitude implied by the optical theorem, establishes bounds on the low-energy coefficients. Ref.~\cite{Adams:2006sv} put forward these so-called ``positivity bounds'' and the general idea that ``not everything goes'' once one has made assumptions about the UV. This has been generalized in many subsequent papers that aim at separating EFT's that have a conventional UV completion from those that do not (see the recent \cite{Bellazzini:2020cot, Caron-Huot:2020cmc, Tolley:2020gtv, Arkani-Hamed:2020blm, Paulos:2016fap, EliasMiro:2022xaa} and references therein). These ideas have been used to constrain, for instance, higher-dimension operators in the Standard Model EFT (e.g.~\cite{EliasMiro:2023fqi}), massive gravity (e.g.~\cite{Bellazzini:2023nqj}) and operators that correct General Relativity (e.g.~\cite{Guerrieri:2022sod}).
 
While this program has progressed under the assumption of (linearly realized) Lorentz invariance -- a natural assumption in particle physics -- a significant portion of physics explores systems in which Lorentz invariance is spontaneously broken. In cosmology, a preferred reference frame always exists -- the one comoving with the energy density -- resulting in the breaking of Lorentz invariance in EFT's describing phenomena such as inflation or dark energy. Condensed matter, almost by definition, constitutes a field of study characterized by the spontaneous breaking of Lorentz invariance. Other examples include field theories at finite temperature and/or chemical potential, like the LHC quark-gluon plasma and ordinary fluids, or EFT's describing localized objects such as black holes or defects. Clearly, it would be valuable to derive bounds on EFT's without relying on Lorentz invariance.

Without Lorentz invariance, generalizing the arguments above is not straightforward (attempts in this direction include \cite{Baumann:2015nta,Grall:2021xxm,Freytsis:2022aho}). The issue is that, in general, the $S$-matrix cannot serve as the bridge between UV and IR. The $S$-matrix that describes the scattering of low-energy excitations might not even exist at high energy \cite{Creminelli:2022onn,Hui:2023pxc,Creminelli:2023kze}. In the absence of Lorentz invariance states cannot be boosted and, generally, low-energy excitations only exist up to a certain energy. Additionally, excitations are typically unstable so that the $S$-matrix is, even at low energies, an approximate concept at best. One can restrict to situations in which the low-energy excitations survive in the UV and are sufficiently stable so that the $S$-matrix exists at low and high energies. A simple example of this situation has been studied in \cite{Hui:2023pxc,Creminelli:2023kze}. Unfortunately, the conclusion is that the $S$-matrix, in the absence of Lorentz invariance, does not enjoy the analytic properties that allowed one to connect UV and IR in the presence of Lorentz invariance. The relation between the $S$-matrix and correlation functions is non-local and microcausality is not sufficient to derive nice properties for the scattering amplitude, like in the Lorentz-invariant case. Without analyticity connecting the UV and the IR becomes challenging, and drawing general conclusions seems difficult.

These difficulties with the $S$-matrix suggest to go back and study a simpler object, in which microcausality is manifest: the (retarded) two-point function. In the absence of Lorentz invariance, but preserving rotations and spacetime translations, the two-point function is a rich object; a function of two variables, $\omega$ and $|\bsk|$, similar to the $S$-matrix in the Lorentz-invariant case. Using the analytic properties of the two-point function of conserved currents and assuming the theory reaches a conformal fixed point in the UV, in \cite{Creminelli:2022onn} it was possible to derive positivity bounds on the operators that describe a conformal superfluid.

The possibility of deriving positivity bounds in the absence of Lorentz invariance should not come as a surprise. Indeed the use of analyticity and dispersion relations to connect UV with the IR goes back to Kramers and Kronig \cite{Kronig1926,Kramers1927} in their study of the electromagnetic response of media, in which Lorentz invariance is clearly broken. In this paper we wish to revisit this old problem, focussing on the case of dielectrics, from the more modern point of view of setting bounds on non-Lorentz-invariant EFT's.

Returning to this well-studied problem (see \cite{Keldysh1989TheDF} for an extended review) requires humility and a clarification of motivation. Firstly, to the best of our knowledge, some of the results, notably the dispersion relations \eqref{transvleon} and \eqref{epsmuleo} on the dielectric permittivity $\varepsilon$ and magnetic permeability $\mu$ at low frequency and momentum, as well as the summary in Fig.~\ref{bound}, are new. Secondly, we wish to describe the problem as an example of settings bounds on Lorentz-breaking EFT's: the assumptions and the language used in the condensed matter community are not immediately extendable to more general setups. For instance, the assumptions about the UV behavior of the system -- and even the very definition of UV -- are markedly different for a normal medium and, say, an EFT that describes cosmic inflation. The derivation of ``macroscopic" Maxwell equations in terms of the in-in effective action should facilitate the extension to other contexts. We sense a similarity with the Lorentz-invariant case: while dispersion relations for the $S$-matrix were routinely used since the `60s, their application to constrain EFT's is much more recent and required a change in perspective.

The Maxwell equations in a medium describe the evolution of the average electromagnetic field in a material and they can be derived using the in-in effective action formalism. We present this derivation in \S\ref{sec:EMmedia}, deferring some background material to Apps.~\ref{ctp} and \ref{linres}. The response of the medium can be parametrized in terms of $\varepsilon(\omega, |\bsk|)$ and $\mu(\omega, |\bsk|)$, where both the electric and magnetic responses depend both on frequency and wave-vector.\footnote{In specific cases it may be a good approximation to neglect the dependence of $\varepsilon$ and $\mu$ on $|\bsk|$, keeping only the one on $\omega$. For instance, in the interaction of visible light with standard matter, the wavelength of the photon is $\alpha^{-1}\simeq137$ times longer than the atomic size, so that one can approximate the response as local in space.} $S$-matrix positivity bounds always stem from the fact that the imaginary part of the amplitude has a definite sign. In the present context, the analogous statement is that the medium can only absorb and not emit energy when perturbed by an external electromagnetic field; we discuss this assumption in \S\ref{positiv}. As we argued above, the crucial ingredient to relate UV and IR is analyticity. In \S\ref{sec:analytic}, we study the domain of analyticity of the photon propagator in the medium: as a consequence of microcausality, the propagator is analytic when the imaginary part of the four-vector $(\omega, \bsk)$ lies in the forward light cone. This generalizes the textbook statement that a retarded Green's function must be analytic in the upper half-plane of complex $\omega$, which is a consequence of retardation only. Microcausality is clearly more restrictive and, correspondingly, one can generalize the celebrated Kramers-Kronig relations to a one-parameter family of relations, first derived by M. Leontovich \cite{Leon}. These relations are a general property of linear response theory (see for example \cite{Chaikin_Lubensky_1995} for an introduction to the subject) and, as such, they should perhaps receive more attention. They are crucial to constrain the electromagnetic response of a medium: the Kramers-Kronig relations only reflect causation, allowing for immediate response at a distance, while, as we will see, bounds on the magnetic response require microcausality.

Leontovich relations, like the Kramers-Kronig ones, require assumptions about the UV behavior. This is the topic of \S\ref{high}. For a concrete model that describes the UV limit of the material, we study in App.~\ref{linhar} the case of degenerate fermions: the so-called Lindhard function and its relativistic generalization. Notice that our study is not confined to condensed matter: for instance one could have a medium of nuclear matter in which the charged constituents are in relativistic motion. Microcausality (Leontovich relations) together with the assumption of a ``passive" medium (sign-definiteness of imaginary parts) can be combined to give bounds on the low-energy and momentum limit of $\varepsilon$ and $\mu$ (\S\ref{sec:bounds}). (Notice that this low-energy limit exists since we confine our analysis to dielectric materials; one would have divergences when studying conductors or superconductors.) We are unable to prove that the one-particle irreducible (1PI) self-energy, the object that directly appears in the macroscopic Maxwell equations, is analytic in the same region as the photon propagator (\S\ref{wider} and App.~\ref{landau}), unless some further physical assumption is added. We are however able to prove a more limited result regarding its domain of analyticity, which is sufficient to re-derive the low-energy bounds on $\varepsilon$ and $\mu$ via another route. Stronger bounds can be derived by making further assumptions (\S\ref{sec:stronger}). One way is to set a lower bound on dissipation, in the same way one does in the $S$-matrix program where some knowledge on the total cross-section sharpens the low-energy bounds. Another possible assumption is that the medium is non-relativistic, so that its response is actually confined in a smaller region compared to what is allowed by relativistic microcausality. We discuss the higher order terms in the low energy expansion of the response functions in \S\ref{conc}, where we conclude the paper with many open generalizations of these methods.

\paragraph{Notation} We work with mostly plus metric signature $(-+++)$. Our convention for Fourier transform is 
\eq{
\nonumber
f(p)=\int\dd[4]{x}e^{-ip\cdot x}f(x)\,,\qquad f(x)=\int\frac{\dd[4]{p}}{(2\pi)^4}e^{ip\cdot x}f(p)\,.
}

%%%%%%%%%%%%%%%%%%%%%%%%%%%%%%%%%%%%%%%%%%%%%%%%%%%%
\section{\label{sec:EMmedia}Effective Maxwell equations in matter}
%%%%%%%%%%%%%%%%%%%%%%%%%%%%%%%%%%%%%%%%%%%%%%%%%%%%
The electromagnetic dynamics in matter is governed by the following action,
\eq{
	S=S_\gamma[a]+S_M\big[a;\psi\big]\,.
	\label{fullaction}
}
The first part is the free photon action
\eq{
	S_\gamma=-\frac{1}{4g^2}\int\dd[4]{x}f_{\mu\nu}^2\,,
}
with $a_\mu$ the vector field and $f_{\mu\nu}=\p_\mu a_\nu-\p_\nu a_\mu$ the field strength, and we have pulled out the dimensionless coupling $g$. The second term in \cref{fullaction} represents the dynamics of matter in which all matter fields are collectively denoted by $\psi$. In what follows, we will not need to specify the precise form of $S_M$ apart from some minimal assumptions (for instance on the high-energy behavior of response functions, see \S\ref{high}). For massless spin-one fields we must have gauge symmetry, i.e.~$a_\mu\to a_\mu+\p_\mu\Lambda$ and $\psi\to\psi \, e^{iq\Lambda}$ for a charge $q$. Most of the time we will suppress dependences on the matter fields.

The action \cref{fullaction} specifies the dynamics but we still need to determine the state. Unlike the usual situation in high energy physics where the vacuum state and its excitations are studied, here we are interested in a many-body state described by a density matrix $\rho$. We assume that the system without external perturbations is in equilibrium, therefore $\rho$ commutes with the Hamiltonian.\footnote{In some cases, for example superfluids, a combination of time translation and an internal symmetry is broken into a diagonal subgroup. In these situations, we call the unbroken symmetry ``time translation". We thank A.~Podo for this comment.} Moreover, we assume that the system is homogeneous therefore it also commutes with the spatial components of the momentum operator. Notice that a generic density matrix, with finite average energy density, breaks Lorentz boosts.\footnote{In fact, Lorentz boosts mix different energy states in the expansion $\rho=\sum_nc_n\ket{n}\bra{n}$ given below and does not commute with the density matrix.} By contrast, we will assume that rotation is a good symmetry of the system.

A generic density matrix of this type can be written as $\rho=\sum_nc_n\ket{n}\bra{n}$ in terms of the eigenstates of the four-momentum operator $P^\mu\ket{n}=p_n^\mu\ket{n}$ and $c_n$ non-negative numbers satisfying $\sum_n c_n = 1$. The discussion will not crucially depend on the form of the density matrix but an example to have in mind is the grand canonical ensemble $\rho=\exp(-\beta(H-\mu Q))/Z$, with $Z=\Tr\exp(-\beta(H-\mu Q))$, $\beta$ the inverse temperature, $H$ the Hamiltonian, $\mu$ the chemical potential and $Q$ a $U(1)$ charge. In covariant form $\rho\propto\exp(\beta u_\mu(P^\mu-\mu J^\mu))$ with $u^\mu$ the normalized ($u_\mu u^\mu=-1$) velocity of the medium and $J^\mu$ the conserved current. In the rest frame of the medium $u^\mu=(1,\boldsymbol{0})$.

\paragraph{Effective action} We are interested in studying the evolution of the average electromagnetic field in the system, i.e.~$F_{\mu\nu}\equiv\ev{f_{\mu\nu}}=\Tr(\rho f_{\mu\nu})$. We emphasize that average fields are defined as an ensemble average rather than spatial average (as for instance in \cite{Jackson:1998nia}) since the latter are not well-defined at arbitrarily high energies. We assume that the amplitude of macroscopic fields is small and therefore that it is sufficient to consider the equation of motion linear in the fields -- equivalently the action is quadratic. However, the coupling to matter is not necessarily weak and therefore for most of the discussion we avoid performing any perturbative expansion in the matter sector.

For the purpose of studying the evolution of the average electromagnetic fields taking into account the presence of matter, the natural object to study is the Closed Time Path (CTP) effective action (also called Schwinger-Keldysh or in-in effective action). Varying the effective action gives the equation of motion for the average fields. We require the in-in, as opposed to in-out, effective action since it is only the former that describes a causal equation of motion. For instance, the equation of motion following from the in-out effective action can have complex solutions for real boundary conditions \cite{Calzetta:2008iqa}. Moreover, dissipation cannot be described in the usual in-out formalism. We review these concepts in App.~\ref{ctp}.

In the following we obtain an expression for the CTP effective action (hereafter simply called effective action) up to quadratic order using the background field method. As explained in App.~\ref{ctp}, this is a straightforward extension of the standard background field method to the CTP formalism. The purpose of this calculation is to obtain Maxwell's equations in matter from a top-down approach. The reader who is not interested in the details can jump directly to \cref{Max}.

A path integral representation of the effective action $\Gamma[A_1,A_2]$ is obtained by combining \cref{wback} and \cref{gamback} as follows:
\eq{
	e^{i\Gamma[A_1,A_2]}=\int^{\rm CTP}_{\rho,{\rm 1PI}} \D a_1\D a_2\D \psi_1\D \psi_2 \, \exp\bigg(iS[a_1+A_1,\psi_1]-iS[a_2+A_2,\psi_2]\bigg)\,.
	\label{Gammaeff}
} 
Here is an explanation of the relation: in the CTP formalism we calculate the path integral on a time contour going from the initial time to the final time, taken to be $-\infty$ and $+\infty$ respectively, and then back to the initial time (see the figure in \cref{phivevrho}). The fields that live on the forward and backward contours are labeled by indices 1 and 2 respectively. We integrate over all the field configurations (for the gauge and matter fields) subject to the boundary condition that at the final time the forward and backward fields match, i.e.~$a_1(t_f)=a_2(t_f)$ and $\psi_1(t_f)=\psi_2(t_f)$ (writing ``CTP" over the integral in \cref{Gammaeff} is a reminder of this final boundary condition). The initial condition is fixed by the density matrix $\rho$. To implement the initial condition we multiply the integrand by $\bra{a_1(t_i),\psi_1(t_i)}\rho\ket{a_2(t_i),\psi_2(t_i)}$.  $S[a,\psi]$ is the total action in \cref{fullaction} for the photon and matter. It is deformed by adding a background $A_1$ and $A_2$ for the photon.

Finally, the effective action, by construction, only produces 1PI diagrams in a perturbative expansion, which we are reminded of by the ``1PI" index in \cref{Gammaeff}. It means that we can drop from the beginning terms that can only generate non-1PI diagrams. That is why we have dropped the external current terms $K_{1,2}$ present in \cref{wback}. We emphasize that by 1PI we mean diagrams that remain connected after cutting an \emph{internal photon} line: it could be that the matter itself has self-interactions for which we must keep all of them even non-1PI ones.

A few comments about gauge invariance are in order. As is usual in the path integral approach to gauge theories, we add a gauge fixing term to the action in \cref{Gammaeff} and integrate over all components of the gauge fields, i.e.~$\D a_1\D a_2=\D a_1^\mu\D a_2^\mu$. We choose the gauge fixing term to be $S_{\rm gf}[a]=-\frac{1}{2}\alpha\int(\p_\mu a^\mu)^2$ for an arbitrary coefficient $\alpha$. The exponent in \cref{Gammaeff} is then $S_\gamma[a_1+A_1]+S_M[a_1+A_1,\psi_1]+S_{\rm gf}[a_1]$, similarly for the $a_2$ fields, without shifting the gauge fixing part. Any dependence on the parameter $\alpha$ must drop out in a physical quantity. It is easy to see that the effective action is invariant under the separate gauge transformations of the background fields
\eq{
	\Gamma[A_1+\p\Lambda_1,A_2+\p\Lambda_2]=\Gamma[A_1,A_2]\,,
}
for arbitrary functions $\Lambda_1$ and $\Lambda_2$ (we assume they vanish at infinity to make sure the boundary conditions are not altered).

We would like to have an expression for the effective action $\Gamma[A_1,A_2]$ up to quadratic order in the background fields (but otherwise non-perturbative). Dropping vertices generating non-1PI terms, we obtain
\eq{
	S[a+A,\psi]=S[a,\psi]+S_\gamma[A]+\int\dd[4]{x}J^\mu(a,\psi) A_{\mu}+\frac{1}{2}\int\dd[4]{x}N^{\mu\nu}(a,\psi) A_\mu A_\nu+\dots\,,
} 
in which we have defined 
\eq{
J^\mu\equiv\fdv{S_M[a,\psi]}{a_\mu(x)}\,,\qquad\qquad N^{\mu\nu}\delta(x-y)\equiv\fdv{J^\mu[a(x),\psi(x)]}{a_\nu(y)}\,.
\label{JNdef}
}
Both terms can depend on the internal photon fields $a_\mu$ and the matter fields $\psi$. In the definition of $N^{\mu\nu}$ we have assumed that the current is a local function of the fields. We provide some examples for clarification. In scalar QED, the relevant part of the matter action is $S_M=\int -|D_\mu\psi|^2$ with $\psi$ a complex scalar and $D_\mu=\p_\mu-ia_\mu$.  Then we have $J^\mu=-i(\psi^\dagger D^\mu\psi-\psi (D^\mu\psi)^\dagger)$ and $N^{\mu\nu}=-2|\psi|^2\eta^{\mu\nu}$. For fermions, $S_M=\int i\bar{\psi}\slashed{D}\psi$ with $\psi$ a Dirac spinor. Then we obtain $J^\mu=\bar{\psi}\gamma^\mu\psi$ and $N^{\mu\nu}=0$.

Since the $S_\gamma[A]$ term only depends on the background field we factor it out in \cref{Gammaeff} and write the effective action as follows,
\eq{
	\Gamma[A_1,A_2]=S_\gamma[A_1]-S_\gamma[A_2]+\Gamma_M[A_1,A_2]\,,
	\label{gammaA1A2}
} 
in which the last term is the contribution of matter and is given by
\eq{
	e^{i\Gamma_M[A_1,A_2]}=\int_{\rho,{\rm 1PI}}^{\rm CTP} \D a_1\D a_2\D \psi_1\D \psi_2 \,\exp(iS[a_1,\psi_1]+i\int J_1^\mu A_{1\mu}+\frac{i}{2}\int N_1^{\mu\nu} A_{1\mu} A_{1\nu} - \{1\to2\})\,.
	\label{gammaM}
}
The leading term, corresponding to $A_{1,2}=0$, vanishes due to the normalization of the effective action. The linear term would be
\eq{
	i\int^{\rm CTP}_{\rho,{\rm 1PI}}e^{i(S_1-S_2)}\int\dd[4]{x}(J_1^\mu A_{1\mu}-J_2^\mu A_{2\mu})=i\int\dd[4]{x}\ev{J^\mu(x)}_{\rm 1PI}(A_{1\mu}-A_{2\mu})\,,
	\label{1ptJ}
}
where we have suppressed the measure of the path integral and $S_1=S[a_1,\psi_1]$, etc. By translation invariance $\ev{J^\mu}_{\rm 1PI}\propto u^\mu$ is independent of the coordinates and can only be proportional to the $u^\mu$, the four-velocity of the medium. This term is either zero, e.g.~for neutral fluids, or it is canceled by a homogeneous background with opposite charge, e.g.~electrons in a solid.\footnote{For electrons in a solid the assumption of homogeneity is not completely correct since the presence of a lattice of ions breaks spatial translations to a discrete subgroup. Most often this effect is ignored by assuming a homogeneous background with the opposite charge without any dynamics, known as the Jellium model. For a discussion of possible granularity effects see \cite{Keldysh1989TheDF}.} We must add a similar term to the effective action
\eq{
\int\dd[4]{x}(A_{1\mu}-A_{2\mu})J_{\rm ext}^\mu
\label{extJ}
}
to model the external current controlled by the experimentalist, e.g.~charges on a capacitor. The quadratic terms can be written as follows:
\eq{
	\Gamma_M[A_1,A_2]=\frac{1}{2}\int\dd[4]{x}\dd[4]{y}\begin{bmatrix}
		A_{1\mu}(x) \,\, A_{2\mu}(x)
	\end{bmatrix}S^{\mu\nu}(x,y)\begin{bmatrix}
		A_{1\nu}(y)\\ A_{2\nu}(y)
	\end{bmatrix}+\dots
\label{GM}
}
in which
\eq{
	S^{\mu\nu}(x,y)=i\begin{bmatrix}
		\ev{TJ^\mu(x)J^\nu(y)}	& -\ev{J^\nu(y)J^\mu(x)}\\
		-\ev{J^\mu(x)J^\nu(y)} & \ev{\bar{T}J^\mu(x)J^\nu(y)}
	\end{bmatrix}_{\rm 1PI}+\ev{N^{\mu\nu}}_{\rm 1PI}\delta(x-y)\begin{bmatrix}1&0\\0&-1\end{bmatrix}\,,
\label{Smunu}
}
where we have used \cref{generalSK} to write the coefficient matrix $S^{\mu\nu}$ in terms of the correlation functions of the current $J^\mu$ plus contact terms.\footnote{It is useful to compare \cref{GM} with similar expressions in App.~\ref{ctp}. First of all, in both \cref{W2} and \cref{Gamma2}, unlike in \cref{GM}, we have used the $r/a$ representation. More importantly, the reader should note that \cref{GM} has similarities and differences with both \cref{W2} and \cref{Gamma2}. It is similar to \cref{W2} because there is an external field $A_\mu$ (similar to $K$ in \cref{W2}) in which we are expanding. The difference is that in \cref{W2} there are external currents, needed to perform the Legendre transform. That is why instead of simple correlation functions of the current in \cref{Smunu} we end up having only the 1PI part. The presence of these external currents is the similarity to \cref{Gamma2} while the difference is that in \cref{GM} we have not included the free part of the action.} Diagrammatically this corresponds to the matter corrections to the effective action with two external legs shown in \cref{1PIblobs}.
\fg{
\includegraphics[scale=1.5]{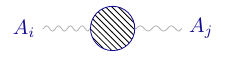}
\includegraphics[scale=1.5]{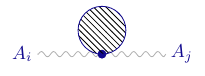}	
	\caption{Schematic diagrams associated to the contribution from matter to the quadratic effective action. The blobs correspond to correlation functions of $J$ (left) and the contact term (right). External propagators are not included in the diagrams.}
	\label{1PIblobs}
}
The effective Maxwell equation is obtained by varying $\Gamma[A_1,A_2]$ with respect to $A_1$ (or equivalently $A_2$) and then setting $A_1=A_2=A$ (see also  \cref{eqgamma}), which will be the value of the average field. Doing so we obtain
\begin{eBox}
\eq{
	\frac{1}{g^2}\p_\alpha F^{\alpha\mu}+\int\dd[4]{y} \Pi^{\mu\nu}(x,y)A_{\nu}(y)=-J_{\rm ext}^\mu(x)\,,
	\label{Max}
}
\end{eBox}
where we have also added the external current as explained in \cref{extJ}.\footnote{\label{fluc}The quadratic CTP effective action contains the information about all the two-point functions including the fluctuations. In fact a standard approach is to solve for the variable $A_a\equiv A_1-A_2$ and obtain a Langevin equation sourced by a noise term. In this work we focus on the average fields and neglect fluctuations. An interesting question would be to study properties of the fluctuations through the fluctuation-dissipation theorem. See \S\ref{conc}.} The influence of matter is captured by the second term,
\eq{
	\Pi^{\mu\nu}(x,y)=i\theta(x^0-y^0)\ev{[J^\mu(x),J^\nu(y)]}_{\rm 1PI}+\ev{N^{\mu\nu}}_{\rm 1PI}\delta(x-y)\,.
	\label{PiJJ}
} 
Eq.~(\ref{Max}) is the Maxwell equation in matter and its solution gives the average fields. As mentioned above, here we have restricted to the quadratic effective action which is a good approximation in situations with weak electromagnetic field compared to, for instance, typical interatomic electromagnetic fields.\footnote{For real materials, the interatomic electric field could be estimated as $\sim e^2/a_0 \sim 10^2$ eV in which $e$ is the charge and $a_0$ is the Bohr radius. This turns out to be large compared to what can be produced experimentally.} By considering cubic or higher-order terms in the effective action one can study nonlinear effects (in the average field) as in \cref{egamma3}. We emphasize that by construction, the effective action contains the information at all-loop orders in a perturbative expansion.

Comparison of \cref{Max} to the general form of the effective equation of motion in \cref{egamma3} shows that the retarded Green's function of the photon,
\eq{
\nonumber
G_\gamma^{\mu\nu}\equiv i\theta(x^0-y^0)\ev{[a^\mu(x),a^\nu(y)]} \,,
}
is expressed as
\eq{
	(G_\gamma^{-1})^{\mu\nu}=\frac{1}{g^2}(\Delta^{-1})^{\mu\nu}-\Pi^{\mu\nu}\,,
	\label{photongreen}
} 
with $\Delta^{\mu\nu}$ the free retarded photon propagator. $\Pi^{\mu\nu}$ is called the photon self-energy tensor.

\paragraph{Self-energy tensor} The self-energy can compactly be expressed as the second variation of the matter effective action,
\eq{
\Pi^{\mu\nu}(x,y)=\frac{\delta^2\Gamma_M[A_r,A_a]}{\delta A_{a,\mu}(x)\delta A_{r,\nu}(y)}\Bigg\vert_{A_a=0,A_r=0}\,,
\label{PiGM}
}
where we have defined $A_r\equiv(A_1+A_2)/2$ and $A_a\equiv A_1-A_2$ (known as the $r/a$ or physical representation, discussed also in App.~\ref{ctp}). Reality of the effective action, $\Gamma_M[A_r,A_a]^*=-\Gamma_M[A_r,-A_a]$, implies that the self-energy is real, $\Pi^{\mu\nu}{}^*=\Pi^{\mu\nu}$. It ensures that solutions to the effective Maxwell equation \cref{Max} with real sources and boundary conditions are real. Gauge invariance of the effective action implies that the self-energy is transverse,
\eq{
	\p_{x^\mu}\Pi^{\mu\nu}(x,y)=\p_{y^\nu}\Pi^{\mu\nu}(x,y)=0\,.
	\label{trans}
}
This ensures that if $A_\mu$ is a solution of \cref{Max} then $A_\mu+\p_\mu\Lambda$ is also a solution. Moreover, by translation symmetry self-energy is only a function of the distance, i.e.~$\Pi^{\mu\nu}(x,y)=\Pi^{\mu\nu}(x-y)$. In Fourier space 
\eq{
	\Pi^{\mu\nu}(p)=\int\dd[4]{x}e^{-ip\cdot x}\Pi^{\mu\nu}(x)\,,
}
with $p^\mu$ the four-momentum vector. As discussed above, Lorentz boosts are broken by the medium so $\Pi^{\mu\nu}$ can be a function of Lorentz-invariant combinations made out of $p^\mu$ and the four velocity $u^{\mu}$ of the medium. Useful combinations are
\eq{
	\omega\equiv -p^\mu u_\mu\,,\qquad k^2\equiv p^\mu p_\mu+(p^\mu u_\mu)^2\,.
	\label{Lsc}
}
In the rest frame of the medium $u^{\mu}=(1,\boldsymbol{0})$ and then $\omega=p^0$ is the energy and $k^2 = \boldsymbol{p}^2$ is the magnitude of spatial momentum-squared. For this reason, most often we will simply write the four-momentum as $p^\mu=(\omega,\vbk)$ and the self-energy as $\Pi^{\mu\nu}(\omega,k)$.  

The condition \cref{trans} in Fourier space implies that $p_\mu\Pi^{\mu\nu}=0$. In the Lorentz-invariant case, this condition fixes the whole tensor structure up to a function known as the vacuum polarization. In the presence of a medium we will have two functions as given below. The projection matrix onto the subspace transverse to $p^\mu$ is defined as
\eq{
	\PP^{\mu\nu}\equiv\eta^{\mu\nu}-\frac{p^\mu p^\nu}{p^2}\,.
}  
We denote the projection of the medium four-velocity onto this subspace by $\bar{u}^\mu\equiv \PP^{\mu\nu}u_\nu=u^\mu+\frac{\omega}{p^2}p^\mu$. We further decompose this subspace into longitudinal and transverse parts, with respect to $\bar{u}^\mu$, using the projection matrices
\eq{
	\PP_L^{\mu\nu}\equiv\frac{\bar{u}^\mu\bar{u}^\nu}{\bar{u}^2}\,,\qquad \PP_T^{\mu\nu}\equiv \PP^{\mu\nu}-\PP_L^{\mu\nu}\,.
} 
In the rest frame of the medium these projection matrices have the following components\footnote{Another way to think about the decomposition is as follows. The condition of gauge invariance fixes $\Pi^{00}$ in terms of $\Pi^{0i}$, and $\Pi^{0i}$ in terms of the $\Pi^{ij}$ components. Then the spatial part of the tensor is decomposed into pieces in the direction of $\boldsymbol{k}$ (longitudinal part) and orthogonal to it (transverse part).}
\eqa{
	&\PP_L^{00}=-\frac{k^2}{p^2}\,,\qquad \PP_L^{0i}=-\frac{\omega k^i}{p^2}\,,\qquad \PP_L^{ij}=-\frac{\omega^2}{p^2}\frac{k^ik^j}{k^2}\,,\label{pLrest}\\
	&\PP_T^{00}=\PP_T^{0i}=0\,,\qquad \PP_T^{ij}=\delta^{ij}-\frac{k^ik^j}{k^2}\label{pTrest}\,.
}
The generic form of the self-energy tensor is then\footnote{\label{p}Another possible term is a projector constructed out of $\varepsilon^{\mu\nu\rho\sigma}$. The presence of this term signals that the medium breaks parity, for instance because of the presence of chiral molecules (e.g.~sugar), and distinguishes between right-handed and left-handed polarizations of the photon. See \cite{Nieves:1988qz}. Materials with this effect are called optically active. In real materials this effect does not survive in the limit $\omega \to 0$, which will be the main focus of this paper. Therefore, we do not consider such terms here. An analogous effect is cosmic birefringence with the coupling $\phi F_{\mu\nu}\tilde{F}^{\mu\nu}$ studied for instance in \cite{Carroll:1998zi,Nakai:2023zdr}. The main difference is that in this case one of the photon polarizations becomes unstable (which is harmless in cosmology since the instability gets regularized by the Hubble scale).}
\eq{
	\Pi^{\mu\nu}=-\pi_L(\omega,k)p^2\PP_L^{\mu\nu}+\pi_T(\omega,k)k^2\PP_T^{\mu\nu}\,,
	\label{Pi}
}
where the prefactors are for later convenience. The two functions $\pi_L(\omega,k)$ and $\pi_T(\omega,k)$ model the response of the medium to an external current according to \cref{Max} at frequency $\omega$ and momentum~$k$.

\paragraph{Phenomenological definitions} The discussion of the previous section shows that the effect of the medium can be described by two functions $\pi_L$ and $\pi_T$ which are related to the correlation function of the current as given by \cref{PiJJ}. Phenomenologically the medium is usually modeled through the electric permittivity and magnetic permeability. In this section we relate these two descriptions. See for instance \cite{Keldysh1989TheDF}. The electric and the magnetic field are defined as\footnote{Alternatively, one can define four-vectors $\tilde{E}^\mu\equiv-u_\alpha F^{\alpha\mu}$ and $\tilde{B}^\mu\equiv-\varepsilon^{\alpha\beta\lambda\mu}F_{\alpha\beta}u_\lambda/2$ which in the rest frame reduce to \cref{EBdef}. However, in a boosted frame $\tilde{E}$ and $\tilde{B}$ depend on both $\boldsymbol{E}$ and $\boldsymbol{B}$. One can show that $F^{\mu\nu}=(u^\mu \tilde{E}^\nu-u^\nu \tilde{E}^\mu)+\varepsilon^{\mu\nu\alpha\beta}\tilde{B}_\alpha u_\beta$.}
\eq{
	E^i=F^{0i}\,,\qquad B^i=\frac{1}{2}\varepsilon^{ijk}F_{jk}\,.
	\label{EBdef}
}
It is easy to check that the electric field has transverse and longitudinal components but the magnetic field is only transverse. The effect of the medium is modeled in terms of an induced current. This is compatible with the effective Maxwell equation obtained in \cref{Max} re-written as
\eq{
	\frac{1}{g^2}\p_\nu F^{\nu\mu}=-(J_{\rm in}^\mu+J_{\rm ext}^\mu)\,,\qquad J_{\rm in}^\mu(x)\equiv\int\dd[4]{y} \Pi^{\mu\nu}(x,y)A_{\nu}(y)\,,
	\label{Maxph}
}
where $J_{\rm in}^\mu=(\rho_{{\rm in}},\boldsymbol{J}_{{\rm in}})$ is the induced current. Since $p_\mu\Pi^{\mu\nu}=0$, the induced current is conserved. The conventional approach is to relate $J^\mu_{\rm in}$ to the electric and magnetic fields, written in the rest frame, as follows\footnote{This is the most general definition assuming invariance under parity. See footnote~\ref{p}.}
\eqa{
	&g^2\rho_{{\rm in}}\equiv(1-\varepsilon)\p_i E^i\,,\label{Jinl}\\ &g^2J_{{\rm in},T}^i\equiv(\tilde{\varepsilon}-1)\p_t E_T^i+\left(1-\frac{1}{\tilde{\mu}}\right)\varepsilon^{ijk}\p_j B_{k}\label{JinT}\,,
}
in which the index $T$ refers to the transverse part of the vector. Notice that there is no need to have an expression for $J_{{\rm in},L}^i$ since it is given by \cref{Jinl} using the current conservation. Moreover, $\varepsilon$, $\tilde{\varepsilon}$ and $\tilde{\mu}$ must be considered as (non-local) operators acting on the right-hand side; in Fourier space they will be general functions of $\omega$ and $k$. We obtain the familiar Maxwell equations in the presence of matter,
\eq{
\varepsilon \, \boldsymbol{\nabla}\cdot\boldsymbol{E}=g^2\rho_{\rm ext}\,,\qquad\qquad \frac{1}{\tilde{\mu}}\boldsymbol{\nabla}\times \boldsymbol{B}-\tilde{\varepsilon}\, \p_t\boldsymbol{E}_T=g^2\boldsymbol{J}_{{\rm ext},T}\,.
}
It is important to note that the definition \cref{JinT} is ambiguous. In fact one can show, using the source-independent Maxwell equations, that the transformation $\tilde{\varepsilon}\to\tilde{\varepsilon}+\delta f$ and $1/\tilde{\mu}\to1/\tilde{\mu}+\delta f\p_t^2/\boldsymbol{\nabla}^2$ for any $\delta f$ leaves the equation invariant.\footnote{The change in the equation will be $\delta f\p_t(\p_t\boldsymbol{\nabla}\times\boldsymbol{B}/\boldsymbol{\nabla}^2-\boldsymbol{E}_T)$ which vanishes using $\boldsymbol{\nabla}\times\boldsymbol{E}+\p_t\boldsymbol{B}=0$.} 
Therefore one has to pick a convention. The two most widely used ones are:
\begin{itemize}
	\item Set $\tilde{\mu}=1$ and only keep $\tilde{\varepsilon}$ which in this context is usually called $\varepsilon_T$. In this case, the entire induced current is written in terms of the electric field. The two quantities $\varepsilon$ and $\varepsilon_T$ are sometimes packed into a tensor called the dielectric tensor.
	\item Set $\tilde{\varepsilon}=\varepsilon$. In this context we simply write $\tilde{\mu}$ as $\mu$. 
\end{itemize}
In the following we will mainly work with the second convention. Using the definition in \cref{Maxph}, after a bit of algebra, one can relate the above quantities to $\pi_L$ and $\pi_T$ as follows\footnote{In the first convention, the transverse part of the dielectric tensor $\varepsilon_T$ can be obtained from \cref{epsmu} by using the transformation rule discussed above. More precisely by choosing $\delta f=k^2/\omega^2(1-\mu^{-1})$ we force $\tilde{\mu}=1$ and get $\varepsilon_T=1+g^2k^2\pi_T/\omega^2$.\label{eT}}:
\begin{eBox}
\eq{
	\varepsilon-1=g^2\pi_L\,,\qquad 1-\frac{1}{\mu}=g^2\left(\pi_T-\frac{\omega^2}{k^2}\pi_L\right)\,.
	\label{epsmu}
}
\end{eBox}
For later usage, we note that $g^2k^2\pi_T=p^2+\omega^2\varepsilon-k^2/\mu$. By using the above relations we can re-write the effective action defined in \cref{gammaA1A2} with matter part given in \cref{GM}. After a bit of algebra, using the $r/a$ representation defined below \cref{PiGM}, we obtain
\eq{
\Gamma[A_r,A_a]=\frac{1}{g^2}\int\frac{\dd[4]{p}}{(2\pi)^4}\left[\varepsilon(\omega,k)\boldsymbol{E}_a(-p)\cdot \boldsymbol{E}_r(p)-\frac{1}{\mu(\omega,k)}\boldsymbol{B}_a(-p)\cdot \boldsymbol{B}_r(p)+\dots\right]\,,
\label{effEMaction}
}
in which we used the definition of electric and magnetic field given in \cref{EBdef} for both $A_a$ and $A_r$ accordingly. We have also resorted to Fourier space for simplicity. The dots in \cref{effEMaction} stand for a bunch of terms. First, terms of the form $\sim A_a A_a$ which capture fluctuations (see footnote~\ref{fluc}). Second, we should add a coupling to the external current which we have suppressed here. Finally, there are also higher-order terms, e.g.~$\propto A^3$, which we neglect. We emphasize  that we define $\varepsilon$ and $\mu$ as the coefficients appearing in the effective action that are in principle well-defined at all energies.
It may happen that in some situations $\varepsilon$ and $\mu$ are not the best variables to work with. For instance, in a conductor it is more convenient to define a conductivity tensor. In those cases, one can easily translate the discussion of this paper in terms of the more appropriate variables.

\paragraph{Photon propagator} We can write an expression for the retarded photon propagator in the medium. The free photon propagator, taking into account the gauge fixing term, is given by
\eq{
	\Delta^{\mu\nu}=\frac{1}{p^2}\left(\PP^{\mu\nu}+\frac{1}{\alpha}\frac{p^\mu p^\nu}{p^2}\right)\,,
	\label{freephoton}
}
where $\alpha$ depends on the gauge choice and we must use the correct $i\epsilon$ prescription for the retarded Green's function, i.e.~$p^2=-(\omega+i\epsilon)^2+k^2$. From \cref{photongreen} we obtain the photon retarded Green's function in the medium\footnote{The inverse of a tensor $a\PP_T^{\mu\nu}+b\PP_L^{\mu\nu}+cp^\mu p^\nu/p^2$ is $\PP_T^{\mu\nu}/a+\PP_L^{\mu\nu}/b+p^\mu p^\nu/cp^2$ because different projectors are orthogonal to each other.}
\eq{
	\frac{1}{g^2}G_\gamma^{\mu\nu}=\frac{\PP_T^{\mu\nu}}{p^2-g^2k^2\pi_T}+\frac{\PP_L^{\mu\nu}}{p^2(1+g^2\pi_L)}+\frac{1}{\alpha}\frac{p^\mu p^\nu/p^2}{p^2}\,.
	\label{fullphoton}
}
Since the electromagnetic field is always coupled to a conserved current the last term in \cref{fullphoton} drops out and therefore the physical part of the propagator is independent of $\alpha$. Dispersion relations for propagating degrees of freedom correspond to poles of the propagator or equivalently, zeros of $(G_\gamma^{\mu\nu})^{-1}$. Calculating the inverse we obtain the following dispersion relations\footnote{Notice that, naively, the longitudinal dispersion relation must read $p^2\varepsilon=0$. However, $p^2$ cancels out by the same factor in the longitudinal projector as given in \cref{pLrest}.}
\eq{
\varepsilon(\omega,k)\omega^2-\frac{1}{\mu(\omega,k)}k^2=0\,,\qquad\qquad
	\varepsilon(\omega,k)=0\,,
}
corresponding to the transverse and longitudinal parts respectively. In the weak coupling limit the latter corresponds to a collective behavior of the charged particles known as plasma oscillations \cite{book:plasmapysics}.  

\paragraph{Linear response} It is useful to discuss a related but slightly different situation. Let us assume that we put the system in a given external electromagnetic field. This induces a current in the system. The question is how this current is related to the applied field at linear order. More formally, we are looking at the action $S_\gamma[a]+S_M\big[a+A_{\rm ext};\psi\big]$ for the applied electromagnetic field $A^\mu_{\rm ext}$ (we drop the kinetic term for $A^\mu_{\rm ext}$ since its dynamics is controlled by the experimentalist) and we are after the response of the system. As discussed in App.~\ref{linres} we have
\eq{
J_{\rm in}^\mu(x)=\int\dd[4]{y}G_J^{\mu\nu}(x,y)A_{{\rm ext},\nu}(y)\,,
\label{resext}
} 
with the response function
\eq{
G_J^{\mu\nu}(x,y)=i\theta(x^0-y^0)\ev{[J^\mu(x),J^\nu(y)]}+\ev{N^{\mu\nu}}\delta(x-y)\,,
\label{Gj}
}
in which $J^\mu$ is the current operator and $N^{\mu\nu}$ captures the dependence of the current on the electromagnetic field as defined in \cref{JNdef}. The response function $G_J^{\mu\nu}$ looks similar to the self-energy $\Pi^{\mu\nu}$ as given in \cref{PiJJ} with the crucial difference that the latter includes only 1PI terms. In fact comparison between \cref{resext} and \cref{Maxph} reveals that $G_J^{\mu\nu}$ relates the induced current to the applied field, while $\Pi^{\mu\nu}$ relates it to the total field. A relation between the two is easily obtained by noting that the total field is given by the external field plus the field produced by the induced current,
\eq{
	A^\mu(x)=A_{\rm ext}^\mu(x)+g^2\int\dd[4]{y}\Delta^{\mu\nu}(x-y)J_{{\rm in},\nu}(y)\,.
	\label{AAextJin1}
} 
Combination of \cref{resext}, \cref{AAextJin1} and \cref{Maxph} shows that\footnote{This relation can also be derived more rigorously using the fact that $\Pi\sim \frac{\delta^2\Gamma_M}{\delta A^2}$ and actually going through the Legendre transform from the generating function $W$ (in the notation of App.~\ref{ctp}) and expanding the currents.} 
\eq{
	(\Pi^{-1})^{\mu\nu}=(G_J^{-1})^{\mu\nu}+g^2\Delta^{\mu\nu}\,,
	\label{mos}
}
where the $\alpha$-dependent part of the photon propagator is dropped because it cancels from \cref{AAextJin1} due to current conservation. In terms of the transverse and longitudinal parts we have
\eq{
	G_J^{\mu\nu}=\frac{p^2k^2\pi_T}{p^2-g^2k^2\pi_T}\PP_T^{\mu\nu}-\frac{p^2\pi_L}{1+g^2\pi_L}\PP_L^{\mu\nu}\,.
	\label{fullJJ}
}
\fg{
	\begin{equation}\nonumber
	\adjustbox{valign=c}{\scalebox{0.85}{\includegraphics{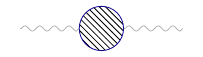}}}+
	\adjustbox{valign=c}{\scalebox{0.85}{\includegraphics{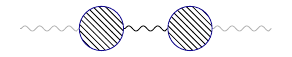}}}+
	\adjustbox{valign=c}{\scalebox{0.85}{\includegraphics{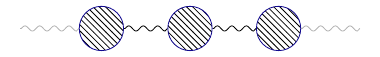}}}+\dots=\dfrac{\adjustbox{valign=c,margin*=-5mm 0mm -4mm 0mm}{\scalebox{1}{\includegraphics{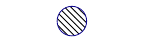}}}}{1-\scalebox{1}{\adjustbox{valign=c,margin*=-3mm 0mm -8mm 0mm}{\includegraphics{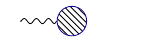}}}}
	\end{equation}
	\caption{The retarded Green's function of the current $G_J^{\mu\nu}$ is given by the resummation of the self-energy tensor $\Pi^{\mu\nu}$ with the free photon propagator $\Delta^{\mu\nu}$. External propagators (with lower opacity) are for shown for clarity and are not included in the expression. The blob corresponds to $\Pi^{\mu\nu}$ (including the contact term).}
	\label{resum}
}
In fact, as indicated in \cref{resum}, this relation is the familiar resummation of the bubble diagrams.
 
%%%%%%%%%%%%%%%%%%%%%%%%%%%%%%%%%%%%%%%%%%%%%%%%%%%% 
\section{Positivity and dissipation}\label{positiv}
%%%%%%%%%%%%%%%%%%%%%%%%%%%%%%%%%%%%%%%%%%%%%%%%%%%%
Applying an external electromagnetic field to the system changes its energy. We have derived the general expression for the change in the energy of the system as a consequence of an applied background field in \cref{deltaH}. For the background field $A_{\rm ext}^\mu$, this gives\footnote{Equivalently we obtain the same result for $\Im G_\gamma^{\mu\nu}$ using $J_{\rm ext}^\mu$ as the external source and $G_\gamma^{\mu\nu}$ as the linear response.}  
\eq{
	\Delta H=\int\frac{\dd[4]{p}}{(2\pi)^4}\omega A_{{\rm ext},\mu}(-p)\Im G_J^{\mu\nu}(p)A_{{\rm ext},\nu}(p)\,,
} 
where $G_J^{\mu\nu}$ is the response function given in \cref{Gj}. Here we are assuming that the system is invariant under spacetime inversion, i.e.~parity and time reversal, and therefore the expression appearing in the integrand of \cref{deltaH}, known as the dissipative part of the response function, is equal to its imaginary part.\footnote{As discussed below \cref{Trev}, this requires some knowledge about the phase factor under spacetime inversion of the operator $J^\mu=\delta S_M/\delta A_\mu$. Since $A^\mu$ describes a spin-one particle which is its own anti-particle, all phases associated to spacetime inversion must be real, i.e.~$\pm1$ (see \cite{Weinberg:1995mt}). For the photon it is +1.} Generally $\Delta H$ could be both positive or negative. However, we restrict ourselves to situations in which the system only absorbs energy from the source and therefore $\Delta H\geq0$. A system with this property is known as passive. This implies that $\omega\Im G_J^{\mu\nu}(p)$ is a positive-(semi)definite matrix. Notice that by reality of the response function, we conclude that $G_J^{\mu\nu}(p)^*=G_J^{\mu\nu}(-p)$ and therefore $\Im G_J^{\mu\nu}(p)$ is odd in $\omega$.

It is instructive to have a microscopic understanding of this result. First of all notice that the contact term in \cref{Gj} does not contribute to the imaginary part. Moreover, we have 
\eq{
\spl{
	\Im G^{\mu\nu}_J(p)&=\frac{1}{2i}\int\dd[4]{x}e^{-ip\cdot x}\bigg(G^{\mu\nu}_J(x)-G^{\mu\nu}_J(-x)\bigg)\\
	&=\frac{1}{2}\int\dd[4]{x}e^{-ip\cdot x}\bigg(\theta(t)\ev{[J^\mu(x),J^\nu(0)]}-\theta(-t)\ev{[J^\nu(-x),J^\mu(0)]}\bigg)\\	
	&=\frac{1}{2}\int\dd[4]{x}e^{-ip\cdot x}\ev{[J^\mu(x),J^\nu(0)]}\,.
	\label{imGab}
}}
In the second line we have used the fact that the response function is symmetric under $\mu\leftrightarrow\nu$ (see \cref{Trev}) and the contact term does not contribute. To obtain the last line we have used translation symmetry and $\theta(t)+\theta(-t)=1$. In other words, the imaginary part of the linear response function in Fourier space is given by the Fourier transform of the commutator of the operators without the theta function. By inserting a resolution of the identity we obtain
\eq{
	\spl{
		2\Im G^{\mu\nu}_J(p)&=\int\dd[4]{x}e^{-ip\cdot x}\Tr(\rho\big(J^\mu(x)J^\nu(0)-J^\nu(0)J^\mu(x)\big))\\
		&=\sum_n\int\dd[4]{x}e^{-ip\cdot x}\bra{n}\rho\big(e^{-iP\cdot x}J^\mu(0)e^{iP\cdot x}J^\nu(0)-J^\nu(0)e^{-iP\cdot x}J^\mu(0)e^{iP\cdot x}\big)\ket{n}\\
		&=\sum_{n,m}c_n\bigg(\int\dd[4]{x}e^{-i(p+p_n-p_m)\cdot x}\bra{n}J^\mu(0)\ket{m}\bra{m}J^\nu(0)\ket{n}-(\mu\leftrightarrow \nu,p_n\leftrightarrow p_m)\bigg)\\
		&=\sum_{n,m}(2\pi)^4\delta(p+p_n-p_m)\bra{n}J^\mu(0)\ket{m}\bra{m}J^\nu(0)\ket{n}(c_n-c_m)\,.
	}
}
In the second line we have used the definition of the trace and the spacetime translation operator $J^\mu(x)=e^{-iP\cdot x}J^\mu(0)e^{iP\cdot x}$. In the third line we have introduced a resolution of the identity between the two currents and used $\rho\ket{n}=c_n\ket{n}$. For the last line we have integrated over $x$ and changed $m\leftrightarrow n$ in the second term. We construct a symmetric object by contracting with a set of real vectors $V^\mu(p)$ and obtain
\eq{
	V_\mu(p)V_\nu(p)\Im G^{\mu\nu}_J(p)=\frac{1}{2}\sum_{n,m}(2\pi)^4\delta(p+p_n-p_m) \, |\bra{n}J^\mu(0)\ket{m}V_\mu(p)|^2(c_n-c_m)\,.
	\label{imGmunu}
}
We can see that for a generic choice of the $c_n$ coefficients $\Im G^{\mu\nu}_J(p)$ does not have a definite sign. However, if we assume that the $c_n$ are monotonically decreasing functions of the energies $E_n$ then we can argue that the above quantity has a definite sign as follows. For $\omega>0$ the argument of the delta function allows only for terms with $E_n<E_m$. Therefore the right-hand side is positive (similarly negative for $\omega<0$). This condition holds for most interesting cases such as finite temperature\footnote{In the presence of a conserved quantity we expand in terms of the eigenstates of $P^\mu-\mu J^\mu$.} with $c_n=e^{-\beta E_n}$.\footnote{Violations of this condition are possible if the system is prepared such that high-energy levels are more populated than low-energy ones: in this case the $c_n$ coefficients will not be monotonic. Such population inversion phenomena happen, for instance, in lasers.} We should note that unlike the positivity condition in the Lorentz-invariant case, which is a consequence of positive norms in the Hilbert space, here it requires certain conditions on the many-body state as well. 

Remember that what appears in the effective Maxwell equations, \cref{Max}, is the self-energy $\Pi^{\mu\nu}$, which is the 1PI part of $G_J^{\mu\nu}$. Here we will argue that $\Im\Pi^{\mu\nu}$ has the same positivity property. First, we take $V^\mu$ to be transverse with respect to $u^\mu$, i.e.~$\PP_T^{\mu\nu}V_\nu=V^\mu$. Therefore, from \cref{fullJJ} we have
\eq{
V_\mu(p)V_\nu(p)\Im G^{\mu\nu}_J(p)=\boldsymbol{V}^2\frac{p^4k^2\Im\pi_T}{|p^2-g^2k^2\pi_T|^2}\,,
}
where without loss of generality we have taken $V^\mu=(0,\boldsymbol{V})$. The positivity condition then implies that $\Im\pi_T>0$ for $\omega>0$ (and negative for $\omega<0$). If instead we take $V^\mu=\bar{u}^\mu$, i.e.~along the longitudinal direction, we obtain  
\eq{
	V_\mu(p)V_\nu(p)\Im G^{\mu\nu}_J(p)=\frac{k^2\Im\pi_L}{|1+g^2 \pi_L|^2}\,,
}
where we have used $\bar{u}^2=-k^2/p^2$. Therefore we conclude that $\Im\pi_L>0$ for $\omega>0$ (and negative for $\omega<0$). We emphasize that the same result could be obtained working with $G_\gamma^{\mu\nu}$. 

While the imaginary parts of $\pi_L$ and $\pi_T$ are sign-definite, the imaginary part of the magnetic permeability $\mu$ is not.\footnote{This last statement, while very well-known in the older literature (for instance \cite{AnriARukhadze1961}), appears to be a controversy in some recent papers \cite{PhysRevE.78.026608,PhysRevE.70.048601,PhysRevE.70.048602}.} This can easily be seen from \cref{epsmu}: the combination $\text{Im}(\pi_T-\omega^2\pi_L/k^2)$ does not appear to have a definite sign. In contrast, the imaginary part of electric permittivity has the same sign as $\Im\pi_L$. 

As a final comment, we conclude from \cref{imGmunu} that generally $\Im G_J^{\mu\nu}$ does not have a gap even if the spectrum of the theory is gapped. By gap in the imaginary part we mean an energy scale below which the imaginary part vanishes. The reason is that the delta function in \cref{imGmunu} has support, focusing on the energy part of the delta function, on $E_n-E_m$ as long as $c_n-c_m$ is nonzero. Even if all energy eigenstates are larger than a mass gap, i.e.~$E_n>m$ for a mass scale $m$, the differences $\delta E_{mn}\equiv E_m-E_n$ could in principle be arbitrarily small. Consider, for instance, a thermal ensemble $c_n=e^{-\beta E_n}$ with temperature much larger than the gap so $m\beta\ll1$. Then there is a nonvanishing contribution to the imaginary part at frequency $\omega=\delta E_{mn}$ with magnitude proportional to $c_n-c_m\simeq\beta\delta E_{mn} e^{-\beta E_n}$ which is not necessarily exponentially suppressed since $\beta E_n$ could be small. 

In contrast, when the temperature is much smaller than the gap, i.e.~$m\beta\gg1$, only the vacuum state will be relevant and therefore $c_n\simeq\delta_{n,0}$. Then the expression \cref{imGmunu} becomes
\eq{
	V_\mu(p)V_\nu(p)\Im G^{\mu\nu}_J(p)=\frac{1}{2}\sum_{n}(2\pi)^4\bigg(\delta(p-p_n)-\delta(p+p_n)\bigg)|\bra{0}J^\mu(0)\ket{n}V_\mu(p)|^2\,.
}
Therefore, one recovers the usual statement that in a gapped theory the imaginary part of the response function vanishes below the gap. 

%%%%%%%%%%%%%%%%%%%%%%%%%%%%%%%%%%%%%%%%%%%%%%%%%%%%
\section{Causality and analyticity}\label{sec:analytic}
%%%%%%%%%%%%%%%%%%%%%%%%%%%%%%%%%%%%%%%%%%%%%%%%%%%%
The principles of quantum mechanics and relativity state that measurements at spacelike separation must not interfere. As a result, the commutator of any two local (bosonic) operators $[\oo_a(x),\oo_b(x')]$ must vanish for $(x-x')^2>0$, a fact known as microcausality. See also \cite{Dubovsky:2007ac,Hartman:2015lfa}. This implies that the current response function $G_J^{\mu\nu}(x)$ and the retarded photon propagator $G_\gamma^{\mu\nu}(x)$ vanish for $t<0$ (because of the theta function) and $x^2>0$. Notice that the contact term in \cref{Gj} is relevant only in the coincident limit and does not affect the conclusion. Moreover, for the photon we must project out the gauge dependent part (equivalently choose $\alpha\to\infty$ in \cref{fullphoton}) since it is not a physical observable.  

By using retardation and microcausality one can argue for the analyticity of the Fourier transform of $G_J^{\mu\nu}(x)$ and $G_\gamma^{\mu\nu}(x)$ in the complex space of $p^\mu=(\omega,\vbk)$. More precisely, 
\eq{
	G_J^{\mu\nu}(p)=\int\dd[4]{x}e^{-ip\cdot x}G_J^{\mu\nu}(x)\,,
	\label{pcomp}
} 
is a holomorphic function in the complex space of the argument, $p^\mu\in\mathbb{C}^4$, if $\omega_{I}>0$ and $p_{I}^2<0$, where we have defined $p^\mu=p^\mu_{R}+ip^\mu_{I}$. We refer to this condition as $p_{I}^\mu\in{\rm FLC}$ in which FLC is short for the forward light cone. This is simple to argue: the integral of \cref{pcomp} is exponentially converging for $p_I^\mu x_\mu=-\omega_{I}t+\boldsymbol{k}_I\cdot \boldsymbol{x}<0$. Due to causality the integrand has support only for $t>0$ and $x^2<0$ which implies that $p_I^\mu x_\mu\leq-t(\omega_I-|\boldsymbol{k}_I|)$. As a result, a sufficient condition for convergence is that $\omega_{I}>|\boldsymbol{k}_I|\geq0$ or $p_{I}^\mu\in{\rm FLC}$. An expression like \cref{pcomp} in the region of convergence defines an analytic function simply because complex derivatives are well-defined. An important assumption here is the polynomial boundedness of the response function at infinity, i.e.~the Green's function does not grow exponentially fast at infinity.\footnote{If the system is unstable such that $G(t)\sim e^{+\gamma t}$ for $t\to+\infty$ then the region of analyticity shrinks to $\omega_{I}\geq\gamma$. Similarly, if the response of the system goes to zero exponentially fast $G(t)\sim e^{-\gamma t}$ for $t\to+\infty$ then it is analytic in a larger region $\omega_{I}\geq-\gamma$.} We emphasize that the region of analyticity does not depend on the real part of the four-momentum $p_R^\mu$. In other words the analyticity region is $\mathbb{R}^4\times{\rm FLC}\subset\mathbb{C}^4$, however most of the time we just refer to it as the FLC. Needless to say that same results hold for the physical part of $G_\gamma^{\mu\nu}(p)$. In passing, we mention that the reality condition implies that $G_J^{\mu\nu}(p)^*=G_J^{\mu\nu}(-p^*)$. Notice that if $p_{I}^\mu\in{\rm FLC}$ then $\text{Im}(-p^\mu{}^*)\in{\rm FLC}$ and therefore both sides are in the region of analyticity.

Since the tensor structures of $G_\gamma^{\mu\nu}$ and $G_J^{\mu\nu}$ are known as in \cref{fullphoton} and \cref{fullJJ}, it is appropriate to study the analyticity of the coefficients. Let us consider a general tensor of the form
\eq{
	f^{\mu\nu}=A(\omega,k)\PP_T^{\mu\nu}+B(\omega,k)\PP_L^{\mu\nu}\,,
	\label{AB}
}
for some functions $A$ and $B$ of the frequency and momentum. If $f^{\mu\nu}$ is analytic in $p_I^\mu\in{\rm FLC}$, one can show that $A$ and $B$ must be analytic in the FLC. Indeed, from \cref{pLrest} and \cref{pTrest} we have $B=-p^2k_ik_jf^{ij}/\omega^2$ which implies that $B$ is analytic in the FLC. Notice that the factor $1/\omega^2$ is harmless since the pole at $\omega=0$ lies outside the FLC after the appropriate $i\epsilon$ prescription. By looking at the trace of the spatial part we get $\delta_{ij}f^{ij}=2A-\omega^2B/p^2$. Once again, the factor $1/p^2$ with appropriate $i\epsilon$ prescription is harmless for analyticity: in fact this is the propagator for a massless free particle (see also App.~\ref{dampwave}). Therefore, we conclude that $A$ must be analytic in the FLC. However, the analyticity of $A$ and $B$ is not sufficient to guarantee that $f^{\mu\nu}$ is analytic. If we look at $f^{\mu\nu}$ for $i\neq j$ we have
\eq{
f^{ij}= - \frac{k^ik^j}{k^2}\left(A+\frac{\omega^2}{p^2}B\right)\,.
	\label{AB2}
} 
The factor $1/k^2$ introduces singularities in the FLC. The reason is that the equation $k^2=k_R^2-k_I^2+2i\boldsymbol{k}_R\cdot\boldsymbol{k}_I=0$ has nontrivial, i.e.~$k^i\neq0$, solutions which are independent of $\omega_I$ and therefore one can find points $(\omega_R+i\omega_I,\boldsymbol{k}_R+i\boldsymbol{k}_I)$ with $k^2=0$ that lie in the forward light cone. (In fact, $1/k^2$ is the Fourier transform of $-1/\boldsymbol{\nabla}^2$ which describes instantaneous action at a distance.) These poles must be removed by zeros of the numerator to ensure analyticity of $f^{ij}$. We conclude that the combination $A+\omega^2B/p^2$ must be zero at $k^2=0$.   

Application of the above results to $G_\gamma^{\mu\nu}$ given in \cref{fullphoton} shows that the combinations $1/(1+g^2\pi_L)$ and $1/(p^2-g^2k^2\pi_T)$ are analytic in the FLC. We will use this fact in \S\ref{sec:bounds} to bound the low-energy limits of $\varepsilon$ and $\mu$. Moreover, from the discussion below \cref{AB2} we require that the combination 
\eq{
\frac{1}{k^2}\left[\frac{1}{p^2-g^2k^2\pi_T}+\frac{\omega^2/p^4}{1+g^2\pi_L}\right]\stackrel{k^2=0}{\sim} \frac{1}{k^2}\frac{g^2(k^2\pi_T-\omega^2\pi_L)}{\omega^2(1+g^2\pi_L)(\omega^2+g^2k^2\pi_T)}\,,
\label{piLpiTcomb}
} 
is regular near $k^2=0$. To obtain the second expression we have used $p^2=-\omega^2$ in the limit $k^2\to0$. We conclude that the combination $k^2\pi_T-\omega^2\pi_L$ must go to zero at least like $k^2$ when $k^2\to0$.\footnote{The reader may wonder how one can exclude the possibility of the denominator of \cref{piLpiTcomb} having a singularity. For instance, if $\pi_L\sim1/k^2$ as $k^2 \to 0$ then it removes the unwanted $1/k^2$ factor in front of \cref{piLpiTcomb}. Notice that, apparently, this possibility is harmless for the analyticity of the combination $1/(1+g^2\pi_L)$ required by causality. However, as we will argue in \S\ref{wider} and App.~\ref{landau}, $\pi_L(\omega,\boldsymbol{q}+\omega\boldsymbol{\xi})$ is analytic in the upper-half plane (UHP) of complex $\omega$ as long as $\boldsymbol{q}\cdot\boldsymbol{\xi}=0$ and $|\boldsymbol{\xi}|<1$. This is a consequence of both causality and positivity. In this parametrization $k^2=q^2+\omega^2\xi^2$, and therefore it is not possible to have $\pi_L\sim1/k^2$ since it implies a pole at $\omega=+iq/\xi$ in the UHP. A similar discussion applies to $k^2\pi_T$.} This indicates that, as a consequence of microcausality, the two functions $\pi_L$ and $\pi_T$ cannot be completely unrelated. Similarly, the coefficients in the expression for $G_J^{\mu\nu}$ given in \cref{fullJJ} are analytic; since these are slightly more complicated and do not give new information we do not report them here. We will deal with the analyticity of $\Pi^{\mu\nu}$ in \S\ref{wider}.  

In order to avoid dealing with several complex variables at the same time we parametrize the FLC by several single-complex-variable subspaces. More precisely, we write
\eq{
p^\mu=(\omega,\vbq+\omega\vbxi)
\label{paraflc}
} 
for complex $\omega$ and real vectors $\vbq$ and $\vbxi$ with the condition that $\xi\equiv|\vbxi|<1$. It is easy to see that every point $p^\mu=(\omega,\vbk)$ in the FLC can be written in the form \eqref{paraflc} for appropriate values of $\vbq$ and $\vbxi$, i.e.~$\vbxi=\frac{\vbk_I}{\omega_I}$ and $\vbq=\vbk_R-\omega_R\vbxi$. As discussed in \cite{Creminelli:2022onn}, the form \eqref{paraflc} is the most general single-variable parametrization. Any other choice for $\vbk(\omega)$ would either change the analyticity region (by mixing the real and imaginary parts) or spoil the behavior at infinity. Then we see that $G_\gamma^{\mu\nu}(\omega,\vbq+\omega\vbxi)$ and $G_J^{\mu\nu}(\omega,\vbq+\omega\vbxi)$, now regarded as functions of $\omega$, are analytic in the UHP of $\omega \in \mathbb{C}$, i.e.~$\omega_I>0$. The same holds for the coefficients appearing in \cref{fullphoton} and \cref{fullJJ} as discussed above.

For a generic function $\chi(z)$ that is analytic in the UHP (and continuous onto the real line) and decays at infinity one can use Cauchy's theorem, applied to $\chi(z)/(z-z_R)$, to prove the following relation
\eq{
\adjustbox{valign=c}{\tikzset{every picture/.style={line width=0.85pt}}

\definecolor{ultramarine}{rgb}{0.07, 0.04, 0.56}

\begin{tikzpicture}

\tikzset{ma/.style={decoration={markings,mark=at position 0.35 with {\arrow[scale=0.8]{>}}},postaction={decorate}}}
\tikzset{ma2/.style={decoration={markings,mark=at position 0.5 with {\arrow[scale=0.8]{<}}},postaction={decorate}}}

\draw[gray,thin] (-2.5,0) -- (2.5,0)[->]; 
\draw[gray,thin] (0,0) -- (0,2.5)[->]; 

%\draw[ma,color=blue] (-2,0) -- (2,0) arc(0:180:2) --cycle;
\draw[ma,color=ultramarine,thin] (2,0) arc (0:180:2);
\draw[ma,color=ultramarine,thin] (-2,0) -- (0.9,0);
\draw[color=ultramarine,thin] (1.1,0) -- (2,0);
\draw[color=ultramarine,thin] (1.1,0) arc (0:180:0.1);

\draw[gray] (2.2,2.2) node{\footnotesize $z$}; 
\draw[gray,thin] (2,2.3)|-(2.3,2);
%\draw[black] (-1.7,1.7) node{\footnotesize $C$};
\draw[black] (1,-0.25) node{\footnotesize $z_R$};  

\filldraw[gray] (1,0) circle (0.6pt);

%\filldraw[blue] (5,0) circle (0.8pt);
%\filldraw[blue] (9,0) circle (0.8pt);
%\filldraw[blue] (7.5,0) circle (0.8pt);
%\draw[black] (1,0) node {\tikzxmark};
%\draw[black] (3,0) node {\tikzxmark};
%\draw[black] (1.5,-0.2) node {\tikzxmark};
%\draw[black] (3.5,-0.2) node {\tikzxmark};
% 
%\draw[black] (4.7,-0.3) node{\footnotesize $C_\infty$}; 
%
%\draw[black] (1,0.3) node{\footnotesize $t_1$}; 
%\draw[black] (2,0.3) node{\footnotesize $\dots$}; 
%\draw[black] (3,0.3) node{\footnotesize $t_m$}; 
%
%\draw[black] (1.5,-0.5) node{\footnotesize $t'_1$};
%\draw[black] (2.25,-0.5) node{\footnotesize $\dots$}; 
%\draw[black] (3.5,-0.5) node{\footnotesize $t'_n$}; 

\end{tikzpicture}}\qquad\qquad 	\chi(z_R)=\frac{1}{i\pi}{\rm PV}\int_{-\infty}^{+\infty}\frac{\dd{z}}{z-z_R}\,\, \chi(z)\,,
\label{rkk11}
}
in which $z_R$ is real, and ${\rm PV}$ means the principal value of the integral for the pole at $z=z_R$.  An important assumption here is that $\chi(z)\to0$ when $|z|\to\infty$, which implies that we can neglect the contour at infinity. Taking the real part of \cref{rkk11} gives an expression for $\Re\chi(z_R)$ in terms of an integral over the $\Im\chi(z)$ over the real axis. We can apply this general result to the analytic functions that we found above by using the parametrization of \cref{paraflc}. Let us denote by $\chi(\omega,\vbq+\omega\vbxi)$ such an analytic function, e.g.~$1/(1+g^2\pi_L)$. Then \cref{rkk11} gives\footnote{One can also derive this relation by noting that $\chi(x)=\chi(x)\theta(t-\boldsymbol{\xi}\cdot\boldsymbol{x})$ for any $\xi<1$ as required by microcausality. Taking the Fourier transform of both sides gives \cref{rkk2}. See \cite{Leon,ggKK} for details.} 

\begin{eBox}
\eq{
	\chi(\omega,\vbq+\omega\vbxi)=\frac{1}{i\pi}{\rm PV}\int_{-\infty}^{+\infty}\frac{\dd{z}}{z-\omega}\,\,\chi(z,\vbq+z\vbxi)\,,
	\label{rkk2}
}	
\end{eBox}
for real $\omega$ as well as real $\boldsymbol{q}$ and $\boldsymbol{\xi}$ with $\xi<1$ as required by \cref{paraflc}. By shifting the spatial momentum $\boldsymbol{q}\to\boldsymbol{q}-\omega\boldsymbol{\xi}$ and relabeling it we can rewrite \cref{rkk2} as
\eq{
	\chi(\omega,\vbk)=\frac{1}{i\pi}{\rm PV}\int_{-\infty}^{+\infty}\frac{\dd{z}}{z-\omega}\,\,\chi(z,\vbk+(z-\omega)\vbxi)\,.
	\label{rkk3}
}

We emphasize that in this relation all independent variables are real. Eq.~(\ref{rkk3}) was initially written by M.~Leontovich \cite{Leon} and we will call it (as well as \cref{rkk2}) Leontovich's relation. This relation is the generalization of the famous Kramers-Kronig formula, derived assuming only retardation, taking into account the condition of microcausality. In fact, for $\boldsymbol{\xi} = \boldsymbol{0}$ one recovers the Kramers-Kronig relations. Taking the real part of the both sides in \cref{rkk3}, as mentioned above, gives an expression for $\Re\chi(\omega,\boldsymbol{k})$ in terms of an integral of $\Im\chi(z,\boldsymbol{k}+(z-\omega)\boldsymbol{\xi})$ for any $\boldsymbol{\xi}$. In \S\ref{sec:bounds} we will apply Leontovich's relation to the various analytic combinations of $\pi_L$ and $\pi_T$. 

Before closing this section let us mention two more points regarding Leontovich's relation. First, as we said, \cref{rkk2} gives the real part of the function in terms of its imaginary part and vice versa. Therefore, we can express the whole function in terms of the imaginary part as follows
\eq{
\spl{
\chi(\omega,\vbk+\omega\boldsymbol{\xi})&=\frac{1}{\pi}{\rm PV}\int\frac{\dd{z}}{z-\omega}\,\,\Im\chi(z,\vbk+z\vbxi)+i\int\dd{z}\delta(z-\omega)\Im\chi(z,\vbk+z\vbxi)\\
&=\frac{1}{\pi}\int\frac{\dd{z}}{z-\omega-i\epsilon}\,\,\Im\chi(z,\vbk+z\vbxi)\,,
}\label{fullim}
} 
where in the first line we have replaced the real part using \cref{rkk2} and to obtain the second line we have used Sokhotski-Plemelj. Although we have derived \cref{fullim} for real $\omega$, it remains valid for complex $\omega$ as well in the domain of analyticity, i.e.~in the UHP.\footnote{We should note that we can write the analog of \cref{fullim} for \cref{rkk3}, i.e.~$\xi$ only appears on the right-hand side, assuming $\omega$ is real. However, this form does not continue to hold for complex $\omega$ simply because the right-hand side involves an integral over $\Im\chi(z,\boldsymbol{k}+(z-\omega)\boldsymbol{\xi})$ which is not an analytic function of $\omega$.} The reason is that the right-hand side of \cref{fullim} is a sum over functions $1/[\omega-(z-i\epsilon)]$ which do not have poles in the UHP.\footnote{More rigorously one can start from $\chi(\omega)$ for complex $\omega$ by writing it in terms of an integral over the real axis using Cauchy's theorem. The function over the real axis can then be written in terms of an integral over the imaginary part as given in \cref{fullim}. After some algebra one recovers \cref{fullim} now with $\omega$ complex.} Eq.~(\ref{fullim}) will be used in \S\ref{wider} and App.~\ref{landau} to study the analyticity of the self-energy tensor $\Pi^{\mu\nu}$.

Finally, we note that the condition of microcausality, equivalently Leontovich's relation, restricts the form of the imaginary part of the function. We have already seen in the previous section that $\omega\Im G^{\mu\nu}_\gamma$ and $\omega\Im G^{\mu\nu}_J$ are positive definite (for passive materials). However, one can ask: does any positive matrix give an eligible imaginary part of a causal function? Interestingly, the answer is no. This can be seen from \cref{rkk3}. The fact that the left-hand side in \cref{rkk3} is independent of $\boldsymbol{\xi}$ is a restriction on the imaginary part from causality. More precisely, the integral over the imaginary part along two different lines in the $(\omega,\boldsymbol{k})$ space associated with two different vectors $\boldsymbol{\xi}_1$ and $\boldsymbol{\xi}_2$, as shown below, must give the same result\footnote{Equivalently, one can show that \cref{imcons} is a direct consequence of the fact that the imaginary part is given by the Fourier transform of a commutator which vanishes outside the full light cone as given in \cref{imGab}.}
\eq{
	\adjustbox{valign=c}{\tikzset{every picture/.style={line width=0.85pt}}

\definecolor{ultramarine}{rgb}{0.07, 0.04, 0.56}

\begin{tikzpicture}

\tikzset{ma/.style={decoration={markings,mark=at position 0.3 with {\arrow[scale=0.8]{>}}},postaction={decorate}}}
\tikzset{ma2/.style={decoration={markings,mark=at position 0.5 with {\arrow[scale=0.8]{<}}},postaction={decorate}}}

\draw[gray,thin] (-0.5,0) -- (2.5,0)[->]; 
\draw[gray,thin] (0,-0.5) -- (0,2.5)[->]; 

\draw[ma,color=ultramarine,thin] (1.2,0.2) -- (1.2,2.4);
\draw[ma,color=ultramarine,thin] (0.65,0.2) -- (1.75,2.4);
\draw[ma,color=ultramarine,thin] (1.75,0.2)-- (0.65,2.4);
\draw[dashed,gray,thin] (0.2,0.3) -- (2.2,2.3);
\draw[dashed,gray,thin] (0.2,2.3) -- (2.2,0.3);
%\draw[color=ultramarine] (1.1,0) -- (2,0);
%\draw[color=ultramarine] (1.1,0) arc (0:180:0.1);

%\draw[gray] (2.2,2.2) node{\footnotesize $z$}; 
%\draw[gray,thin] (2,2.3)|-(2.3,2);
%\draw[black] (-1.7,1.7) node{\footnotesize $C$};
\draw[black] (2.7,0) node{\footnotesize $\vec{k}$};  
\draw[black] (0,2.7) node{\footnotesize $\omega$}; 
\draw[black] (1.9,1.3) node{\footnotesize $(\omega,k)$}; 
\draw[black] (1.2,2.6) node{\footnotesize $\xi=0$}; 
\draw[black] (0.5,2.6) node{\footnotesize $\xi_1$}; 
\draw[black] (1.9,2.6) node{\footnotesize $\xi_2$}; 

\filldraw[gray] (1.2,1.3) circle (0.6pt);

%\filldraw[blue] (5,0) circle (0.8pt);
%\filldraw[blue] (9,0) circle (0.8pt);
%\filldraw[blue] (7.5,0) circle (0.8pt);
%\draw[black] (1,0) node {\tikzxmark};
%\draw[black] (3,0) node {\tikzxmark};
%\draw[black] (1.5,-0.2) node {\tikzxmark};
%\draw[black] (3.5,-0.2) node {\tikzxmark};
% 
%\draw[black] (4.7,-0.3) node{\footnotesize $C_\infty$}; 
%
%\draw[black] (1,0.3) node{\footnotesize $t_1$}; 
%\draw[black] (2,0.3) node{\footnotesize $\dots$}; 
%\draw[black] (3,0.3) node{\footnotesize $t_m$}; 
%
%\draw[black] (1.5,-0.5) node{\footnotesize $t'_1$};
%\draw[black] (2.25,-0.5) node{\footnotesize $\dots$}; 
%\draw[black] (3.5,-0.5) node{\footnotesize $t'_n$}; 

\end{tikzpicture}}\qquad {\rm PV}\int\dd{z}\frac{\Im\chi(z,\vbk+(z-\omega)\vbxi_1)}{z-\omega}={\rm PV}\int\dd{z}\frac{\Im\chi(z,\vbk+(z-\omega)\vbxi_2)}{z-\omega}\,.
	\label{imcons}
} 
Notice that this restriction does not exist if we only consider retardation ($\boldsymbol{\xi} = \boldsymbol{0}$) as we do for Kramers-Kronig relations. A more detailed study of this set of constraints is beyond the scope of this paper and will be explored elsewhere.

%%%%%%%%%%%%%%%%%%%%%%%%%%%%%%%%%%%%%%%%%%%%%%%%%%%%
\section{High-energy behavior}\label{high}
%%%%%%%%%%%%%%%%%%%%%%%%%%%%%%%%%%%%%%%%%%%%%%%%%%%%
One of the assumptions required for the Leontovich relation \cref{rkk3} to be valid is that the arc at infinity can be neglected. This requires some knowledge about the high-energy behavior of the two-point function of $J^\mu$, or equivalently of $A^\mu$. We stress that we require the high-energy behavior for {\em complex} $p^\mu$ in the domain of analyticity. As was argued in \cite{Creminelli:2022onn}, the behavior at high energies in the complex plane is dictated by the short distance behavior of the two-point function in position space.
  
Unfortunately, as far as we know, there is no universal bound on the asymptotic behavior of the two-point function. By contrast, for the $S$-matrix such a bound exists \cite{Froissart:1961ux,Martin:1962rt}, at least for gapped theories, only by requiring unitarity, causality and Lorentz invariance. One possible assumption, as studied in \cite{Creminelli:2022onn}, is that at high energies the system is well-described by a conformal field theory which implies that the two-point function of a conserved current behaves as $\sim p^{d-2}$ in $d$ spacetime dimensions (and as $\sim p^2\log p^2$ in $d=4$). 

In condensed matter systems one can usually assume that the Green's functions vanish at high energies \cite{Jackson:1998nia,lan84,book:plasmapysics}, i.e.~that the medium becomes irrelevant at high frequencies. The main idea is that at high energies one can think of the system as a collection of charged particles (say, electrons). In this limit, i.e.~$\omega\gg\nu_c$ in which $\nu_c$ is the average collision frequency \cite{book:plasmapysics}, one can also neglect interactions among the particles. Therefore, the system is described as a gas of free electrons. The dielectric response function of this system has been calculated long ago by Lindhard \cite{lindhard} in the non-relativistic limit. We provide a first-principle derivation, including relativistic effects, in App.~\ref{linhar}; the calculation is basically the one-loop correction to the photon propagator (see \cref{loop}) assuming a finite chemical potential for electrons. As explained in \cref{Pidec}, there are two contributions to the response function: one is the standard quantum electrodynamics correction and the other is the contribution from the finite electron density. The finite density contribution, as given in \cref{piLpiThigh}, at high energies $\omega,k\to\infty$, behaves as
\eq{
g^2\pi_L\to-\frac{\omega_p^2}{\omega^2}+\dots\,,\qquad g^2\pi_T\to-\frac{\omega_p^2}{k^2}+\dots\,,
\label{plasmapi}
}
in which $\omega_p^2 \equiv g^2 n/m$ is called the plasma frequency with $n$ the number density and $m$ the mass of the electrons. Notice that from \cref{plasmapi} we conclude that the imaginary part decays even faster, as $1/\omega^3$. Eq.~(\ref{plasmapi}) implies that $\varepsilon_{(T)}=1-\omega_p^2/\omega^2+\dots$ at high frequencies both for the longitudinal and transverse parts of the dielectric tensor (see footnote~\ref{eT}) and we will refer to it as ``plasma behavior". We should note that in general there are different components present in the medium that have different plasma frequencies. In \cref{plasmapi} we take $\omega_p^2$ to be the largest one, usually associated with the lightest particles.

What about the Lorentz-invariant contribution? As given in \cref{Piqed}, at energies below the mass $\omega,k\ll m$, its value is very small, $\pi_{L,{\rm qed}} \approx -p^2/(60\pi^2m^2)$ and negligible. However, for energies above the mass it grows as $\log p^2$. Therefore we are forced to close the contour at moderately high energies in which the description by means of a non-relativistic plasma is a good approximation, and avoid extremely high energies in which relativistic effects become important. This is shown in \cref{piLTlind} for the real parts of $\pi_L$ and $\pi_T$. There is a window of energies in which relativistic corrections are not important and we can trust the plasma behavior of non-relativistic fermions.
\fg{
	\includegraphics[width=0.4\textwidth]{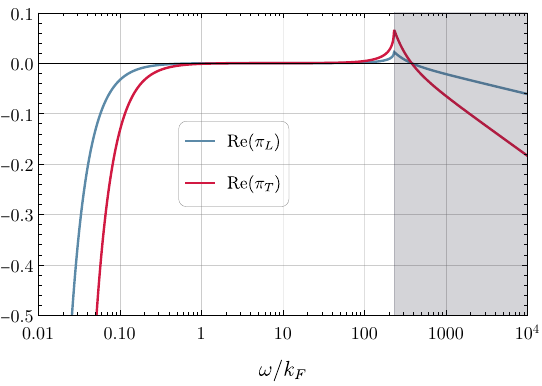}\hfil
	\includegraphics[width=0.385\textwidth]{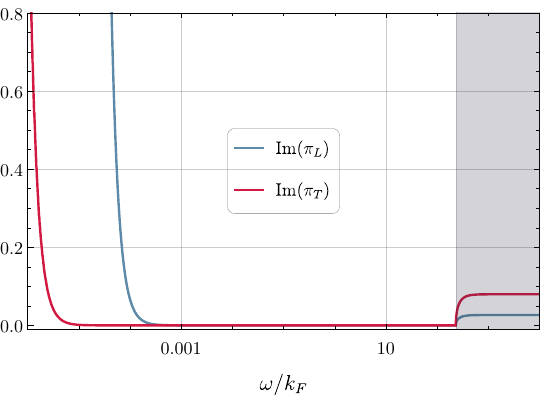}
	\caption{Real (left) and imaginary (right) parts of the functions $\pi_L$ and $\pi_T$ obtained from \cref{Pidec} and \cref{pidecLT} for the system of non-interacting fermions at leading order in the coupling to the photon. We have chosen $\boldsymbol{k} = \omega\boldsymbol{\xi}$ and set $|\boldsymbol{\xi}|=0.5$ but other values result in qualitatively similar behavior. We have chosen $m/k_F=100$ which is a typical value in materials. We observe that the relativistic growth starts from energies $\omega\sim m$ which is much larger than the non-relativistic plasma regime. The region above the pair production threshold $p^2=-4m^2$ is shaded. There is a guaranteed range of energies where \cref{plasmapi} applies since $\omega_p^2/m^2=g^2k_F^3/3\pi^2m^3\ll1$ for $k_F/m\ll1$.}
	\label{piLTlind}
}
Moreover, the error in neglecting the relativistic corrections is controlled by the ratio $\omega_p^2/m^2$ which is of order $10^{-12}$ for typical materials. If we close the contour of integration at, say, $\sim\order{10^2}\omega_p$ the contribution from \cref{plasmapi} is negligible.

For media of ``higher energies'', for example nuclear matter, one cannot close the contour below the electron mass. One can still make use of the Kramers-Kronig and Leontovich relations, however, as we will discuss in \S\ref{sec:stronger}.

%%%%%%%%%%%%%%%%%%%%%%%%%%%%%%%%%%%%%%%%%%%%%%%%%%%%
\section{Bounds on the low-energy behavior}\label{sec:bounds}
%%%%%%%%%%%%%%%%%%%%%%%%%%%%%%%%%%%%%%%%%%%%%%%%%%%%
In this section we bound $\varepsilon(0,0)$ and $\mu(0,0)$ using the Leontovich relation \cref{rkk3} and positivity conditions on the imaginary parts. Notice that both $\varepsilon$ and $\mu$ can in principle have singularities like $1/\omega^n$ as $\omega \to 0$ without violating causality. For instance, in the case of conductors we have $\varepsilon\sim i\sigma/\omega$ where $\sigma$ denotes the conductivity. For superconductors, we have $\varepsilon,\mu\sim1/(\omega^2-c_s^2k^2)$ associated to the propagation of the Goldstone mode with speed of sound $c_s$ (see \cite{1986PThPS8643W,schakel1998boulevard}). In this paper, we restrict ourselves to cases in which there is no singularity at the origin. In other words, our results will be applicable only to insulators (dielectrics) where at low energies there is no other dynamical degree of freedom than photons and leave other -- perhaps more interesting -- cases to future studies.\footnote{See \cite{Landry:2022nog} for a discussion of the classification of low-energy degrees of freedom based on the derivative expansion.} From \cref{epsmu} we see that $g^2\pi_L=\varepsilon(0,0)-1$ and $g^2k^2\pi_T=0$ at low energies. Moreover, the combination of \cref{epsmu} and the causality condition \cref{piLpiTcomb} implies that $g^2(k^2\pi_T-\omega^2\pi_L)=(1-\mu(0,0)^{-1})k^2$ at low energies.

In the following we use the longitudinal and transverse parts of $G_\gamma^{\mu\nu}$ to write dispersion relations involving $\varepsilon(0,0)$ and $\mu(0,0)$. It is possible to obtain similar (and equivalent) relations using $G_J^{\mu\nu}$. We do not apply the Leontovich relation directly to $\Pi^{\mu\nu}$ since we still need to argue for its analyticity. This will be discussed in \S\ref{wider}. We note that many of these results have been derived in the literature on this subject \cite{losyakov,Kirzhnits:1989,Keldysh1989TheDF}.

\paragraph{Longitudinal part} Applying the discussion around \cref{AB} to the longitudinal part of $G_\gamma^{\mu\nu}$ given in \cref{fullphoton} implies that $1/(p^2\varepsilon(\omega,\boldsymbol{k}))$ is analytic when $\text{Im } p$ lies in the FLC, where, recall, $\varepsilon=1+g^2\pi_L$. Let us consider the function $1/\varepsilon-1$. This function is analytic in the same region and, according to \cref{plasmapi}, goes to zero at infinity. Eq.~\eqref{rkk2} implies
\eq{
\frac{1}{\varepsilon(\omega,\boldsymbol{q}+\omega\vbxi)}-1 = -\frac{1}{i\pi}{\rm PV}\int_{-\infty}^{+\infty}\frac{\dd{z}}{z-\omega}\,\,\frac{g^2 \pi_L(z,\boldsymbol{q}+z\vbxi)}{1+g^2 \pi_L(z,\boldsymbol{q}+z\vbxi)} \,.
\label{long1}
} 
Setting $\omega = \boldsymbol{q} =0$ we find
\eq{
\frac{1}{\varepsilon(0,0)}-1=-\frac{2g^2}{\pi}\int_{0}^{+\infty}\frac{\dd{z}}{z}\,\,\frac{\Im\pi_L(z,z\vbxi)}{|1+g^2\pi_L(z,z\vbxi)|^2}\,.
\label{long}
}
To obtain \cref{long} we used the reality condition $\pi_L(-z,-z\boldsymbol{\xi}) = \pi_L(z,z\boldsymbol{\xi})^*$, from which it follows that $\Im \pi_L(z,z\boldsymbol{\xi}) = \Im\varepsilon(z,z\vbxi)$ is odd in $z$ and so also that $\varepsilon(0,0)$ is real. In this way we have written $1/\varepsilon$ in terms of $\Im\pi_L$ which has a positivity property as discussed in \S\ref{positiv}. A similar relation can be obtained for nonzero $\boldsymbol{q}$.\footnote{Notice that $\varepsilon(0,\boldsymbol{q})$ is also real due to rotational invariance, $\varepsilon(0,\boldsymbol{q})^*=\varepsilon(0,-\boldsymbol{q})=\varepsilon(0,\boldsymbol{q})$.} We should note that for our purposes $\vbxi$ on the right-hand side is not necessary and can be set to zero; as discussed in \cref{imcons}, by causality the integral turns out to be independent of $\vbxi$ (for $\xi<1$). It may be useful, however, to keep $\vbxi$ in case something is known about the imaginary part at finite momentum. Finally, from \cref{long} one can argue that $1/\varepsilon(0,0)\leq1$ which implies
\begin{eBox}
\eq{
 \varepsilon(0,0)\geq1\,. 
 \label{eps1}
}
\end{eBox}
The latter possibility of $\varepsilon(0,0)<0$ is ruled out following the discussion of \S\ref{wider}. Noticeably, since $\varepsilon(0,0)\geq1$, there is no electric analog of paramagnetism.

\paragraph{Transverse part} The coefficient of the transverse part of $G_\gamma^{\mu\nu}$ in \cref{fullphoton} is $1/(p^2-g^2k^2\pi_T)$, which is analytic following the discussion in \S\ref{sec:analytic}. Since this is divergent at low energies when $\boldsymbol{k} = \omega \boldsymbol{\xi}$, we instead consider the combination
\eq{
\chi_1\equiv\frac{\omega^2}{p^2-g^2k^2\pi_T}=\frac{1}{-\varepsilon+\frac{k^2}{\omega^2}\frac{1}{\mu}}\,.
\label{chi1}
}
Next, we parametrize the momentum according to \cref{paraflc}, $\chi_1(\omega,\vbq+\omega\vbxi)$. By using \cref{plasmapi}, which implies that $k^2\pi_T\to\text{const.}$ at infinity, we see that in the limit $\omega\to\infty$ the function approaches $-1/(1-\xi^2)$. Therefore, for the combination $\chi_1+1/(1-\xi^2)$, we are able to neglect the contour at infinity in the Leontovich relation. Using \cref{rkk2} we obtain
\eq{
\frac{1}{-\varepsilon(\omega,\boldsymbol{q}+\omega\vbxi)+\frac{(\boldsymbol{q}+\omega\vbxi)^2}{\omega^2\mu(\omega,\boldsymbol{q}+\omega\vbxi)}}+\frac{1}{1-\xi^2}=\frac{1}{i\pi}{\rm PV}\int_{-\infty}^{+\infty}\frac{\dd{z}}{z-\omega}\left[\frac{z^2}{-z^2+(\boldsymbol{q}+z\vbxi)^2(1-g^2\pi_T(z,\boldsymbol{q}+z\vbxi))} +\frac{1}{1-\xi^2}\right] \,.
\label{trans1}
}
Setting $\boldsymbol{q} = \boldsymbol{0}$ and taking the limit $\omega \to 0$, from the real part of \cref{trans1} we obtain 
\eq{
	\frac{1}{-\varepsilon(0,0)+\frac{\xi^2}{\mu(0,0)}}+\frac{1}{1-\xi^2}=\frac{2g^2\xi^2}{\pi}\int_{0}^{+\infty}\frac{\dd{z}}{z}\,\,\frac{\Im\pi_T(z,z\vbxi)}{|-1+\xi^2-g^2\xi^2\pi_T(z,z\vbxi)|^2}\,.
	\label{transvleon}
}
Like before we can restrict the integration range to positive $z$ because $\pi_T(-z,-z\vbxi) = \pi_T(z,z\vbxi)^*$. In the limit $\xi\to0$, this equation reproduces the longitudinal dispersion relation. The reason is that, as argued below \cref{piLpiTcomb}, by causality the combination $k^2 \pi_T-\omega^2\pi_L \sim k^2$ as $k\to0$. Therefore, $\xi^2\pi_T(z,z\vbxi) \sim \pi_L(z,z\vbxi)$ as $\xi\to0$ at fixed $z$. However, for finite $\xi$ this equation has new information. By the positivity condition on the imaginary part of $\pi_T$ the right-hand side must be positive. Therefore we obtain
\eq{\label{eq:xibound}
\varepsilon(0,0)\geq\frac{\xi^2}{\mu(0,0)} + 1 - \xi^2 \,.
}
Another possibility would be $\varepsilon(0,0)<\xi^2\mu(0,0)$ but this cannot happen as we will argue in \S\ref{wider}. Optimal bounds are found by setting $\xi = 0$ and $\xi = 1$. Again we observe that $\boldsymbol{\xi} = \boldsymbol{0}$ yields the longitudinal bound given in \cref{eps1}. On the other hand at $\xi=1$ we obtain
\begin{eBox}
\eq{
\varepsilon(0,0)\geq\frac{1}{\mu(0,0)}\,.
}
\end{eBox}
One can also consider, instead of $\chi_1$ in \cref{chi1}, another function defined as 
\eq{
	\chi_2\equiv\frac{k^2}{p^2-g^2k^2\pi_T}=\frac{1}{\frac{1}{\mu}-\frac{\omega^2}{k^2}\varepsilon}\,,
}
which is analytic in the FLC since $1/(p^2-g^2k^2\pi_T)$ is analytic in that region. Once again we parametrize it as $\chi_2(\omega,\vbq+\omega\vbxi)$ and use the Leontovich relation in the symmetric form given in \cref{rkk2}. In the limit $\omega\to\infty$, $\chi_2\to-\xi^2/(1-\xi^2)$. Therefore, we use Leontovich's relation for the combination $\chi_2+\xi^2/(1-\xi^2)$ and get
\eq{
\frac{1}{\frac{1}{\mu(\omega,\boldsymbol{q}+\omega\vbxi)}-\frac{\omega^2\varepsilon(\omega,\boldsymbol{q}+\omega\vbxi)}{(\boldsymbol{q}+\omega\vbxi)^2}}+\frac{\xi^2}{1-\xi^2}=\frac{1}{i\pi}{\rm PV}\int_{-\infty}^{+\infty}\frac{\dd{z}}{z-\omega}\left[\frac{(\boldsymbol{q}+z\vbxi)^2}{-z^2+(\boldsymbol{q}+z\vbxi)^2(1-g^2\pi_T(z,\boldsymbol{q}+z\vbxi))} +\frac{\xi^2}{1-\xi^2}\right]\,.
	\label{trans2}
}
Setting $\boldsymbol{q} = \boldsymbol{0}$ and then taking $\omega\to0$ limit gives \cref{transvleon} where both sides are multiplied by $\xi^2$. However, we can also take the limit $\omega\to0$ but keep $\vbq$ finite:
\eq{
\mu(0,\vbq)+\frac{\xi^2}{1-\xi^2}=\frac{g^2}{\pi}{\rm PV}\int_{-\infty}^{+\infty}\frac{\dd{z}}{z}\,\,\frac{(\vbq+z\vbxi)^4\Im\pi_T(z,\vbq+z\vbxi)}{|-z^2+(\vbq+z\vbxi)^2-g^2(\vbq+z\vbxi)^2\pi_T(z,\vbq+z\vbxi)|^2}\,.
\label{mudisp1}
} 
Since the right-hand side is positive, this gives a lower bound on the magnetic permeability at finite momentum $\boldsymbol{q}$ in the static limit ($\omega=0$). Setting $\xi = 0$ and then taking\footnote{The limit $q\to0$ of \cref{mudisp1} is subtle, since the RHS is discontinuous at $\boldsymbol{q} = \boldsymbol{0}$. The correct procedure is to take the limit after integration.} $\boldsymbol{q} \rightarrow \boldsymbol{0}$ we conclude
\begin{eBox}
\begin{equation}
	\mu(0,0) \geq 0 \,.
\end{equation}
\end{eBox}

The bounds obtained above on $\varepsilon(0,0)$ and $\mu(0,0)$ are summarized in Fig.~\ref{bound} in which the shaded region is allowed. The horizontal and vertical boundaries correspond to the analyticity and positivity of the longitudinal and transverse parts of the Green's function respectively.
\fg{
	\includegraphics[width=0.45\textwidth]{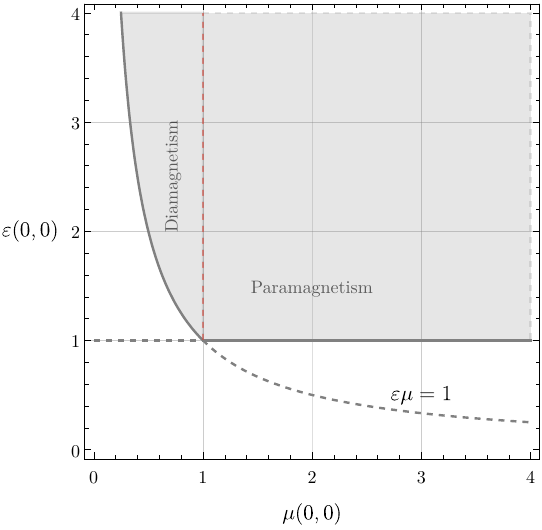}
	\caption{The allowed region for $\varepsilon$ and $\mu$ from positivity and analyticity. The horizontal boundary results from \cref{long} (or equivalently \cref{long2}) and the positivity of $\Im\pi_L$ while the vertical boundary results from \cref{transvleon} (or equivalently \cref{epsmuleo}) and the positivity of $\Im\pi_T$ in the limit $\xi\to1$. While $\varepsilon$ is forced to be larger than unity, $\mu$ could be both less than 1, in which the medium is called diamagnetic, or larger than 1, for a paramagnetic medium. Notice that only diamagnetism is constrained by causality and positivity.}
	\label{bound}
}
Notice that the inequality $\varepsilon(0,0)\mu(0,0)\geq1$ is nothing but the condition of subluminal speed of propagation of photons inside the medium, since the transverse photon propagator at low energies can be written as $1/(-\varepsilon(0,0)\omega^2+k^2/\mu(0,0))$. Moreover, $\varepsilon(0,0) > 0$ is required to avoid a ghost, and $\mu(0,0) > 0$ to avoid a gradient instability. The fact that one can recover bounds from subluminality condition has been observed in the Lorentz-invariant context as well \cite{Adams:2006sv}. However, we emphasize that here we obtain dispersion relations which in principle could be estimated or measured, and thereby, the bounds could get stronger. We will see an example below. Moreover, it should be pointed out that the vertical boundary cannot be obtained only by using retardation, i.e.~the Kramers-Kronig relation, since that corresponds to setting $\boldsymbol{\xi} = \boldsymbol{0}$ while to obtain the vertical boundary we needed to take $\xi \rightarrow 1$. Therefore, while Lorentz symmetry is broken by the presence of the medium, Lorentz invariance of the theory has nontrivial consequences.

%%%%%%%%%%%%%%%%%%%%%%%%%%%%%%%%%%%%%%%%%%%%%%%%%%%%
\section{Analyticity of $\Pi^{\mu\nu}$}\label{wider}
%%%%%%%%%%%%%%%%%%%%%%%%%%%%%%%%%%%%%%%%%%%%%%%%%%%%
In the previous section we used the photon Green's function to bound the values of $\varepsilon$ and $\mu$ at low energies. Since these quantities are defined through the self-energy tensor $\Pi^{\mu\nu}$, one may wonder why we have not used $\Pi^{\mu\nu}$ directly. The reason is that we do not have a general argument for the analyticity of $\Pi^{\mu\nu}$, and we will elaborate on this issue here.

The argument for the analyticity of $G_J^{\mu\nu}(p)$ and $G_\gamma^{\mu\nu}(p)$ is that microcausality requires commutators to vanish for spacelike separated points. However, there is no representation of $\Pi^{\mu\nu}$ in terms of a commutator of local operators. We emphasize that \cref{PiJJ} is not such an expression, because we restrict to the 1PI part of the commutator (up to a contact term) and it is not clear that this has the same property. In fact, we remind the reader that 1PI correlation functions are calculated from the effective action, which requires adding a background external current which is in principle a non-local function of the fields. As a result, it is not obvious that the $\ev{[J(x),J(y)]}_{\rm 1PI}$ vanishes for spacelike separated points $x,y$. More physically, $\Pi^{\mu\nu}$, in contrast to $G_\gamma^{\mu\nu}$ and $G_J^{\mu\nu}$, does not describe the reaction of the system to an external source and as such its microcausality is not guaranteed (cf. the contribution of Kirzhnitz in \cite{Keldysh1989TheDF}).

In fact, by definition, the self-energy tensor is given by the difference of the inverse of the photon propagator in the full and the free theory as given in \cref{photongreen}. From this relation any pole of $\Pi^{\mu\nu}$ corresponds to a zero of $G_\gamma^{\mu\nu}$ which seems to be harmless. On the other hand, a pole in $\Pi^{\mu\nu}$ might imply a singularity in $G_\gamma^{\mu\nu}$ at a different point. The reason is that near a pole $\Pi^{\mu\nu}$ diverges and therefore the equation $\frac{1}{g^2}(\Delta^{-1})^{\mu\nu}-\Pi^{\mu\nu}=0$ likely has a solution. By \cref{photongreen} this means that $G_\gamma^{\mu\nu}$ is singular. However, we could not find a reason for this singularity to necessarily lie in the FLC (and therefore not be allowed).\footnote{As an example take $\Pi^{\mu\nu}$ of the form $\Pi^{\mu\nu}=\frac{1}{g^2}(\Delta^{-1})^{\mu\nu}+{\rm pole}$. Then everything is perfectly consistent for $G_\gamma^{\mu\nu}$.} 

However, in the perturbative limit this additional pole is always near the singularity of $\Pi^{\mu\nu}$ and is thus not allowed. More precisely, in the limit that one can expand perturbatively in the coupling, it easy to see that the self-energy tensor must have the same analyticity region as the photon propagator by the following argument. Manipulating \cref{photongreen} yields
\eq{
	g^2\Pi=\Delta^{-1}-\Bigg[1+\sum_{n=1}g^{2n}\Delta^{-1}\cdot G_{(n)}\Bigg]^{-1}\cdot\Delta^{-1} \;,
}
where we have written $G_\gamma=g^2\Delta+\sum_{n=1}g^{2(n+1)}G_{(n)}$ and for brevity we have resorted to matrix notation for the spacetime indices with dot being matrix multiplication. By expansion we get
\eqa{
	\Pi_{(0)}&=\Delta^{-1}\cdot G_{(1)}\cdot\Delta^{-1} \,, \\ \Pi_{(1)}&=\Delta^{-1}\cdot G_{(2)}\cdot \Delta^{-1}-\Delta^{-1}\cdot G_{(1)}\cdot\Delta^{-1}\cdot G_{(1)}\cdot\Delta^{-1}\,,\quad\text{etc.}\,,
	\label{Pii}
} 
where we have written $\Pi=\sum_{n=0}g^{2n}\Pi_{(n)}$. 
Notice that since the full photon propagator $G^{\mu\nu}_\gamma$ vanishes outside the light cone in real space, and so is analytic when its imaginary part lies in the FLC in Fourier space, all the terms in the perturbative expansion are separately causal. This must be true since in the limit that the coupling goes to zero each term is parametrically smaller than the previous order, i.e.~if one of them is acausal this cannot be removed by the addition of other terms. As a result, by looking at \cref{Pii} and higher orders, we see that at each order $\Pi_{(i)}$ is constructed out of causal contributions and therefore it must be causal. 

So a singularity of $\Pi^{\mu\nu}$ in the FLC, if it exists, must be non-perturbative in the coupling. Notice that this singularity cannot be of the form of a branch point or essential singularity. The presence of a branch point or essential singularity in $\Pi^{\mu\nu}$ spoils the analyticity of $G_\gamma^{\mu\nu}$. As a result, if there is a singularity in the FLC it must be a pole.

The situation is different if we take into account the condition of positivity. In App.~\ref{landau} we show that a response function, parametrized as $(\omega,\vbq+\omega\vbxi)$ with the condition $\vbq\cdot\vbxi=0$, cannot have any zeros in the region of analyticity, i.e.~$\omega \in $ UHP. Applying this to $G_\gamma^{\mu\nu}$, we conclude that $G_\gamma^{-1}$ is analytic and therefore, by using \cref{photongreen}, so is $\Pi^{\mu\nu}(\omega,\vbq+\omega\vbxi)$ with $\vbq\cdot\vbxi=0$. Unfortunately, the theorem is not conclusive beyond the condition $\vbq\cdot\vbxi=0$. In the following, we will exploit this fact to write other forms of dispersion relations directly using $\Pi^{\mu\nu}$. 

\paragraph{Self-energy dispersion} Although we are not able to prove the analyticity of $\Pi^{\mu\nu}$ in ${\rm FLC}\subset\mathbb{C}^4$, even taking into account positivity, one can still write dispersion relations directly using $\Pi^{\mu\nu}$. For instance, we concluded above that the coefficient of the longitudinal part, $p^2\pi_L$,  is analytic in the UHP of $\omega$ assuming the parametrization $(\omega,\vbq+\omega\vbxi)$ with  $\vbq\cdot\vbxi=0$. Therefore, we can apply Leontovich's relation \cref{rkk2} to the combination $g^2\pi_L=\varepsilon-1$, and get
\eq{
\varepsilon(\omega,\vbq+\omega\vbxi)-1=\frac{1}{i\pi}{\rm PV}\int_{-\infty}^{+\infty}\frac{\dd{z}}{z-\omega}\,\,g^2\pi_L(z,\boldsymbol{q}+z\vbxi)\,,\qquad(\vbq\cdot\vbxi=0)\,.
\label{long2}
}  
Therefore, setting $\boldsymbol{q} = \boldsymbol{0}$ and $\omega =0$, we obtain
\begin{eBox}
\eq{
	\varepsilon(0,0)-1=\frac{2g^2}{\pi}\int_{0}^{+\infty}\frac{\dd{z}}{z}\,\,\Im\pi_L(z,z\vbxi)\,.
	\label{epsleo}
}
\end{eBox}
This relation is analogous to \cref{long} which was obtained from the analyticity of the photon propagator. In particular, from the positivity of the right-hand side we conclude that $\varepsilon(0,0)\geq1$; this rules out the other possibility $\varepsilon(0,0)<0$ discussed after \cref{long}.

Similarly, from the coefficient of the transverse part of $\Pi^{\mu\nu}$ we conclude that $k^2\pi_T$ is analytic with the same parametrization as above. Therefore, we can consider the combination\footnote{This combination is actually $\varepsilon_T-1$.}
\eq{
\chi_3\equiv\frac{g^2k^2\pi_T}{\omega^2}=(\varepsilon-1)+\frac{k^2}{\omega^2}\left(1-\frac{1}{\mu}\right)\,.
}
Assuming $\varepsilon \to 1$ and $\mu \to 1$ at high energy (see \cref{plasmapi}) this function goes to zero at infinity. By using Leontovich's relation we obtain 
\eq{
(\varepsilon(\omega,\vbq+\omega\vbxi)-1)+\frac{q^2+\omega^2\xi^2}{\omega^2}\left(1-\frac{1}{\mu(\omega,\vbq+\omega\vbxi)}\right)=\frac{1}{i\pi}{\rm PV}\int_{-\infty}^{+\infty}\frac{\dd{z}}{z-\omega}\,\,\frac{g^2(q^2+z^2\xi^2)}{z^2}\pi_T(z,\boldsymbol{q}+z\vbxi)\,,
} 
where we have used $\vbq\cdot\vbxi=0$. For nonzero $q$ the limit $\omega\to0$ is not well-defined. Setting $\boldsymbol{q} = \boldsymbol{0}$ we obtain
\eq{
	(\varepsilon(0,0)-1)+\xi^2\left(1-\frac{1}{\mu(0,0)}\right)=\frac{2g^2\xi^2}{\pi}\int_{0}^{+\infty}\frac{\dd{z}}{z}\,\,\Im\pi_T(z,z\vbxi)\,.
}
This is analogous to \cref{transvleon}. Once again, taking the limit $\xi\to0$ gives the same relation as \cref{epsleo}. However, sending $\xi\to1$ we obtain
\begin{eBox}
\eq{
	\varepsilon(0,0) -\frac{1}{\mu(0,0)} =\frac{2g^2}{\pi}\int_{0}^{+\infty}\frac{\dd{z}}{z}\,\,\Im\pi_T(z,z)\,.
	\label{epsmuleo}
}
\end{eBox}
Using the positivity of $\Im\pi_T$, we find $\varepsilon(0,0)\geq1/\mu(0,0)$ which is compatible with \cref{transvleon}.

%%%%%%%%%%%%%%%%%%%%%%%%%%%%%%%%%%%%%%%%%%%%%%%%%%%%
\section{Examples and improvements}\label{sec:stronger}
%%%%%%%%%%%%%%%%%%%%%%%%%%%%%%%%%%%%%%%%%%%%%%%%%%%%
In this section we discuss further the results derived above, focussing on the possibility of saturating the bounds and improving them under some stronger assumptions.

\paragraph{Models living on the boundaries} One can ask: is it possible, at least conceptually, to lie on the boundaries of the allowed region shown in Fig.~\ref{bound}? The horizontal boundary with $\varepsilon(0,0)=1$ corresponds to a medium without any electric response at low energies. Therefore, it would suffice to set $\pi_L \equiv 0$. Different points along the boundary $\varepsilon(0,0)=1$ then correspond to different limiting values of $\pi_T$. To have a mathematically consistent example let us consider the following form:
\eq{
	g^2\pi_T(\omega,\bsk)=\frac{\omega_T^2}{-\omega^2-i\gamma\omega+c_s^2k^2+\omega_0^2}\,.
	\label{piThor}
}
As discussed in App.~\ref{dampwave}, as long as $\gamma>0$, $0 \le c_s^2 \le 1$, $\omega_0^2>0$ and $\omega_T^2>0$ this is an analytic function when $\text{Im}(\omega,\bsk)$ lies in the FLC, with the correct positivity condition. Moreover, from $\mu^{-1}=1-g^2\pi_T$ we see that the condition $\mu(0,0)>0$ implies that $\omega_T^2<\omega_0^2$. {More physically, an ensemble of magnetic dipoles does not contribute to $\varepsilon$ and results in paramagnetism. The paramagnetic response can be arbitrarily large as we approach the Curie point, $\mu \to \infty$.}

The vertical boundary of Fig.~\ref{bound} corresponds to $\varepsilon(0,0)\mu(0,0)=1$. Therefore, we need both an electric and magnetic response to lie on the vertical boundary. Since it implies that the speed of light is unity, perhaps the easiest way is to start from imposing Lorentz invariance on $\pi_L$ and $\pi_T$: $\pi_L=-\frac{k^2}{p^2}\pi_T$. Therefore, we can consider 
\eq{
g^2\pi_L(\omega,k)=-\frac{k^2}{p^2}g^2\pi_T=\frac{\omega_L^2}{-\omega^2-i\varepsilon\omega+k^2+\omega_0^2}\,,
\label{piLver}
} 
which for $\omega_0^2>0$ and $\omega_L^2>0$ has the correct analyticity and positivity properties. In the Lorentz-invariant case we have $\varepsilon=\mu^{-1}=1+g^2\pi_L$ at all energies. One may wonder whether in \cref{piLver} any other function, with the correct properties, will also work. However, one must be careful about the positivity of $\pi_T=-p^2\pi_L/k^2$ since the prefactor $p^2$ changes sign. For instance, adding a subluminal speed of sound $c_s$ or finite decay width $\gamma$ to \cref{piLver}, while being consistent for $\pi_L$, is not consistent for $\pi_T$. It is, in fact, possible to provide a more physical example. Consider the theory
\eq{
\mathcal{L}=-\frac{1}{4g^2}F_{\mu\nu}F^{\mu\nu}-\frac{1}{2}(\p_\mu\phi)^2-\frac{1}{2}m^2\phi^2+\frac{\alpha}{\Lambda^2}\phi^2F_{\mu\nu}F^{\mu\nu}\,,
}   
in which $\phi$ is a massive scalar field. Integrating it out, say in some nontrivial background, gives corrections to the photon kinetic term that are always proportional to $F_{\mu\nu}F^{\mu\nu}$. As a result, the condition $\varepsilon=\mu^{-1}$ is satisfied. {The Lagrangian above describes a scalar particle with electric and magnetic polarizabilities which are equal and opposite, since $F_{\mu\nu}F^{\mu\nu} \propto E^2 - B^2$. This situation is rather common, since it corresponds to the dimension-6 operator above. In order to deviate from this relation one has to consider higher-order operators like $F^{\mu\alpha}F_\mu^\beta\partial_\alpha\phi\partial_\beta\phi$, which are naturally suppressed. For instance pions and kaons have polariazabilities which are approximately equal and opposite, see for example\cite{Holstein:1990qy}.}

\paragraph{Lower bound on dissipation} The bounds of Fig.~\ref{bound} can be improved if one has some knowledge about the right-hand side of \cref{epsleo} and \cref{epsmuleo} (equivalently \eqref{long} and \eqref{transvleon}). In particular one needs a lower bound on dissipation. One possibility is to assume the ``plasma" behavior at large $\omega$ as discussed in Sec.~\ref{high}. At high energy one has $\pi_L \sim - \omega_p^2/\omega^2$. This implies that the function $\omega^2 \pi_L(\omega,0)+\omega_p^2$ is analytic in the UHP of $\omega$ and decays at infinity. Writing the dispersion relation \eqref{rkk11} for $z_R =0$ gives
the well-known sum rule (see for instance \cite{Jackson:1998nia}):
\eq{
	\omega_p^2= \frac{2 g^2}{\pi} \int_0^{\infty} \omega \dd\omega \Im \pi_L(\omega,0)\,.
	\label{eq:sumplasma}
}  
One can use this equality to give a lower bound on the right-hand side of \cref{epsleo}. If dissipation happens at higher and higher frequencies, one can have an arbitrarily small right-hand side of  \cref{epsleo} while preserving the constraint \cref{eq:sumplasma}. If we assume that the imaginary part is zero above some $\omega_{\rm UV}$ then \cref{epsleo}, for $\boldsymbol{\xi} = \boldsymbol{0}$, becomes
\eq{
	\varepsilon(0,0)-1=\frac{2g^2}{\pi}\int_{0}^{\omega_{\rm UV}}\frac{\dd{z}}{z}\,\,\Im\pi_L(z,0) \geq \frac{2g^2}{\pi}\int_{0}^{\omega_{\rm UV}} \frac{z\dd{z}}{\omega_{\rm UV}^2}\,\,\Im\pi_L(z,0) = \frac{\omega_p^2}{\omega_{\rm UV}^2}\,.
	\label{epsleobound}
}
In reality the imaginary part will not vanish exactly above the frequency $\omega_{\rm UV}$, but it will decay typically as $\omega^{-3}$. Anyway the integrals over frequency are convergent, so the bound above will only receive relative corrections of order unity. In real materials this ratio is roughly of order unity.
{One can follow the same logic for $\pi_T$, which at high energy goes as $\pi_T \sim - \omega_p^2/(\xi^2 \omega^2)$, see \cref{plasmapi}. One obtains another sum rule for $\omega_p^2$, viz.
\eq{
\omega_p^2=\frac{2g^2\xi^2}{\pi}\int_{0}^{+\infty}z\dd{z} \Im\pi_T(z,z\vbxi) \;.
}
This can be used in \cref {epsmuleo} to give
\eq{
	(\varepsilon(0,0)-1)+\xi^2\left(1-\frac{1}{\mu(0,0)}\right) \ge \frac{\omega_p^2}{\omega_{\rm UV}^2}\,.
	\label{epstrbound}
}
Notice this does not give a bound on the low-frequency speed of light in the medium, given by $1/\sqrt{\varepsilon(0,0) \mu(0,0)}$.}

In passing it is worthwhile to mention another relation for $\mu(0,\boldsymbol{q})$ (other than \cref{mudisp1}) which involves the plasma frequency. Let us consider the function 
\eq{
\chi_4\equiv g^2k^2\pi_T=\omega^2(\varepsilon-1)+k^2\left(1-\frac{1}{\mu}\right)\,.
}   
From \cref{plasmapi}, in the limit $\omega\to\infty$ we have $\chi_4\to-\omega_p^2$. Therefore, we can apply Leontovich's relation to $\chi_4+\omega_p^2$ which gives
\eq{
\omega^2(\varepsilon(\omega,\vbq+\omega\vbxi)-1)+(q^2+\omega^2\xi^2)\left(1-\frac{1}{\mu(\omega,\vbq+\omega\vbxi)}\right)+\omega_p^2=\frac{1}{i\pi}{\rm PV}\int_{-\infty}^{+\infty}\frac{\dd{z}}{z-\omega}\left[g^2(q^2+z^2\xi^2)\pi_T(z,\boldsymbol{q}+z\vbxi)+\omega_p^2\right]\,,
}
assuming $\vbq\cdot\vbxi=0$. For nonzero $q$, taking the $\omega\to0$ limit gives
\eq{
q^2\left(1-\frac{1}{\mu(0,\boldsymbol{q})}\right)+\omega_p^2=\frac{2g^2}{\pi}\int_{0}^{+\infty}\frac{\dd{z}}{z}(q^2+z^2\xi^2)\Im\pi_T(z,\boldsymbol{q}+z\vbxi)\,.
\label{mudisp}
}
Restricting to $\boldsymbol{\xi} = \boldsymbol{0}$ we obtain
\eq{
\frac{q^2+\omega_p^2}{q^2}-\frac{1}{\mu(0,\boldsymbol{q})}=\frac{2g^2}{\pi}\int_{0}^{+\infty}\frac{\dd{z}}{z}\Im\pi_T(z,q)\,.
}
This relation was derived in \cite{Keldysh1989TheDF}. In particular, assuming the positivity condition, it implies $\mu(0,\boldsymbol{q})>q^2/(q^2+\omega_p^2)$.

\paragraph{Wider ``light"cone} Most materials consist of non-relativistic particles, i.e.~their typical velocity is much smaller than the speed of light. Does this fact have any implications for the allowed values for electric and magnetic response of the medium? The answer is probably yes. Let us consider a system of charged particles. In certain limits, the behavior of the system is described, in kinetic theory, by a single-particle distribution function $f(\boldsymbol{V})$ (see \cite{book:plasmapysics} for details). In this limit one can calculate the dielectric tensor, defined above \cref{epsmu}, as follows
\eq{
\varepsilon^{ij}(\omega,\boldsymbol{k})\equiv1+\frac{g^2}{\omega^2}\Pi^{ij}=\left(1-\frac{\omega_p^2}{\omega^2}\right)\delta^{ij}+\frac{\omega_p^2}{\omega^2}\int\dd[3]{\boldsymbol{V}}\frac{V^iV^j}{\omega+i\epsilon-\boldsymbol{V}\cdot\boldsymbol{k}}\,k^\ell\pdv{f(\boldsymbol{V})}{V^\ell}\,.
\label{plasmaeps}
} 
Notice that the dielectric tensor contains all the information about the electromagnetic properties of the medium. Eq.~(\ref{plasmaeps}) has the remarkable property that it is analytic in a larger region than the FLC; poles of \cref{plasmaeps} are located at $\omega_I=\boldsymbol{V}\cdot\boldsymbol{k}_I$ which are far from the FLC for non-relativistic velocities $|\boldsymbol{V}|\ll1$. Intuitively, the reason is that a change in the total electromagnetic field at a point $\boldsymbol{x}$ can induce a current in a different point $\boldsymbol{x}'$, mainly due to the movement of charged particles from $\boldsymbol{x}$ to $\boldsymbol{x}'$, and this occurs at low velocity. More precisely, in the weak field limit, one has to solve the linearized equation of motion for $f(\boldsymbol{V})$ perturbations, i.e.~the Vlasov equation, to obtain \cref{plasmaeps}. This equation predicts a slower propagation of information; the factor $(\omega+i\epsilon-\boldsymbol{V}\cdot\boldsymbol{k})^{-1}$ is in fact the propagator of the linearized Vlasov equation. 

The above description is only valid if one can neglect (thermal or quantum) fluctuations in the system. Indeed in order to be able to solve for the single-particle distribution function, $f(\boldsymbol{V})$, one generally requires the knowledge of multi-particle correlation functions. It turns out that for sufficiently dense systems, all the multi-particle correlation functions are expressible in terms of $f(\boldsymbol{V})$, as for instance in \cref{plasmaeps}, which is the two-point function of the current. In particular the process of exchanging a photon from $\boldsymbol{x}$ to $\boldsymbol{x}'$, which propagates relativistically, can be neglected.\footnote{In the condensed matter literature, this is usually called the random phase approximation (RPA).}

In a situation where the above effects are negligible, the analyticity region is effectively larger. This means that the parameter $\xi$ used in the dispersion relations can take on values larger than unity. Equivalently, in real space, it corresponds to a response function that vanishes outside a narrower cone than the relativistic cone. For a non-relativistic system with typical velocity $v\ll1$, the parameter $\xi$ can be as large as $1/v$. The allowed region for $\varepsilon(0,0)$ and $\mu(0,0)$ shrinks as the vertical boundary is modified as
\eq{
	\varepsilon(0,0)\geq\frac{1}{v^2\mu(0,0)}-\frac{1-v^2}{v^2}\,,
	\label{newbound}
} 
where we have set $\xi=1/v$ in \cref{eq:xibound}. The modified vertical boundary is shown in Fig.~\ref{bound}. Notice that this affects the allowed region only on the diamagnetic side while the paramagnetic side remains unchanged. This is in particular very interesting since the experimentally measured values for $|\mu-1|$ for diamagnetic materials, in most cases, are extremely small $\sim10^{-5}$.  
\fg{
	\includegraphics[width=0.43\textwidth]{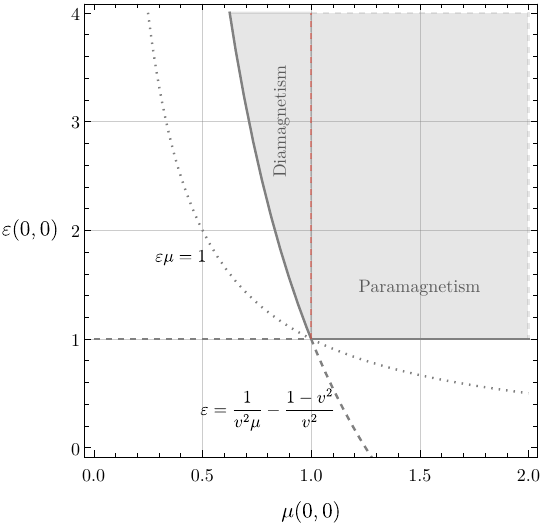}\hfil
	\includegraphics[width=0.45\textwidth]{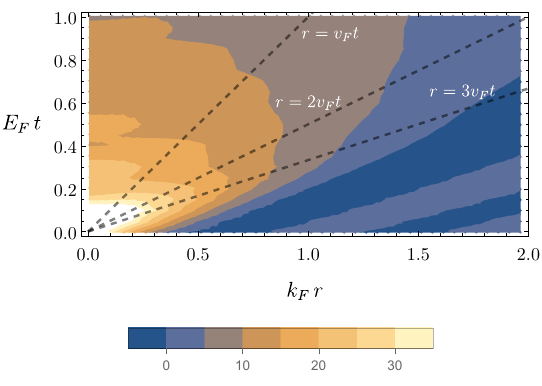}
	\caption{Left: Improved (speculative) bound on $\varepsilon$ and $\mu$ from lower speed of particles. The new curve is for $v^2=0.1$. For more realistic values like $v^2\sim10^{-4}$ the diamagnetic region will be extremely small. Right: Inverse Fourier transform of the longitudinal Lindhard function, given in \cref{piLNR}. The plot shows that the response is very small outside a narrow cone (controlled by $v_F$) compared to the relativistic cone (corresponding to $c=1$).}
	\label{bound4}
}

As an example, we can consider the non-relativistic Lindhard function given in \cref{piLNR} and \cref{piTNR}. For small momenta $k\ll k_F$, the function involves $\log(\omega\pm v_Fk)$. Therefore, it has  larger region of analyticity; the parameter $\xi$ can be increased up to values $\sim1/v_F$. However, for larger momenta quantum mechanical effects become important and the above statement is no longer true; remember that, as discussed in App.~\ref{linhar}, the relativistic expressions \cref{piLR} and \cref{piTR} have the correct analyticity properties. We have checked explicitly for the Lindhard function that, although the region of analyticity is not strictly speaking larger, Leontovich's relation is still satisfied for values $1 \ll \xi \lesssim 1/(3v_F)$, to a very good approximation. This is consistent with the fact that the Lindhard response function in real space is very small outside the $r=3v_Ft$ cone as depicted in \cref{bound4}.

\paragraph{Closure of the contour at high energy} 
As we discussed above, for condensed matter media one can close the arc in the upper half-plane of Fig.~\eqref{rkk11} at energies below the electron mass and thus disregard the vacuum loops of electrons and other charged particles. However, this is not possible in general: in the case of nuclear matter, for instance, only going to energies well above MeV the medium becomes negligible and the contour in the complex plane can be closed.\footnote{We assume to remain below the energies that characterise the spontaneous breaking of the electroweak symmetry, otherwise one should take into account the full $SU(2) \cross U(1)$ structure. For a non-Abelian group the discussion is qualitatively different, as for instance in asymptotic freedom; this is related to the fact the current is no longer a gauge-invariant operator.}  The vacuum polarisation due to loops of electrons and other charged particles is effectively a medium, with the only difference that the response is now Lorentz-invariant. This symmetry implies that quantities can only depend on $p^2$, and not separately on $\omega$ and $k$, and enforces the relation $\pi_L=-k^2\pi_T/p^2$. For energies well above the mass of the electron the longitudinal response reads \cite{Peskin:1995ev}
\eq{
\pi_{L,{\rm qed}}(p^2) = \frac{1}{12 \pi^2} \left[\log (\frac{-(\omega+ i \epsilon)^2 +k^2}{m^2}) - \frac53 +{\cal O}\left(\frac{m^2}{p^2}\right) \right] \;,
\label{QEDlong}
}
where we took the $i \epsilon$ prescription appropriate for the retarded Green's function. This expression does not decay at infinity so one cannot neglect the integration over the large circle, which is a necessary step to derive the Kramers-Kronig and Leontovich relations. One is forced to limit the integration over the real line up to a maximum frequency, $|\omega| < \omega_{\rm UV}$, and at the same time keep the contribution of the semicircle with radius $\omega_{\rm UV}$. Notice that the effects of vacuum loops are perturbative, i.e.~suppressed by the QED coupling $g^2$, so one can disregard them if the effect of the medium of interest gives corrections which are parametrically larger. It is however quite simple and physically instructive to take into account the effects which are enhanced by the potentially large logarithm. 

Let us study for instance how \cref{epsleo} is modified including the contribution from the arc at large energies
\eq{
	\varepsilon(0,0)-1=\frac{2g^2}{\pi}\int_{0}^{\omega_{\rm UV}}\frac{\dd{z}}{z}\,\,\Im\pi_L(z,0) + \frac{g^2}{\pi} \Im \int_{\cap_{{\rm UV}}} \frac{\dd{z}}{z}\,\,\pi_L(z,0)\,,
} 
where we took $\boldsymbol{\xi} = \boldsymbol{0}$ for simplicity. {The last term evaluates to $\frac{g^2}{12 \pi^2}\left[\log(\omega_{\rm UV}^2/m^2)-\frac53 \right]$. This equals $\Re\varepsilon(\omega_{\rm UV},0) -1$ (see Eq.~\eqref{QEDlong}) and gives the QED coupling at the scale $\omega_{\rm UV}$ (notice that one usually defines the running coupling for Euclidean momenta, while here we have timelike momenta and this is the reason why we also have an imaginary part).} Therefore one can rewrite the dispersion relation as 
\eq{
	\varepsilon(0,0)-\Re\varepsilon(\omega_{\rm UV},0)=\frac{2g^2}{\pi}\int_{0}^{\omega_{\rm UV}}\frac{\dd{z}}{z}\,\,\Im\pi_L(z,0) \,.
}
This result makes perfect sense physically. In the presence of vacuum loops, the value of $\varepsilon$ runs with the energy: the dispersion relation gives the increase of $\varepsilon$ compared to the UV, as a consequence of the medium.\footnote{The right-hand side of the equation above is always positive both in vacuum, where only loops of charged particles are present, and when extra matter is present. However one cannot, in general, separate the two contributions and argue that each one gives a positive contribution to the imaginary part. In particular there is no guarantee that the right-hand side increases when adding matter to the vacuum.} One expects this to hold at any order in perturbation theory. Going back to Fig.~\ref{bound}, one can say that the horizontal boundary becomes energy-dependent, since it corresponds to the value of $\varepsilon$ at the UV scale. The other boundary, which corresponds to the speed of light, is not affected by vacuum loops, because they are Lorentz-invariant. 

%%%%%%%%%%%%%%%%%%%%%%%%%%%%%%%%%%%%%%%%%%%%%%%%%%%%
\section{Conclusions and future directions}\label{conc}
%%%%%%%%%%%%%%%%%%%%%%%%%%%%%%%%%%%%%%%%%%%%%%%%%%%%
In this paper we studied bounds on the electromagnetic properties of a homogeneous, isotropic and passive medium by using the requirements of microcausality and assuming that the effect of the medium is negligible at high energies. This is a classic topic in condensed matter physics and we revisited it with a scope and a language  which are more connected with high-energy physics and the recent activity in the $S$-matrix bootstrap. From this point of view, our results are constraints on the leading operators in the low-energy effective field theory of photons after integrating out the medium. The main results are the dispersion relations for $\varepsilon$ and $\mu$, Eqns. \eqref{epsleo} and \eqref{epsmuleo}, and Fig.~\ref{bound}.

To conclude we comment on various possible extensions of our work.

\paragraph{Derivatives} What about higher-order terms in the effective theory? In the Lorentz-invariant case, in a weakly coupled theory with a gap, all the higher-order terms in the low-energy expansion of the amplitude are positive \cite{Adams:2006sv}. As we will argue below, this is not generally true in the present case. Let us consider a generic response function $\chi(\omega)$, where for simplicity we have ignored spatial dispersion (or set $\boldsymbol{\xi} = \boldsymbol{0}$). By using \cref{fullim} we can write
\eq{
\dv[n]{\chi}{\omega}\bigg\vert_{\omega=0}=\frac{n!}{\pi}\int_{-\infty}^{+\infty}\dd{z}\frac{\Im\chi(z)}{(z-i\epsilon)^{n+1}}\,.
\label{chid}
}
If the imaginary part has a gap, i.e.~$\Im\chi(z)=0$ for $|z| < m$, then \cref{chid} implies that the derivatives are zero for odd $n$ and positive for even $n$. However, as discussed in \S\ref{positiv}, generally this is not the case and the imaginary part is nonzero at all energies down to $\omega\to0$ where it goes to zero. (This is similar to what happens in the $S$-matrix case, when considering loops in the EFT, see for instance \cite{Bellazzini:2020cot}.) More explicitly, for $n=1$, we conclude from \cref{chid}, using the fact that the imaginary part is an odd function, that
\eq{
\chi'(0)=\frac{4i}{\pi}\int_0^{+\infty}\dd{z}\frac{\epsilon z}{(z^2+\epsilon^2)^2}\Im\chi(z)\,,
}
which is of the form $i\times{\rm positive}$. The positivity of $\chi'(0)/i$ is consistent with the fact that we can write $i\Im\chi(\omega) = \chi'(0)\omega+\dots$ as $\omega \rightarrow 0$. On the other hand, for $n=2$ we obtain 
\eq{
\chi''(0)=\frac{4}{\pi}\int_0^{+\infty}\dd{z}\frac{z(z^2-3\epsilon^2)}{(z^2+\epsilon^2)^3}\Im\chi(z)\,,
}     
from which we cannot deduce a definite sign for $\chi''(0)$. The same conclusion holds for higher orders. Notice that for this result, it is crucial to take $\epsilon\to0$ only after performing the integral. More precisely, one can split the integral into two pieces $(0,R)$ and $(R,+\infty)$ for some finite value $R$. For the interval $(R,+\infty)$ the $\epsilon$ terms can be neglected and the result is positive. We choose $R$ such that the imaginary part can be approximated linearly in the first interval. Therefore, we obtain
\eq{
\chi''(0)=\frac{4}{\pi}\left( -\frac{\chi'(0)/i}{R}+\int_R^{+\infty}\dd{z}\frac{\Im\chi(z)}{z^3}\right) \,.
\label{chidd}
}     
The first term is always negative, so $\chi''(0)$ is either positive or negative depending on which term is dominant. One arrives at the same conclusion by considering the ``arc" variables introduced in \cite{Bellazzini:2020cot} as follows. Consider integrating the function $\chi(z)/z^3$ along the contour shown below. Neglecting the arc at infinity we obtain
\eq{
\adjustbox{valign=c}{\tikzset{every picture/.style={line width=0.85pt}}

\definecolor{ultramarine}{rgb}{0.07, 0.04, 0.56}
\definecolor{jasper}{rgb}{0.84, 0.23, 0.24}

\begin{tikzpicture}

\tikzset{ma/.style={decoration={markings,mark=at position 0.35 with {\arrow[scale=0.8]{>}}},postaction={decorate}}}
\tikzset{ma2/.style={decoration={markings,mark=at position 0.5 with {\arrow[scale=0.8]{<}}},postaction={decorate}}}

\draw[gray,thin] (-2.5,0) -- (2.5,0)[->]; 
\draw[gray,thin] (0,0) -- (0,2.5)[->]; 

%\draw[ma,color=blue] (-2,0) -- (2,0) arc(0:180:2) --cycle;
\draw[ma,color=ultramarine,thin] (2,0) arc (0:180:2);
\draw[ma,color=jasper,thin] (0.5,0) arc (0:180:0.5);
\draw[ma,color=ultramarine,thin] (-2,0) -- (-0.5,0);
\draw[ma,color=ultramarine,thin] (0.5,0) -- (2,0);

\draw[gray] (2.2,2.2) node{\footnotesize $z$}; 
\draw[gray,thin] (2,2.3)|-(2.3,2);
%\draw[black] (-1.7,1.7) node{\footnotesize $C$};
\draw[black] (0.6,0.6) node{\footnotesize $C_R$};  

%\filldraw[gray] (1,0) circle (0.6pt);

%\filldraw[blue] (5,0) circle (0.8pt);
%\filldraw[blue] (9,0) circle (0.8pt);
%\filldraw[blue] (7.5,0) circle (0.8pt);
%\draw[black] (1,0) node {\tikzxmark};
%\draw[black] (3,0) node {\tikzxmark};
%\draw[black] (1.5,-0.2) node {\tikzxmark};
%\draw[black] (3.5,-0.2) node {\tikzxmark};
% 
%\draw[black] (4.7,-0.3) node{\footnotesize $C_\infty$}; 
%
%\draw[black] (1,0.3) node{\footnotesize $t_1$}; 
%\draw[black] (2,0.3) node{\footnotesize $\dots$}; 
%\draw[black] (3,0.3) node{\footnotesize $t_m$}; 
%
%\draw[black] (1.5,-0.5) node{\footnotesize $t'_1$};
%\draw[black] (2.25,-0.5) node{\footnotesize $\dots$}; 
%\draw[black] (3.5,-0.5) node{\footnotesize $t'_n$}; 

\end{tikzpicture}}\qquad \qquad
\int_{C_R}\dd{z}\frac{\chi(z)}{z^3}=2i\int_R^{+\infty}\dd{z}\frac{\Im\chi(z)}{z^3}\,.
\label{arc}
}
For small enough $R$, let us say below a cutoff $\Lambda$, an effective description exists such that we can expand $\chi(z)\simeq\chi(0)+\chi'(0)z+\chi''(0)z^2/2+\dots$ which can be used to evaluate the integral over the arc $C_R$ in \cref{arc}. One can check that, except for $\chi''(0)$, the contribution of all the even derivatives vanishes while for the odd derivatives, except for $\chi'(0)$, they are suppressed by powers of $R/\Lambda$ which can be neglected. Rearranging these terms results in \cref{chidd}. Even though Eq.~\eqref{chidd} does not fix the sign of $\chi''(0)$, the inequality $\chi''(0)+4/(\pi R) \cdot \chi'(0)/i >0$ contains valuable information. If $\chi''(0) <0$, it puts an upper bound on $R$ and therefore on $\Lambda$: for larger values of $R$ the EFT must break down so that the expansion of $\chi$ used to derive Eq.~\eqref{chidd} is not valid anymore.
Eq.~\eqref{chidd} also implies that even higher derivatives are positive in cases where dissipation at low energies can be made parametrically small.\footnote{This is analogous to neglecting loops in the EFT for $S$-matrix calculations \cite{Bellazzini:2020cot}.} As an example, let us consider the Drude model studied in App.~\ref{dampwave} (see also \cref{piThor}). One can check that for small friction $\gamma\ll \omega_0$ all even higher derivatives are positive, while for $\gamma\gg \omega_0$ there is no definite sign for the even derivatives. 

One interesting direction to explore is based on the knowledge of the sign of the imaginary part in the complex plane, see Eq.~\eqref{land2}. This property is related to the so-called typically real functions, which have been instrumental to the $S$-matrix bootstrap, see for instance \cite{Raman:2021pkf}. One might be able to derive two-sided bounds on the relative coefficients in the above low-energy expansion.\footnote{We thank A.~Zahed for discussion about this point.}

\paragraph{Several complex variables} While the retarded Green's function is analytic in an open domain of $\mathbb{C}^4$, we have only used single-variable complex analysis. It would be interesting to try to use the complete power of complex analysis with multiple variables, e.g.~writing dispersion relations involving integration over the boundary of the forward light cone. Among others this could be useful for completing the proof of analyticity of $\Pi^{\mu\nu}$, deriving more powerful results on the domain of analyticity (see \cite{Sommer:1970mr} for the Lorentz-invariant case) and understanding the constraints on the imaginary part from causality (\cref{imcons}). Notice that constraints on the imaginary part of the response can be converted into statements about the fluctuations of the system via the fluctuation-dissipation theorem.

\paragraph{Other systems} In the discussion above we stuck to the case of dielectrics. It would be natural to apply the same methods to media with a different low-energy limit, namely conductors and superconductors, or to crystals where translations are broken to a discrete subgroup. Studying the two-point function of the stress-energy tensor one should be able to derive constraints on transport and on the EFT that describes fluids (see \cite{Hartman:2017hhp,Delacretaz:2021ufg,Heller:2022ejw,Heller:2023jtd} for progress in this regard). In this case one could test the results with many examples of fluids derived from strongly coupled theories via the AdS/CFT correspondence. The bounds derived above can also be interpreted, in the limit in which the system is made of a dilute collection of particles, as bounds on the electric and magnetic polarizability of single particles. It would be interesting to explore the connection between these statements and the positivity bounds derived using the $S$-matrix. Eventually the same logic can be applied to the interaction with the gravitational field.

\paragraph{Inflation} We have been discussing systems in which there is a spontaneous breaking of Lorentz invariance. In inflation one is interested in the breaking of de Sitter isometries preserving rotations, spacial translations and (approximate) dilations. As a starting point in the study of the bounds in this setup, one should investigate the general properties of two-point functions imposed by microcausality and positivity.

\paragraph{} We hope to make progress in all these directions.

\paragraph{Acknowledgements} We thank A.~Altland, B.~Holstein, L.~Delacrétaz,  G.~La Rocca, A.~Longo, M.~Mirbabayi, G.~Villadoro and A.~Zhiboedov for useful discussions. We thank L.~Delacrétaz, A.~Podo and L.~Santoni for comments on the manuscript. L.S. is supported by the SNSF grant 200021 213120.

\appendix

%%%%%%%%%%%%%%%%%%%%%%%%%%%%%%%%%%%%%%%%%%%%%%%%%%%%
\section{Review of the Closed Time Path formalism}\label{ctp}
%%%%%%%%%%%%%%%%%%%%%%%%%%%%%%%%%%%%%%%%%%%%%%%%%%%%
The Closed Time Path (CTP) formalism is useful for studying the evolution of expectation values of operators. We assume the initial state of the system is specified by a density matrix $\rho$. The expectation value of a local Hermitian operator $\phi$ is then given by
\eq{
	\ev{\phi(x)}_\rho\equiv\Tr[\rho\phi(x)]=\Tr[ U(t_i,t)\phi(\vbx)U(t,t_i)\rho]\,.
} 
In the first expression $\phi(x)$ is the Heisenberg operator at $x = (t,\vbx)$ while in the last expression $\phi(\vbx)$ is the Schr\"{o}dinger operator defined at some initial time $t_i$. The operator $U(t_2,t_1)$ accounts for the time evolution from $t_1$ to $t_2$. Using the definition of trace and introducing resolutions of the identity $\mathbbm{1} = \int \mathcal{D}\phi\ket{\phi}\bra{\phi}$ -- the integration is over field configurations at constant time -- this can be expressed as 
\eq{
	\ev{\phi(x)}_\rho=\int \mathcal{D}\phi \mathcal{D}\phi' \mathcal{D}\phi'' \bra{\phi''}U(t_i,t)\ket{\phi} \phi(\vbx) \bra{\phi}U(t,t_i)\ket{\phi'} \bra{\phi'}\rho\ket{\phi''}\,.
	\label{evphi}
} 
This expression is useful because we have a path integral representation for the amplitudes,
\eq{
	\bra{\phi}U(t,t_i)\ket{\phi'}=\int^{\vp(t)=\phi}_{\vp(t_i)=\phi'}\D \vp\, e^{iS[\vp]{\big\vert}^t_{t_i}}\,,
}
similarly for the other factor. Plugging this back into \cref{evphi} we obtain
\eq{
	\ev{\phi(x)}_\rho=\int^{\phi_1(t)=\phi_2(t)}\D \phi_1\,\D\phi_2\, \bra{\phi_1(t_i)}\rho\ket{\phi_2(t_i)} \phi_1(t,\vbx) ~ e^{i(S[\phi_1]-S[\phi_2]){\big\vert}^t_{t_i}} \,.
	\label{phictp}
} 
Note that in the integrand the factor $\phi_1(t,\vbx)$ can be replaced by $\phi_2(t,\vbx)$ since the two variables are forced to be equal at time $t$. Also notice there are no restrictions on the initial conditions of the variables; this information is encoded in $\rho$. This expression has a convenient graphical representation as follows:
\eq{
	\ev{\phi(x)}_\rho=\adjustbox{valign=c}{\tikzset{every picture/.style={line width=0.75pt}}

\definecolor{ultramarine}{rgb}{0.07, 0.04, 0.56}

\begin{tikzpicture}

\tikzset{ma/.style={decoration={markings,mark=at position 0.4 with {\arrow[scale=0.8]{>}}},postaction={decorate}}}
\tikzset{ma2/.style={decoration={markings,mark=at position 0.6 with {\arrow[scale=0.8]{<}}},postaction={decorate}}}

\draw[color=ultramarine] (-0.1,0) -- (-0.1,0.1);  
\draw[ma,color=ultramarine] (-0.1,0) -- (3,0); 
\draw[color=ultramarine] (3,0) -- (3,-0.2); 
\draw[ma2,color=ultramarine] (-0.1,-0.2)-- (3,-0.2);  
\draw[color=ultramarine] (-0.1,-0.2) -- (-0.1,-0.3); 
%\filldraw[ultramarine] (0,0) circle (0.8pt);
%\filldraw[ultramarine] (0,-0.2) circle (0.8pt);
\draw[black] (3,0) node {\tikzxmark}; 
\draw[black] (0,-0.5) node{\footnotesize $t_i$}; 
\draw[black] (3,-0.5) node{\footnotesize $t$}; 
\draw[black] (-0.3,-0.1) node{\footnotesize $\rho$}; 
\draw[black] (3,0.3) node{\footnotesize $\phi(t,\vbx)$}; 
%\draw[blue,densely dotted] (0,0) .. controls(-0.1,-0.1) .. (0,-0.2);
\end{tikzpicture}}\,.
}
Every portion of the graph corresponds to a factor in the path integral \cref{phictp}. A better way to express this is to add a forward and backward integration to some final time $t_f$,
\eq{
	\spl{
		\ev{\phi(x)}_\rho&=\Tr[ U(t_i,t_f)U(t_f,t)\phi(\vbx)U(t,t_i)\rho]\\
		&=\int^{\phi_1(t_f)=\phi_2(t_f)}\D \phi_1\,\D\phi_2\, \bra{\phi_1(t_i)}\rho\ket{\phi_2(t_i)} \phi_1(t,\vbx) \, e^{i(S[\phi_1]-S[\phi_2]){\big\vert}^{t_f}_{t_i}} \\
		&=\adjustbox{valign=c}{\tikzset{every picture/.style={line width=0.75pt}}

\definecolor{ultramarine}{rgb}{0.07, 0.04, 0.56}

\begin{tikzpicture}

\tikzset{ma/.style={decoration={markings,mark=at position 0.5 with {\arrow[scale=0.8]{>}}},postaction={decorate}}}
\tikzset{ma2/.style={decoration={markings,mark=at position 0.5 with {\arrow[scale=0.8]{<}}},postaction={decorate}}}

\draw[color=ultramarine] (-0.1,0) -- (-0.1,0.1);   
\draw[ma,color=ultramarine] (-0.1,0) -- (3,0); 
\draw[ma,color=ultramarine] (3,0) -- (5,0); 
\draw[color=ultramarine] (5,0) -- (5,-0.2); 
\draw[ma2,color=ultramarine] (-0.1,-0.2)-- (5,-0.2) ;
\draw[color=ultramarine] (-0.1,-0.2) -- (-0.1,-0.3);  
%\filldraw[blue] (0,0) circle (0.8pt);
%\filldraw[blue] (0,-0.2) circle (0.8pt);
\draw[black] (3,0) node {\tikzxmark}; 
\draw[black] (0,-0.5) node{\footnotesize $t_i$}; 
\draw[black] (5,-0.5) node{\footnotesize $t_f$}; 
\draw[black] (-0.3,-0.1) node{\footnotesize $\rho$}; 
\draw[black] (3,0.3) node{\footnotesize $t$}; 
%\draw[blue,densely dotted] (0,0) .. controls(-0.1,-0.1) .. (0,-0.2);
\end{tikzpicture}}\\
		&=\adjustbox{valign=c}{\tikzset{every picture/.style={line width=0.75pt}}

\definecolor{ultramarine}{rgb}{0.07, 0.04, 0.56}

\begin{tikzpicture}

\tikzset{ma/.style={decoration={markings,mark=at position 0.5 with {\arrow[scale=0.8]{>}}},postaction={decorate}}}
\tikzset{ma2/.style={decoration={markings,mark=at position 0.5 with {\arrow[scale=0.8]{<}}},postaction={decorate}}}

\draw[color=ultramarine] (-0.1,0) -- (-0.1,0.1);    
\draw[ma,color=ultramarine] (-0.1,0) -- (5,0); 
\draw[color=ultramarine] (5,0) -- (5,-0.2); 
\draw[ma2,color=ultramarine] (-0.1,-0.2)-- (3,-0.2) ; 
\draw[ma2,color=ultramarine] (3,-0.2)-- (5,-0.2) ;
\draw[color=ultramarine] (-0.1,-0.2) -- (-0.1,-0.3); 
     
%\filldraw[blue] (0,0) circle (0.8pt);
%\filldraw[blue] (0,-0.2) circle (0.8pt);
\draw[black] (3,-0.2) node {\tikzxmark}; 
\draw[black] (0,-0.5) node{\footnotesize $t_i$}; 
\draw[black] (5,-0.5) node{\footnotesize $t_f$}; 
\draw[black] (-0.3,-0.1) node{\footnotesize $\rho$}; 
\draw[black] (3,-0.5) node{\footnotesize $t$}; 
%\draw[blue,densely dotted] (0,0) .. controls(-0.1,-0.1) .. (0,-0.2);
\end{tikzpicture}}\,.
	}
\label{phivevrho}
}
In this way the only dependence on $t$ appears in the location of the operator insertion and one can send the initial and final times to infinity. Finally, notice that the insertion may be put either in the forward or the backward path. In fact, more generally, as long as we do not pass through other insertions we can slide points from forward to backward paths and vice versa.

Adding more points is straightforward. For instance for two points we have
\eq{
	\spl{
		\ev{\phi(x)\phi(x')}_\rho&=\Tr[ U(t_i,t)\phi(\vbx)U(t,t')\phi(\vbx')U(t',t_i)\rho]\\
		&=\begin{cases}
			&t>t'\quad\adjustbox{valign=c}{\tikzset{every picture/.style={line width=0.75pt}}

\definecolor{ultramarine}{rgb}{0.07, 0.04, 0.56}

\begin{tikzpicture}

%\tikzset{ma/.style={decoration={markings,mark=at position 0.5 with {\arrow[scale=0.8]{>}}},postaction={decorate}}}
%\tikzset{ma2/.style={decoration={markings,mark=at position 0.5 with {\arrow[scale=0.8]{<}}},postaction={decorate}}}
 
\draw[color=ultramarine] (-0.1,0.1)--(-0.1,0) -- (4,0)-- (4,-0.2)--(-0.1,-0.2)--(-0.1,-0.3) ; 

%\filldraw[blue] (0,0) circle (0.8pt);
%\filldraw[blue] (0,-0.2) circle (0.8pt);
\draw[black] (1.5,0) node {\tikzxmark};
\draw[black] (2.5,0) node {\tikzxmark};

\draw[black] (-0.3,-0.1) node{\footnotesize $\rho$}; 
 
\draw[black] (0,-0.5) node{\footnotesize $t_i$}; 
\draw[black] (4,-0.5) node{\footnotesize $t_f$}; 

\draw[black] (1.5,0.3) node{\footnotesize $t'$}; 
\draw[black] (2.5,0.3) node{\footnotesize $t$}; 

\end{tikzpicture}}\quad\text{or}\quad\adjustbox{valign=c}{\tikzset{every picture/.style={line width=0.75pt}}

\definecolor{ultramarine}{rgb}{0.07, 0.04, 0.56}

\begin{tikzpicture}

%\tikzset{ma/.style={decoration={markings,mark=at position 0.5 with {\arrow[scale=0.8]{>}}},postaction={decorate}}}
%\tikzset{ma2/.style={decoration={markings,mark=at position 0.5 with {\arrow[scale=0.8]{<}}},postaction={decorate}}}
 
\draw[color=ultramarine] (-0.1,0.1)--(-0.1,0) -- (4,0)-- (4,-0.2)--(-0.1,-0.2)--(-0.1,-0.3) ; 

%\filldraw[blue] (0,0) circle (0.8pt);
%\filldraw[blue] (0,-0.2) circle (0.8pt);
\draw[black] (1.5,0) node {\tikzxmark};
\draw[black] (2.5,-0.2) node {\tikzxmark};

\draw[black] (-0.3,-0.1) node{\footnotesize $\rho$}; 
 
\draw[black] (0,-0.5) node{\footnotesize $t_i$}; 
\draw[black] (4,-0.5) node{\footnotesize $t_f$}; 

\draw[black] (1.5,0.3) node{\footnotesize $t'$}; 
\draw[black] (2.5,-0.5) node{\footnotesize $t$}; 

\end{tikzpicture}}\\
			&t<t'\quad\adjustbox{valign=c}{\tikzset{every picture/.style={line width=0.75pt}}

\definecolor{ultramarine}{rgb}{0.07, 0.04, 0.56}

\begin{tikzpicture}

%\tikzset{ma/.style={decoration={markings,mark=at position 0.5 with {\arrow[scale=0.8]{>}}},postaction={decorate}}}
%\tikzset{ma2/.style={decoration={markings,mark=at position 0.5 with {\arrow[scale=0.8]{<}}},postaction={decorate}}}
 
\draw[color=ultramarine] (-0.1,0.1)--(-0.1,0) -- (4,0)-- (4,-0.2)--(-0.1,-0.2)--(-0.1,-0.3) ; 

%\filldraw[blue] (0,0) circle (0.8pt);
%\filldraw[blue] (0,-0.2) circle (0.8pt);
\draw[black] (1.5,-0.2) node {\tikzxmark};
\draw[black] (2.5,0) node {\tikzxmark};
 
\draw[black] (-0.3,-0.1) node{\footnotesize $\rho$}; 
 
\draw[black] (0,-0.5) node{\footnotesize $t_i$}; 
\draw[black] (4,-0.5) node{\footnotesize $t_f$}; 

\draw[black] (1.5,-0.5) node{\footnotesize $t$}; 
\draw[black] (2.5,0.3) node{\footnotesize $t'$}; 

\end{tikzpicture}}\quad\text{or}\quad\adjustbox{valign=t}{\tikzset{every picture/.style={line width=0.75pt}}

\definecolor{ultramarine}{rgb}{0.07, 0.04, 0.56}

\begin{tikzpicture}

%\tikzset{ma/.style={decoration={markings,mark=at position 0.5 with {\arrow[scale=0.8]{>}}},postaction={decorate}}}
%\tikzset{ma2/.style={decoration={markings,mark=at position 0.5 with {\arrow[scale=0.8]{<}}},postaction={decorate}}}
 
\draw[color=ultramarine] (-0.1,0.1)--(-0.1,0) -- (4,0)-- (4,-0.2)--(-0.1,-0.2)--(-0.1,-0.3) ; 

%\filldraw[blue] (0,0) circle (0.8pt);
%\filldraw[blue] (0,-0.2) circle (0.8pt);
\draw[black] (1.5,-0.2) node {\tikzxmark};
\draw[black] (2.5,-0.2) node {\tikzxmark};

\draw[black] (-0.3,-0.1) node{\footnotesize $\rho$}; 
 
\draw[black] (0,-0.5) node{\footnotesize $t_i$}; 
\draw[black] (4,-0.5) node{\footnotesize $t_f$}; 

\draw[black] (1.5,-0.5) node{\footnotesize $t$}; 
\draw[black] (2.5,-0.5) node{\footnotesize $t'$}; 

\end{tikzpicture}}
		\end{cases}\\
		&=\int^{\phi_1(t_f)=\phi_2(t_f)}\D \phi_1\,\D\phi_2\,\bra{\phi_1(t_i)}\rho\ket{\phi_2(t_i)} \, \, e^{i(S[\phi_1]-S[\phi_2]){\big\vert}^{t_f}_{t_i}}\\
		&\qquad\qquad\qquad\times\begin{cases}
			&t>t'\quad \phi_1(x')\phi_1(x) \quad\text{or}\quad \phi_1(x')\phi_2(x)\\
			&t<t'\quad \phi_1(x')\phi_2(x) \quad\text{or}\quad \phi_2(x')\phi_2(x)
		\end{cases}\,.
	}
}
The important point here is whether the evolution $U(t,t')$ is forward or backward, and notice we have used the sliding trick to obtain the other representation. The $\phi_2$'s correspond to anti time-ordering. Moreover, the fields associated to $\phi_2$'s always appear to the left of the $\phi_1$'s. More generally we have\footnote{To get an even more general observable one can add more forward and backward paths.}
\eq{
	\spl{
		&\ev{\bar{T}(\phi(x'_1)\dots\phi(x'_n)){T}(\phi(x_1)\dots\phi(x_m))}_\rho=\adjustbox{valign=c}{\tikzset{every picture/.style={line width=0.75pt}}

\definecolor{ultramarine}{rgb}{0.07, 0.04, 0.56}

\begin{tikzpicture}

%\tikzset{ma/.style={decoration={markings,mark=at position 0.5 with {\arrow[scale=0.8]{>}}},postaction={decorate}}}
%\tikzset{ma2/.style={decoration={markings,mark=at position 0.5 with {\arrow[scale=0.8]{<}}},postaction={decorate}}}
\draw[color=ultramarine] (-0.1,0.1)--(-0.1,0) -- (5,0)-- (5,-0.2)--(-0.1,-0.2)--(-0.1,-0.3) ;  
%\draw[color=blue] (0,0) -- (5,0)-- (5,-0.2)--(0,-0.2) ; 

%\filldraw[blue] (0,0) circle (0.8pt);
%\filldraw[blue] (0,-0.2) circle (0.8pt);

\draw[black] (-0.3,-0.1) node{\footnotesize $\rho$}; 
\draw[black] (1,0) node {\tikzxmark};
\draw[black] (3,0) node {\tikzxmark};
\draw[black] (1.5,-0.2) node {\tikzxmark};
\draw[black] (3.5,-0.2) node {\tikzxmark};
 
\draw[black] (0,-0.5) node{\footnotesize $t_i$}; 
\draw[black] (5,-0.5) node{\footnotesize $t_f$}; 

\draw[black] (1,0.3) node{\footnotesize $t_1$}; 
\draw[black] (2,0.3) node{\footnotesize $\dots$}; 
\draw[black] (3,0.3) node{\footnotesize $t_m$}; 

\draw[black] (1.5,-0.5) node{\footnotesize $t'_1$};
\draw[black] (2.25,-0.5) node{\footnotesize $\dots$}; 
\draw[black] (3.5,-0.5) node{\footnotesize $t'_n$}; 

\end{tikzpicture}}\\
		&\qquad\qquad=\int_\rho \D\phi_1\,\D \phi_2 \, e^{i(S[\phi_1]-S[\phi_2])}\, \phi_2(x'_1)\dots\phi_2(x'_n) \phi_1(x_1) \dots \phi_1(x_m)\,.
	}
	\label{generalSK}
}
For brevity we have suppressed the initial and final boundary conditions of the path integral.

\paragraph{Generating functional} A convenient way to deal with this type of observable is to define a generating functional\footnote{We used $K_{1,2}$ as the argument of the generating function to save $J$ as the electromagnetic current in the main text.},
\eq{
	\spl{
		Z[K_1,K_2]&\equiv\int \D \phi_1\,\D \phi_2 \, e^{i(S[\phi_1]-S[\phi_2])}\,e^{i\int_x(K_1(x)\phi_1(x)-K_2(x)\phi_2(x))}\\
		&=\ev{\bar{T}\Big(e^{-i\int_xK_2(x)\phi(x)}\Big)T\Big(e^{i\int_xK_1(x)\phi(x)}\Big)}\,,
}}
where in the second line we have used \cref{generalSK} (from now on we suppress the dependence on the density matrix). Assuming $K$ is real, $\oo_K\equiv T\Big(e^{i\int_xK(x)\phi(x)}\Big)$ is unitary and it follows that $Z[K,K]=\Tr[\rho \, \oo_K^\dagger\oo_K]=1$ is normalized. Taking variations of $Z$ with respect to the arguments gives various correlation functions. More specifically, we have
\eq{
	\spl{
		&\ev{\bar{T}(\phi(x'_1)\dots\phi(x'_n)){T}(\phi(x_1)\dots\phi(x_m))}\\
		&\qquad\qquad\qquad=i\fdv{}{K_2(x_1')}\dots i\fdv{}{K_2(x_n')}\frac{1}{i}\fdv{}{K_1(x_1)}\dots\frac{1}{i}\fdv{}{K_1(x_m)}Z[K_1,K_2]\big\vert_{K_1=K_2=0}\,.
	}
	\label{corr}
}
We can interpret the current term as a deformation of the action, i.e.~$S_K\equiv S+\int_x K(x)\phi(x)$ and therefore $Z[K_1,K_2]=\Tr[U_{K_2}^\dagger U_{K_1}\rho]$ in which $U_K$ is the time evolution with the deformed action. As a result, we can obtain correlation functions in the deformed theory by varying $Z$ and setting $K_1=K_2=K$ rather than zero. For instance
\eq{
	\frac{1}{i}\fdv{}{K_1(x)}Z\big\vert_{K_1=K_2=K}=i\fdv{}{K_2(x)}Z\big\vert_{K_1=K_2=K}=\ev{\phi(x)}_K\,,
} 
similarly for higher orders.

Another useful object is the logarithm of the generating function, i.e.~$Z[K_1,K_2]\equiv \exp(iW[K_1,K_2])$. Similar to the standard case, $W[K_1,K_2]$ generates connected correlations functions. Notice that 
\eq{
	Z[K_1,K_2]^*=\ev{\oo_{K_2}^\dagger\oo_{K_1}}^*=\ev{\oo_{K_1}^\dagger\oo_{K_2}}=Z[K_2,K_1]\,,
	\label{Zconj}
}  
which implies $W[K_1,K_2]^*=-W[K_2,K_1]$. More generally we can consider complex currents. In that case we have $W[K_1,K_2]^*=-W[K_2^*,K_1^*]$.

\paragraph{Effective action} Another useful object is the effective action. As is well-known, the effective action is the generating functional for 1PI diagrams. As a result, at tree level it may be used to give all loop corrections to an observable. In particular, it yields the quantum equation of motion. (For a review in standard QFT see \cite{Jackiw:1974cv,Abbott:1981ke} and in the context of CTP see \cite{Chou:1984es,Wang:1998wg,Calzetta:2008iqa}.) The effective action is defined as the Legendre transform of $W$, as follows:
\eq{
	\Gamma[\phi_1,\phi_2]\equiv W[K_1,K_2]-\int(K_1\phi_1-K_2\phi_2)\,,
	\label{effac}
} 
when the currents satisfy the following;
\eq{
	\fdv{W}{K_1(x)}-\phi_1(x)=0 \qquad\text{and} \qquad\fdv{W}{K_2(x)}+\phi_2(x)=0\,.
	\label{wleg}
}
It follows that
\eq{
	\fdv{\Gamma}{\phi_1(x)}=-K_1(x) \qquad\text{and} \qquad\fdv{\Gamma}{\phi_2(x)}=K_2(x)\,.
	\label{Gleg}
}
Moreover, the second variation of $\Gamma$ gives the inverse of the full connected propagator as follows:
\eq{
	\frac{\delta^2\Gamma}{\delta\phi^i\delta\phi^j}=-\fdv{K_i}{\phi^j}=-\left(\fdv{\phi^j}{K_i}\right)^{-1}=-\left(\frac{\delta^2W}{\delta K_i\delta K_j}\right)^{-1}\,,
	\label{dyson}
} 
where we have used \cref{wleg} and \cref{Gleg}. We also used the notation $\phi^1=\phi_1$ and $\phi^2=-\phi_2$ to account for minus signs. Taking more derivatives reveals that any connected correlation function can be constructed out of tree diagrams with edges corresponding to the full propagator and vertices corresponding to $\Gamma$ as the action, hence the name quantum effective action. 

Taking the complex conjugate of \cref{wleg} and using the relation below \cref{Zconj} we conclude that if the pair $(\phi_1,\phi_2)$ corresponds to $(K_1,K_2)$ then $(\phi_2,\phi_1)$ is associated to $(K_2^*,K_1^*)$. Therefore, from the definition we have $\Gamma[\phi_1,\phi_2]^*=-\Gamma[\phi_2,\phi_1]$. In addition, since $\phi_1=\phi_2$ corresponds to $K_1=K_2$ we have $\Gamma[\phi,\phi]=0$ (note that by definition $W[K,K]=0$). Finally, notice that 
\eq{
	\fdv{\Gamma}{\phi_1}\Big\vert_{\phi_1=\phi_2=\phi}=-\fdv{\Gamma}{\phi_2}\Big\vert_{\phi_1=\phi_2=\phi}=-K\,,
	\label{eqgamma}
}
is the quantum-corrected equation of motion for the field expectation value in the presence of a current $K$, i.e.~$\ev{\phi}_K=\phi$. It is worthwhile to contrast this with the expectation value we calculate in the standard in-out formalism. In the latter case, the assumption is that interactions turn on and off adiabatically in the far past and the far future. Therefore, the initial and final states only are equal up to a phase. This is a good approximation for particle physics but it is incapable of taking into account out-of-equilibrium processes, e.g.~dissipation.

For perturbative calculations one has to add all 1PI diagrams for a given number of external legs. Another approach, more useful for non-perturbative results, is the background field method. Consider the original theory in the presence of a given background. We have
\eq{
	\exp(iW[K_1,K_2]_{\bar{\phi}_1,\bar{\phi}_2})\equiv\int \D \phi_1\,\D \phi_2 \, e^{i(S[\phi_1+\bar{\phi}_1]-S[\phi_2+\bar{\phi}_2])}\,e^{i\int_x(K_1(x)\phi_1(x)-K_2(x)\phi_2(x))}\,.
	\label{wback}
}
From the generating function in the presence of the background we can define the effective action as before,
\eq{
	\Gamma[\phi_1,\phi_2]_{\bar{\phi}_1,\bar{\phi}_2}=W[K_1,K_2]_{\bar{\phi}_1,\bar{\phi}_2}-\int(K_1\phi_1-K_2\phi_2)\,,
	\label{effbakdef}
}
with $K_1$ and $K_2$ defined by the same equations as in \cref{wleg} with $W$ replaced by $W_{\bar{\phi}_1,\bar{\phi}_2}$. Now it is a straightforward exersise to show that 
\eq{
	\Gamma[\phi_1,\phi_2]_{\bar{\phi}_1,\bar{\phi}_2}=\Gamma[\phi_1+\bar{\phi}_1,\phi_2+\bar{\phi}_2]\,,
	\label{backeff}
} 
where the right-hand side is the effective action of the theory without the background field as defined in \cref{effac}. An important consequence of this is that we can use the theory with the background to calculate the effective action. In other words we have
\eq{
	\Gamma[\phi_1,\phi_2]=\Gamma[0,0]_{\phi_1,\phi_2}=W[K_1,K_2]_{\phi_1,\phi_2}\,,
	\label{gamback}
}
where the last equality follows from \cref{effbakdef}. Notice that by means of a field redefinition in \cref{wback} we have $W[K_1,K_2]_{\phi_1,\phi_2}=W[K_1,K_2]-\int(K_1\phi_1-K_2\phi_2)$, and therefore \cref{gamback} is consistent with \cref{effac}.

\paragraph{Perturbation in the background field} It is useful to write a perturbative expansion, in the background field, for the effective action. As discussed in the main text, this is useful for situations in which the average field is weak (compared to some microscopic scale in the problem). We start by expanding $W[K_1,K_2]$ perturbatively in the currents. It is very convenient to work, instead of $K_i$ for $i=1,2$, with linear combinations
\eq{
	K_r\equiv\frac{1}{2}(K_1+K_2)\,,\qquad K_a\equiv K_1-K_2\,.
	\label{Kra}
}
(Working with these new variables is sometimes referred to as the $r/a$ or physical representation.) Normalization and reality conditions imply $W[K_r,K_a=0]=0$ and $W[K_r,K_a]^*=-W[K_r,-K_a]$. Expanding $W[K_r,K_a]$ gives
\eq{
W[K_r,K_a]=\frac{1}{2}\int\dd{x}\dd{y}\frac{\delta^2W}{\delta K_I(x)\delta K_J(y)}\Big\vert_{K=0}K_I(x)K_J(y)+\dots\,,
	\label{Wexp2}
} 
in which the indices $I,J,\dots$ are either $r$ or $a$ as introduced above. Notice that the zeroth order term vanishes by normalization. The coefficient of the linear term, i.e.~$\fdv{W}{K_I}\propto\fdv{Z}{K_I}$, is related to the expected value of the field in the absence of any external sources, which we assume vanishes (or is compensated for by a counter-term). Therefore, the leading term is quadratic as indicated above. It requires a little algebra to express the coefficients of the expansion in terms of the correlation functions of the fields:
\eq{
W[K_r,K_a]=\frac{1}{2}\int\dd{x}\dd{y}\begin{bmatrix}
		K_r(x) K_a(x)
	\end{bmatrix}\begin{bmatrix}
		0 & G_A(x,y)\\
		G_R(x,y)	& iG_S(x,y)
	\end{bmatrix}\begin{bmatrix}
		K_r(y)\\K_a(y)
	\end{bmatrix}+\dots\,,
	\label{W2}
}
where we have defined 
\eq{
G_R(x,y)=i\theta(t_x-t_y)\ev{[\phi_x,\phi_y]}\,,\quad G_A(x,y)=-i\theta(t_y-t_x)\ev{[\phi_x,\phi_y]}\,,\quad G_S(x,y)=\frac{1}{2}\ev{\{\phi_x,\phi_y\}}\,,
\label{Green's}
}
referred to as retarded, advanced and symmetric Green's functions. Note that the $\theta$-functions come from the (anti-)time ordering in the closed time path. The fact that the coefficient of $K_r^2$ vanishes is a consequence of the normalization condition. More generally the coefficient of $K_r^n$ is identically zero. The coefficient of $K_a^2$ is purely imaginary which is a consequence of the reality condition. More generally, the coefficient of $K_r^nK_a^m$ is purely imaginary for even $m$. It is possible to show that all the coefficients in the above expansion can be written in terms of nested commutators and/or anti-commutators. For details we refer the reader to \cite{PhysRevB.22.3385,Chou:1984es}.

Similar to the (connected graphs) generating function $W[K_r,K_r]$, we can expand $\Gamma[\phi_r,\phi_a]$ in terms of the fields,
\eq{
\phi_r\equiv\frac{\phi_1+\phi_2}{2}\,,\qquad\phi_a\equiv\phi_1-\phi_2\,.
\label{raphi}
} 
In the $r/a$ representation, the normalization and reality condition read $\Gamma[\phi_r,\phi_a=0]=0$ and $\Gamma[\phi_r,\phi_a]^*=-\Gamma[\phi_r,-\phi_a]$. Coefficients of the expansion can be obtained by using the same trick as in \cref{dyson}. Up to quadratic order, by using \cref{W2} and \cref{dyson}, we have
\eq{
	\Gamma[\phi_r,\phi_a]=\frac{1}{2}\int\dd{x}\dd{y}\begin{bmatrix}
		\phi_r(x) \phi_a(x)
	\end{bmatrix}\begin{bmatrix}
		0 & -G_A^{-1}(x,y)\\
		-G_R^{-1}(x,y)	& iG_R^{-1}G_SG_A^{-1}(x,y)
	\end{bmatrix}\begin{bmatrix}
		\phi_r(y)\\\phi_a(y)
	\end{bmatrix}+\dots\,,
	\label{Gamma2}
}
in which it is was used that $\int\dd{y}G_R(x,y)G_R^{-1}(y,z)=\delta(x-z)$. Once again, the vanishing of $\phi_r^2$ term results from normalization and the coefficient of $\phi_a^2$ is purely imaginary because of reality, with similar generalizations to higher orders. The effective equation of motion in the presence of an external current $K$, given in \cref{eqgamma}, is now obtained:
\eq{
\fdv{\Gamma}{\phi_a(x)}\Big\vert_{\phi_a=0,\phi_r=\phi}=\frac{\delta^2\Gamma}{\delta \phi_a(x) \delta \phi_r(y)}\Big\vert_{\phi=0}\phi(y)+\frac{1}{2}\frac{\delta^3\Gamma}{\delta \phi_a(x) \delta \phi_r(y) \delta \phi_r(z)}\Big\vert_{\phi=0}\phi(y)\phi(z)+\dots=-K(x)\,,
\label{egamma3}
}
where we have included the next-to-leading order for illustration. The solution is the expectation value of the field in the presence of the external current $K$, that is
\eq{
\ev{\phi}_K(x)=\fdv{W}{K_a(x)}\Big\vert_{K_a=0,K_r=K}=\int\dd{y}G_R(x,y)K(y)+\int\dd{y}\dd{z}C_R(x,y,z)K(y)K(z)+\dots\,,
\label{nnlinres}
}  
with $G_R$ given in \cref{Green's} and $C_R(x,y,z)=-\theta(t_x-t_y)\theta(t_y-t_z)\ev{[[\phi_x,\phi_y],\phi_z]}$. More generally, all the higher-order terms involve nested commutators \cite{PhysRevB.22.3385,Chou:1984es}. This result can also be obtained from a direct in-in calculation \cite{Weinberg:2005vy}. In this paper, we focus on the leading term which is linear in the external current. In the statistical mechanics literature, the linear approximation is usually referred to as the Kubo formula \cite{kubo}.

%%%%%%%%%%%%%%%%%%%%%%%%%%%%%%%%%%%%%%%%%%%%%%%%%%%%
\section{Linear response}\label{linres}
%%%%%%%%%%%%%%%%%%%%%%%%%%%%%%%%%%%%%%%%%%%%%%%%%%%%
In this section we review some standard results in linear response theory. Consider an action $S$ which has been deformed by adding a set of background fields $K_n$ to be called $S_K$. We do not restrict the way the action is deformed but we assume that it remains local. We define a set of composite operators as
\eq{
	\oo_n(x)\equiv\fdv{S_K}{K_n(x)}\,.
	\label{Odef}
}
A simple example is the addition of external currents $S[\phi]\to S[\phi]+\int K(x)\phi(x)$ as discussed below \cref{corr} in which case the operator $\oo$ is the dynamical field $\phi$. In general $\oo_n$ can be a function of the dynamical fields and their (spatial and time) derivatives as well as the background fields. For real background fields, the operators are Hermitian. We would like to study the evolution of the expectation value $\ev{\oo_n(x)}$, in some given density matrix, up to linear order in the background fields $K_n$. Notice that $\oo_n(x)$ is the Heisenberg operator evolved with the full Hamiltonian including the effect of background fields. From \cref{generalSK}, $\ev{\oo_n(x)}_K$ is given by 
\eq{
\ev{\oo_n(x)}_K=\int \D\phi_1\D \phi_2 \, e^{i(S_K[\phi_1]-S_K[\phi_2])}\,\oo_n(x)\,,
\label{evo}
}
in which $\phi$ collectively represents the dynamical degrees of freedom in the system. We notice that up to linear order in the background fields
\eq{
\oo_n(x)=\oo_n^{(0)}(x)+\int\dd[4]{y}\oo_{nm}^{(1)}(x,y)K_m(y)+\dots
\label{On}
}
in which $\oo_n^{(0)}(x)=\fdv{S_K}{K_n(x)}\big\vert_{K=0}$ and $\oo_{nm}^{(1)}(x,y)=\frac{\delta^2S_K}{\delta K_n(x)\delta K_m(y)}\big\vert_{K=0}$. Although we have used a more general notation here, there is an implicit delta function (or its derivatives) hidden in the second term of \cref{On} resulting from the locality assumption. In addition, up to linear order, the deformed action is $S_K=S+\int\dd[4]{y}\oo_n^{(0)}(y)K_n(y)+\dots$. Therefore, expansion of \cref{evo} up to linear order gives
\eq{
\ev{\oo_n(x)}_K-\ev{\oo_n(x)}_{K=0}=\int\dd[4]{x'}G_{nm}(x,x')K_m(x')\,,
\label{linO}
}
with
\eq{
G_{nm}(x,x')\equiv i\theta(t-t')\ev{[\oo_n^{(0)}(x),\oo_m^{(0)}(x')]}+\ev{\oo_{nm}^{(1)}(x,x')}\,.
\label{Gij}
}
Equivalently, this result can be derived from the standard in-in formalism using the interaction picture. The linear term in \cref{nnlinres} is a special case of \cref{linO} when the deformation is an external current. The relevant example in this work is matter coupled to an external electromagnetic field $A^\mu_{\rm ext}$. This is equivalent to considering the following deformation of the action we started with in \cref{fullaction}: $S_\gamma[a]+S_M[a+A_{\rm ext},\psi]$. Then it is simply seen that $\oo^{(0)}$ and $\oo^{(1)}$ corresponds to $J^\mu$ and $N^{\mu\nu}$ respectively defined in \cref{JNdef} and the analogue of \cref{Gij} will be \cref{Gj}.  

Introducing background fields changes the energy of the system. At each moment in time, the total energy of the system is the expectation value of the Hamiltonian $\Tr(\rho H_K)$ where $H_K$ is the Hamiltonian modified by the background fields $K_n$. Let us assume that the background fields vanish at $t=\pm\infty$ and are turned on for some period of time. The change in the energy of the system is calculated by $\Delta H=\int_{-\infty}^{+\infty}\dd{t}\dv{}{t}\Tr(\rho H_K)$. The time derivative of the density matrix, switching to the Schr\"odinger picture, is $i\dot{\rho}=[H_K,\rho]$, so that its contribution vanishes after taking the trace. The remaining term comes from the explicit time dependence in $H_K$ associated to the background fields $K_n$. We note that
\eq{
\pdv{\mathcal{H}_K}{K_n}=\pi\pdv{\dot{\phi}}{K_n}-\left[\pdv{\mathcal{L}_K}{\dot{\phi}}\pdv{\dot{\phi}}{K_n}+\pdv{\mathcal{L}_K}{K_n}\right]=-\pdv{\mathcal{L}_K}{K_n}\,,
\label{HLK}
}
in which $\mathcal{H}_K$ and $\mathcal{L}_K$ are Hamiltonian and Lagrangian densities and $\pi \equiv \p\mathcal{L}_K/\p\dot{\phi}$ is the momentum conjugate to $\phi$. For simplicity of notation we have only used one dynamical field $\phi$, with an obvious generalization to more complicated systems. We should note that for constructing the Hamiltonian we regard $\dot{\phi}(\phi,\pi,K)$ as a function of $\pi$ as well as the background fields. Therefore, we can write
\eq{\spl{
\Delta H&=\int\dd[4]{x}\Tr(\rho\dot{\mathcal{H}}_K)=\int\dd[4]{x}\Tr(\rho\pdv{\mathcal{H}_K}{K_n})\dot{K}_n=-\int\dd[4]{x}\ev{\oo_n(x)}_K\dot{K}_n\\
&=-\int\frac{\dd[4]{p}}{(2\pi)^4}i\omega K_n(-p)G_{nm}(p)K_m(p)\\
&=\int\frac{\dd[4]{p}}{(2\pi)^4}\omega K_n(-p)\left(\frac{G_{nm}(p)-G_{mn}(-p)}{2i}\right)K_m(p)\,.\label{deltaH}
}}
In the first line, we have used \cref{HLK} and \cref{Odef}. In the second line, we have restricted to the linear response formula \cref{linO} assuming, as it is relevant for the context of this paper, that $\ev{\oo_n(x)}_{K=0}=0$ and switched to Fourier space. In the last line, we have used the reality of the integral to replace $G_{nm}$ with the expression in the parenthesis. Notice that for Hermitian $\oo_n$, the Green's function is real implying that $G_{nm}(-p)=G_{nm}(p)^*$. The expression inside the parenthesis is called the dissipative part of the response function. This can be simplified assuming spacetime inversion (parity and time reversal) symmetry of the state. First of all, note that the contribution of the contact term in \cref{Gij} is symmetric under $n\leftrightarrow m$. The reason is that it is symmetric under $(n,x)\leftrightarrow(m,x')$ but because of the delta function it is also symmetric under $x\leftrightarrow x'$. Focusing on the first term of \cref{Gij} we have (see also \cite{Hartnoll:2009sz})
\eq{
	\spl{
		G_{mn}(x)&=i\theta(t)\Tr(\rho[\oo^{(0)}_m(x),\oo^{(0)}_n(0)])=-i\theta(t)\Tr(\rho[\oo^{(0)}_n(-x),\oo^{(0)}_m(0)])\\
		&=-i\theta(t)(\eta_n\eta_m)^*\Tr(\Theta\rho[\oo^{(0)}_n(x),\oo^{(0)}_m(0)]\Theta^{-1})\\
		&=-i\theta(t)(\eta_n\eta_m)^*\Tr((\rho[\oo^{(0)}_n(x),\oo^{(0)}_m(0)])^\dagger)\\
		&=(\eta_n\eta_m)^*G_{nm}(x)\,.
	}\label{Trev}
}
In the first line we have used spacetime translation invariance to shift the argument inside the commutator. In the second line we have used the spacetime inversion operator $\Theta$ and that the operator transforms as $\Theta\oo^{(0)}_m(x)\Theta^{-1}=\eta_m\oo^{(0)}_m(-x)$ with some phase $\eta_m$ which depends on the operator. Moreover, we have used invariance of the density matrix under spacetime inversion, i.e.~$\Theta\rho\Theta^{-1}=\rho$.\footnote{If the set of operators are in the trivial representation under spacetime transformations, i.e. scalar operators, then it is enough that the system is invariant under time-reversal and the role of parity is played by rotation symmetry. Otherwise, we also need invariance under parity to rewrite \cref{deltaH} in terms of $\Im G_{mn}$. For instance, in the case of the current operator $J^\mu$, it is possible to have a parity violating term in the two point function (see \cref{p}). This term, while contributing to the expression for the imaginary part, does not contribute to the dissipative part in \cref{deltaH}.} Finally, we use the fact that $\Tr(\Theta A\Theta^{-1})=\Tr(A^\dagger)$ for any linear operator $A$.\footnote{Remember that for an antiunitary operator $\Theta$, if we define $\ket{\tilde{a}}\equiv\Theta\ket{a}$ and $|\tilde{b} \rangle \equiv \Theta\ket{b}$, we have $\braket{a}{b}=\langle \tilde{b} | \tilde{a} \rangle$. Then it is easy to see that for any linear operator $A$, we have $\bra{b}A^\dagger\ket{a}=\bra{\tilde{a}}\Theta A\Theta^{-1} | \tilde{b} \rangle$. The trace $\Tr A^\dagger=\sum_n\bra{n}A^\dagger\ket{n}$ can be written as $\sum_{\tilde{n}}\bra{\tilde{n}}\Theta A\Theta^{-1}\ket{\tilde{n}}$, which is $\Tr(\Theta A\Theta^{-1})$. See also \cite{Uhlmann_2016}.} Hence we conclude that the response function is symmetric in its indices up to some phase factors. In cases in which the phase factors are unity, the expression for the dissipative part of the response function in \cref{deltaH} is simply $\Im G_{nm}(p)$.

%%%%%%%%%%%%%%%%%%%%%%%%%%%%%%%%%%%%%%%%%%%%%%%%%%%%
\section{Lindhard function}\label{linhar}
%%%%%%%%%%%%%%%%%%%%%%%%%%%%%%%%%%%%%%%%%%%%%%%%%%%%
In this section we study the electromagnetic properties of a gas of fermions, e.g.~electrons, at finite density and zero temperature. For finite temperature calculations see for example \cite{Weldon:1982aq}. The condition of finite density of charged particles can be modeled by adding a chemical potential to the Lagrangian. See also \cite{Shuryak:1980tp,Nicolis:2023pye,Podo:2023ute}. Therefore a Dirac fermion $\psi$ at finite chemical potential $\mu$ can be described by
\eq{
	\mathcal{L}=i\bar{\psi}(\slashed{\p}+i\mu\slashed{u})\psi-m\bar{\psi}\psi
	\label{Ldirmu}
}
in which $u^\mu=(1,0)$ is the rest frame velocity of the fermions and $m$ their mass. We assume $\mu>m>0$. Then the ground state of the system consists of fermions filling up to energy $E_F=\sqrt{m^2+k_F^2}=\mu$, known as the Fermi energy, where $k_F$ is called the Fermi momentum.

The photon coupling is similar to the Lorentz-invariant case, i.e.~$\p_\mu\to\p_\mu-iA_\mu$. We want to calculate the photon (retarded) self-energy which is given by $\ev{[J^\mu,J^\nu]}_{\rm 1PI}$ with $J^\mu=\bar{\psi}\gamma^\mu\psi$, perturbatively in the coupling to the photon. At leading order we can can drop the 1PI index, corresponding to the diagram shown in \cref{loop}. Moreover, it is easier to first calculate the diagram using the usual Feynman rules for the $S$-matrix (which gives the time ordered correlation function) and then obtain the retarded response function by moving the singularities with an appropriate $i\epsilon$ prescription \cite{2015imbpbookC}.
\fg{
	\adjustbox{valign=c}{\includegraphics{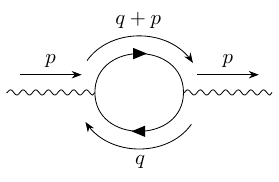}}\hfil
	\adjustbox{valign=c}{\scalebox{1.25}{\tikzset{every picture/.style={line width=0.85pt}}

\definecolor{ultramarine}{rgb}{0.07, 0.04, 0.56}
\definecolor{orange-red}{rgb}{1.0, 0.27, 0.0}
\definecolor{tropicalrainforest}{rgb}{0.0, 0.46, 0.37}

\begin{tikzpicture}
	
	\tikzset{ma/.style={decoration={markings,mark=at position 0.5 with {\arrow[scale=0.8]{>}}},postaction={decorate}}}
	\tikzset{ma2/.style={decoration={markings,mark=at position 0.5 with {\arrow[scale=0.8]{<}}},postaction={decorate}}}
	
	\draw[gray,thin] (-2.7,0) -- (2.7,0)[->]; 
	\draw[gray,thin] (0,-1) -- (0,1)[->]; 
	
	%\draw[ma,color=blue] (-2,0) -- (2,0) arc(0:180:2) --cycle;
	%\draw[ma,color=ultramarine] (2,0) arc (0:180:2);
	\draw[ma,color=ultramarine] (-2.5,-0.2) -- (1.5,-0.2);
	\draw[color=ultramarine] (1.5,-0.2) -- (1.5,0.2);
	\draw[color=ultramarine] (1.5,0.2) -- (2.5,0.2);
	\draw[rounded corners=2,color=tropicalrainforest,dash dot] (0.04,-0.16) rectangle (1.46,0.16);
	\draw[ma,color=orange-red,dashed] (-2.5,-0.2) -- (0,-0.2);
	\draw[color=orange-red,dashed] (0,-0.2) -- (0,0.2);
	\draw[color=orange-red,dashed] (0,0.2) -- (2.5,0.2);
	%\draw[color=ultramarine] (1.1,0) arc (0:180:0.1);

	\draw[gray] (2.2,1.2) node{\footnotesize $q^0$}; 
	\draw[gray,thin] (2,1.3)|-(2.3,1);
	%\draw[black] (-1.7,1.7) node{\footnotesize $C$};
	\draw[black] (1.7,-0.25) node{\footnotesize $\mu$};  
	
	%\filldraw[gray] (1,0) circle (0.6pt);
	
	%\filldraw[blue] (5,0) circle (0.8pt);
	%\filldraw[blue] (9,0) circle (0.8pt);
	%\filldraw[blue] (7.5,0) circle (0.8pt);
	%\draw[black] (1,0) node {\tikzxmark};
	%\draw[black] (3,0) node {\tikzxmark};
	%\draw[black] (1.5,-0.2) node {\tikzxmark};
	%\draw[black] (3.5,-0.2) node {\tikzxmark};
	% 
	%\draw[black] (4.7,-0.3) node{\footnotesize $C_\infty$}; 
	%
	%\draw[black] (1,0.3) node{\footnotesize $t_1$}; 
	%\draw[black] (2,0.3) node{\footnotesize $\dots$}; 
	%\draw[black] (3,0.3) node{\footnotesize $t_m$}; 
	%
	%\draw[black] (1.5,-0.5) node{\footnotesize $t'_1$};
	%\draw[black] (2.25,-0.5) node{\footnotesize $\dots$}; 
	%\draw[black] (3.5,-0.5) node{\footnotesize $t'_n$}; 
	
\end{tikzpicture}}}
	\caption{Left: Leading order correction to the photon self-energy from the gas of fermions at finite chemical potential. Right: The contour of integration in the presence of the chemical potential.}
	\label{loop}
}
The Feynman propagator of the modified Lagrangian \cref{Ldirmu} is 
\eq{
	S(x-y)\equiv i\ev{T\psi(x)\bar{\psi}(y)}=\int\frac{\dd[4]{p}}{(2\pi)^4}e^{ip \cdot (x-y)}\frac{-\slashed{p}'+m}{p'{}^2+m^2-i\epsilon \, {\rm sgn}(\omega(\omega+\mu))}\,,
} 
in which $p'{}^\mu=p^\mu+\mu u^\mu=(\omega+\mu,\boldsymbol{k})$. Notice that in the absence of the chemical potential we recover the Lorentz-invariant $i\epsilon$ prescription for the propagator. Therefore, by using the standard Feynman rules of quantum electrodynamics, for the diagram of \cref{loop} we get
\eq{
	i\Pi^{\mu\nu}=-\int\frac{\dd[4]{q}}{(2\pi)^4}\Tr(\gamma^\mu S(q)\gamma^\nu S(q+p))\,.
}
After some algebra one finds that
\eq{
	i\Pi^{\mu\nu}=-\int\frac{\dd[4]{q}}{(2\pi)^4}\frac{4I^{\mu\nu}(q)}{[-(q^0+i\epsilon \,{\rm sgn}(q^0-\mu))^2+E_{\boldsymbol{q}}^2][-(q^0+\omega+i\epsilon \,{\rm sgn}(q^0+\omega-\mu))^2+E_{\boldsymbol{q}+\boldsymbol{k}}^2]}\,,
	\label{Pilindmid}
}
with $E_{\boldsymbol{q}}=\sqrt{m^2+\boldsymbol{q}^2}$. To obtain \cref{Pilindmid} we have changed the variable of integration $q^0\to q^0-\mu$ and used properties of the gamma matrices, e.g.~$\{\gamma^\mu,\gamma^\nu\}=-2\eta^{\mu\nu}$ in our metric signature, to simplify the numerator as 
\eq{
	I^{\mu\nu}(q)=q^\mu(p+q)^\nu+q^\nu(p+q)^\mu-(q \cdot (p+q)+m^2)\eta^{\mu\nu}\,.
}
The denominator has poles at $q^0=\pm E_{\boldsymbol{q}}-i\epsilon \,{\rm sgn}(q^0-\mu)$ and $q^0+\omega=\pm E_{\boldsymbol{q}+\boldsymbol{k}}-i\epsilon\,{\rm sgn}(q^0+\omega-\mu)$. One can think of the effect of the nonzero chemical potential in terms of deforming the $q^0-$integration contour. This is shown in \cref{loop} with the solid line. An illustrative way to perform the integral is to decompose it into a contour used in the Lorentz-invariant case with $\mu=0$ (dashed line in \cref{loop}) plus a closed curve which captures the effect of the chemical potential (dash-dotted line). Thus we can write the result as follows 
\eq{
	\Pi^{\mu\nu}=\Pi^{\mu\nu}_{\rm qed}+\Pi^{\mu\nu}_F\,.
	\label{Pidec}
}
To extract the longitudinal and transverse parts we use definition \cref{Pi} which implies that
\eq{
	\pi_L=\frac{1}{k^2}\Pi^{00}\,,\qquad\pi_T=\frac{1}{2k^2}(\delta_{ij}\Pi^{ij}-\omega^2\pi_L)\,.
	\label{pidecLT}
}
The longitudinal and transverse parts can be decomposed similarly to \cref{Pidec}. The first term is the Lorentz-invariant result which can be found in any quantum field theory textbook. The condition of Lorentz invariance implies that there is no distinction between longitudinal and transverse directions, i.e.~$\pi_L=-k^2\pi_T/p^2$, therefore $\Pi^{\mu\nu}_{\rm qed}=\pi_{L,{\rm qed}}(p^\mu p^\nu-p^2\eta^{\mu\nu})$, with\footnote{This is the result of on-shell renormalization scheme, $\Pi^{\mu\nu}_{\rm qed}(0)=0$.} 
\eq{
	\pi_{L,{\rm qed}}=-\frac{k^2}{p^2}\pi_{T,{\rm qed}}=-\frac{1}{2\pi^2}\int_0^1\dd{x}x(1-x)\log(\frac{x(1-x)p^2+m^2}{m^2})\,.
	\label{Piqed}
}
The next term comes from the finite chemical potential and can be calculated as
\eq{
	\Pi^{\mu\nu}_F(\omega,k)=-2\int\frac{\dd[3]{\boldsymbol{q}}}{(2\pi)^3}\frac{\theta(E_F-E_{\boldsymbol{q}})}{E_{\boldsymbol{q}}}\left[\frac{I^{\mu\nu}(q^0=E_{\boldsymbol{q}},\boldsymbol{q})}{(E_{\boldsymbol{q}}+\omega)^2-E_{\boldsymbol{q}+\boldsymbol{k}}^2}+\big(\omega\to-\omega,\boldsymbol{k}\to-\boldsymbol{k}\big)\right]\,.
	\label{PiF}
}
We can now obtain the retarded correlation functions by imposing the appropriate $i\epsilon$ prescription; in both expressions \cref{Piqed} and \cref{PiF}, it is understood that $\omega\to\omega+i\epsilon$ to have the correct analyticity properties. 

For the longitudinal part we can perform the angular part of the integral and obtain
\eq{
	\pi_{L,F}=\int_0^{k_F}\frac{q^2\dd{q}}{2\pi^2k^2E_q}\left[1-\frac{(\omega+2E_q)^2-k^2}{4kq}\log(\frac{\omega(\omega+2E_q)+2kq-k^2}{\omega(\omega+2E_q)-2kq-k^2})\right]+\big(\omega\to-\omega\big)\,.
	\label{piLR}
}
Similarly, for the transverse component we obtain
\eq{
	\spl{
		\pi_{T,F}=\int_0^{k_F}\frac{q^2\dd{q}}{4\pi^2k^2E_q}\Bigg[-\left(1+\frac{\omega^2}{k^2}\right)+&\frac{\omega^2(\omega+2E_q)^2-k^2(k^2+4q^2+4\omega E_q)}{4k^3q}\times\\&\log(\frac{\omega(\omega+2E_q)+2kq-k^2}{\omega(\omega+2E_q)-2kq-k^2})\Bigg]+\big(\omega\to-\omega\big)\,.
	}
	\label{piTR}
}
The above integrals can also be performed analytically but the expressions are not very illuminating so we do not report them here.

It is instructive to check the conditions of analyticity and positivity for these functions. Let us first focus on the Lorentz-invariant piece in \cref{Piqed}. Since this is the outcome of a perturbative calculation, following the discussion in \S\ref{wider}, $\pi_{L,{\rm qed}}$ (and similarly $k^2\pi_{T,{\rm qed}}$) must be analytic in the FLC. The only source of non-analyticity is the branch cut of the $\log$ function which for the principal branch locates along the negative real axis of the argument. Therefore, from \cref{Piqed} it is easy to see that the location of the branch cut is $p^2+m^2/(x(1-x))\leq0$ or equivalently $-(\omega+i\epsilon)^2+k^2+M^2=0$ for any $M^2\geq4m^2$. As we will argue in detail in App.~\ref{dampwave}, solutions to this equation always lie outside the FLC and therefore $\pi_{L,{\rm qed}}$ is analytic in the FLC. We can calculate the imaginary part of \cref{Piqed} for real $\omega$ and $k$ as follows
\eq{
\spl{
\Im\pi_{L,{\rm qed}}&=-\frac{1}{2\pi^2}\int_0^1\dd{x}x(1-x)\theta\big[-x(1-x)p^2-m^2\big]\pi{\rm sgn}(-\omega)\\
&=\frac{1}{24\pi}\left[1+\left(1+\frac{2m^2}{-p^2}\right)\sqrt{1-\frac{4m^2}{-p^2}}\right]{\rm sgn}(\omega)\theta(-p^2-4m^2)\,.
}\label{implqed}
}
In the first line we have used $\log(-|x|\pm i\epsilon)=\log(|x|)\pm i\pi$ and the fact that the imaginary part of the argument of the $\log$ in \cref{Piqed} comes with the combination $\sim-i\epsilon\,\omega$ which gives the sign function. As is well-known, the imaginary part is nonzero only for timelike momenta larger than the pair production threshold, i.e.~$-p^2>4m^2$, after which it grows positive and saturates to $1/12\pi$ for $\omega>0$.

The contribution of the finite chemical potential in \cref{piLR} and \cref{piTR} is slightly more involved. First we deal with the $\log$ which has a similar structure both for the longitudinal and transverse parts. Notice that one can re-write the argument of the logarithm, in a more suggestive way, as $(\omega\pm E_q+i\epsilon)^2-(k\pm q)^2-m^2$. Therefore, the location of the branch cut is similar to the Lorentz-invariant case and therefore outside the FLC. Notice that the shifts $\omega\to\omega\pm E_q$ and $k\to k\pm q$ do not affect this conclusion as $E_q$ and $q$ are real. Another source of non-analyticity are the factors $1/k^2$. However, it is straightforward to take the $k^2\to0$ limit of the integrand 
\eqa{
\pi_{L,F} &=\int_0^{k_F}q^2\dd{q}\left[\frac{4(-3E_q^2+q^2)}{3\pi^2E_q(4E_q^2-\omega^2)\omega^2}+\order{k^2}\right]\label{piLexp}\\
k^2\pi_{T,F} & =\int_0^{k_F}q^2\dd{q}\left[\frac{4(-3E_q^2+q^2)}{3\pi^2E_q(4E_q^2-\omega^2)}+\order{k^2}\right]\label{piTexp}\,,
} 
which is regular. Further, one could also worry about the appearance of $k=\sqrt{k^2}$ in \cref{piLR} and \cref{piTR}. The absence of an $\order{k}$ term in the expansion above shows that $k=0$ is not a branch point. Therefore, both $\pi_{L,F}$ and $k^2\pi_{T,F}$ are analytic in the FLC as required by causality, as discussed in \S\ref{sec:analytic}. In addition, we see from \cref{piLexp} and \cref{piTexp} that the combination $k^2\pi_{T,F}-\omega^2\pi_{T,F}=\order{k^2}$, a consequence of causality. The imaginary parts of $\pi_{L,F}$ and $\pi_{T,F}$ can be obtained by a similar calculation presented in \cref{implqed}, although the result will be more complicated. We will calculate the imaginary part in the non-relativistic limit below.   

\paragraph{Non-relativistic limit} We note that if all the energies are much smaller than the mass, i.e.~$\omega,k\ll m$, one can neglect the Lorentz-invariant piece and the integrand of the second term can be expanded as follows
\eq{
	\Pi_F^{\mu\nu}\simeq-\int\frac{\dd[3]{\boldsymbol{q}}}{(2\pi)^3}\frac{\theta(\tilde{E}_F-\tilde{E}_{\boldsymbol{q}})}{m^2}\left[\frac{I^{\mu\nu}(m+\tilde{E}_{\boldsymbol{q}},\boldsymbol{q})}{\tilde{E}_{\boldsymbol{q}}-\tilde{E}_{\boldsymbol{q}+\boldsymbol{k}}+\omega}+\big(\omega\to-\omega,\boldsymbol{k}\to-\boldsymbol{k}\big)\right]\,,
}
where we have defined $E_{\boldsymbol{q}}=m+\tilde{E}_{\boldsymbol{q}}$ with $\tilde{E}_{\boldsymbol{q}}\approx q^2/2m$ in the non-relativistic limit and $\tilde{E}_F\equiv E_F-m\approx k_F^2/2m\ll m$. For the longitudinal component we have $I^{00}\simeq2m^2$ and therefore we can do the integrals and obtain
\eq{
	\pi_L=\frac{mk_F}{2\pi^2k^2}\left[\frac{1}{2}-\frac{(k^2-2m\omega)^2-4k_F^2k^2}{8k_Fk^3}\log(\frac{k^2+2k_Fk-2m\omega}{k^2-2k_Fk-2m\omega})\right]+\big(\omega\to-\omega\big)\,.
	\label{piLNR}
}
This result has famously been obtained by Lindhard \cite{lindhard} and is known as Lindhard function. The prefactor is sometimes written in terms of the plasma frequency $\omega_p^2 \equiv g^2n/m$ and the Fermi velocity $v_F\equiv k_F/m$, i.e.~$\frac{mk_F}{2\pi^2}=\frac{3\omega^2_p}{2g^2v_F^2}$. To derive this we use the fact that the number density of fermions is $n=k_F^3/3\pi^2$. Similarly, for the transverse piece we obtain $\delta_{ij}I^{ij}\simeq3m\omega+\boldsymbol{q}\cdot(2\boldsymbol{q}-\boldsymbol{k})$. Notice that in the non-relativistic approximation we keep terms like $m\omega\sim k^2\sim kk_F$. The rest of the calculation is very similar to the longitudinal case. The final answer is
\eq{
	\spl{	\pi_T=\frac{k_F}{32\pi^2mk^4}\Bigg[&\frac{1}{2}(3k^4-12m^2\omega^2-4k_F^2k^2)\\&-\frac{(4k^4-[(k^2-2m\omega)^2-4k_F^2k^2])[(k^2-2m\omega)^2-4k_F^2k^2]}{8k_Fk^3}\log(\frac{k^2+2k_Fk-2m\omega}{k^2-2k_Fk-2m\omega})\Bigg]\\&+\big(\omega\to-\omega\big)\,.
	}
	\label{piTNR}
}
The above expression differs slightly from what was given by Lindhard \cite{lindhard}; the reason is likely due to the fact that the spin of fermions has been neglected in his calculation. It is useful to look at the above results in the low-energy and high-energy limits. We are mainly interested in the functions parametrized as $\pi_L(\omega,\omega\xi)$ and $\pi_T(\omega,\omega\xi)$ for $\xi<1$ (see \S\ref{sec:analytic}). Taking $\omega\to0$ in this parametrization we obtain
\eq{
g^2\pi_L(\omega,\omega\xi)=-\frac{\omega_p^2}{\omega^2}+\dots\,,\qquad g^2\pi_T(\omega,\omega\xi)=-\frac{\omega_p^2}{\omega^2\xi^2}+\dots\,.
\label{pilow}
}
Equivalently, in terms of electric permittivity and magnetic permeability we get
\eq{
\varepsilon=1-\frac{\omega_p^2}{\omega^2}+\dots\,,\qquad\frac{1}{\mu}=1-\left(\frac{2k_F^2}{5m^2}\right)\frac{\omega_p^2}{\omega^2}+\dots\,,\quad(\omega\to0)\,.
\label{lowpiLT}
}
To obtain the expansion for $\mu$ one needs to be careful about the expansion of the combination $\pi_T-\omega^2\pi_L/k^2$ as naively the leading order terms given in \cref{lowpiLT} vanish. Moreover, the transverse part of dielectric tensor, as defined in footnote~\ref{eT}, is $\varepsilon_T=1-\omega_p^2/\omega^2+\dots$. One can understand this result by noting that all the charged particles in this system are freely moving. Therefore, at low energies it behaves like a conductor which in this case is very well modeled by the simple Drude model (see the discussion in \S\ref{sec:stronger}). We remind the reader that the $i\epsilon$ prescription is understood here as $(\omega+i\epsilon)^2=\omega^2+i\epsilon\omega$. The effect of interaction with a background lattice or with the other fermions can be modeled by adding a finite decay width, i.e.~$\omega^2+i\gamma\omega$. In that case, a useful quantity is conductivity $\sigma_{(T)}\equiv i\omega(1-\varepsilon_{(T)})$ which in the present case gives, at low frequencies, $\sigma_{(T)}=\omega_p^2/\gamma$. 

For high energies, we note that the main interest is in the range $k_F\ll \omega\ll m$ for the non-relativistic limit. In that case one obtains
\eq{
g^2\pi_L(\omega,\omega\xi)=-\frac{\omega_p^2}{\omega^2}+\dots\,,\qquad g^2\pi_T(\omega,\omega\xi)=-\frac{\omega_p^2}{\omega^2\xi^2}+\dots\,,\quad(\omega\to\infty)\,.
\label{piLpiThigh}
}  
Moreover, we have $g^2(\pi_T-\omega^2\pi_L/k^2)=2k_F^2\omega_p^2/(5m^2\omega^2)+\dots$ relevant for $\mu$.  We will use these relations in \S\ref{high} and \S\ref{sec:bounds}. The fact that the low and high energy limits in \cref{pilow} and \cref{piLpiThigh} turn out to be the same is a consequence of taking the non-relativistic limit, i.e. $k_F\ll m$ or $v_F\ll1$. Setting $k=\omega\xi$ for $\xi\ll1/v_F$ and for energies $\omega\ll m$ is effectively equivalent to taking $k\to0$ limit for all $\omega$ in this range. 

Finally, let us calculate the imaginary part of the Lindhard function in the non-relativistic limit. A nonzero imaginary part comes from the logs in \cref{piLNR} and \cref{piTNR}. We can see that
\eq{
\spl{
\Im\log(\frac{k^2+2k_Fk-2m\omega}{k^2-2k_Fk-2m\omega})&=-\pi\Big[\theta\big(2m\omega-k^2-2k_Fk\big)-\theta\big(2m\omega-k^2+2k_Fk\big)\Big]\\
&=\pi\theta\big(2k_Fk-|2m\omega-k^2|\big)\,.
}
\label{imlog}
}
To obtain the first line we have used $\log(\cdots) \to \log(\cdots-i\epsilon)$. To obtain the second line we note that if $|2m\omega+k^2|>2k_Fk$ both of the theta functions are either zero or one. Notice that for the part $(\omega\to-\omega)$ in both \cref{piLNR} and \cref{piTNR}, we have $\log(\dots+i\epsilon)$ and therefore the imaginary part has a minus sign compared to \cref{imlog}. After some algebra we obtain
\eq{
\Im\pi_L=\frac{m}{16\pi k^5}\Big[[4k_F^2k^2-(k^2-2m\omega)^2]\theta\big(2k_Fk-|2m\omega-k^2|\big)-[4k_F^2k^2-(k^2+2m\omega)^2]\theta\big(2k_Fk-|2m\omega+k^2|\big)\Big]\,,
\label{impiL}
}
and for the transverse part
\eq{
\spl{
\Im\pi_T=\frac{1}{256\pi mk^7}\Big[&(4k^4+[4k_F^2k^2-(k^2-2m\omega)^2])[4k_F^2k^2-(k^2-2m\omega)^2]\theta\big(2k_Fk-|2m\omega-k^2|\big)\\
&-(4k^4+[4k_F^2k^2-(k^2+2m\omega)^2])[4k_F^2k^2-(k^2+2m\omega)^2]\theta\big(2k_Fk-|2m\omega+k^2|\big)\Big]\,.
}
\label{impiT}
}
\paragraph{Gapped Lindhard} As mentioned above, at low energies, the Lindhard function describes a conductor rather than an insulator. However, starting from the Lindhard function, it is possible to build a model for an insulator by introducing a gap. Following \cite{PhysRevB.25.6310}, we modify the imaginary part of the Lindhard function as follows
\eq{
	\Im\tilde{\pi}_{L}(\omega,\boldsymbol{k})=\begin{cases}
		\Im\pi_{L}\Big({\rm sgn}(\omega)\sqrt{\omega^2-E_g^2},\boldsymbol{k}\Big)\,,\qquad&|\omega|\geq E_g \,, \\
		0\,,\qquad&|\omega|< E_g\,,
	\end{cases}
	\label{gaplin}
}
with $E_g$ the energy gap. A similar modification can be done for the transverse part. Next, one can construct the modified real part from \cref{gaplin} by using the Kramers-Kronig relations (or equivalently Leontovich's relation with $\boldsymbol{\xi} = \boldsymbol{0}$). One can verify that, for instance, we obtain $\varepsilon(0,0)=1+\omega_p^2/E_g^2$. See \cite{PhysRevB.25.6310} for details. Notice that this construction is non-relativistic and it is not obvious how it can be generalized for the relativistic cases \cref{piLR} and \cref{piTR}. The modification must be such that the modified imaginary part satisfies the condition \cref{imcons}.

%%%%%%%%%%%%%%%%%%%%%%%%%%%%%%%%%%%%%%%%%%%%%%%%%%%%
\section{The absence of zeros}\label{landau}
%%%%%%%%%%%%%%%%%%%%%%%%%%%%%%%%%%%%%%%%%%%%%%%%%%%%
In this section we argue that a response function $\chi(\omega)$ cannot have zeros in the UHP of complex $\omega$. Notice that by causality, the response $\chi(t)$ must vanish for $t\leq0$ which implies that $\chi(\omega)$ is analytic in the upper half $\omega$-plane. We also assume that it vanishes in the high-energy limit, $\chi(\omega)\to0$ for $|\omega| \to \infty$. Moreover, $\chi$ must satisfy a positivity condition, that is, $\omega\Im\chi \geq 0$ for $\omega \in \mathbb{R}$. Finally, by reality of the response function $\chi(t)$ we have $\chi(\omega)^*=\chi(-\omega^*)$. In particular, this implies that $\chi$ is real on the imaginary axis. Using the above assumptions it is possible to argue that $\chi(\omega)$ does not have zeros in the UHP. A stronger version of this was provided by Landau \cite{Landau:1980mil} (apparently it was first proposed by N.~N.~Me\v{\i}man): the equation $\chi(\omega)=a$ for real $a$ has, at most, one solution in the UHP. Therefore, $\chi(\omega)$ along the imaginary axis is monotonically decreasing from $\chi(0)$ to zero and elsewhere is always complex. This argument is very well explained in \cite{Landau:1980mil} (\S 123) and we do not repeat it here. Instead we give a simpler version of the proof and discuss a possible extension when we also have nonzero momentum.

It has been argued in \cref{fullim} that we can always write the response function in the domain of analyticity as an integral over its imaginary part. Restricting to the single-variable case we get 
\eq{
\chi(\omega)=\frac{1}{\pi}\int_{0}^{\infty}\frac{\dd{z^2}}{z^2-(\omega+i\epsilon)^2}\,\,\Im\chi(z)\,,
\label{fullim2}
}
where we have used the fact that $\Im\chi(\omega)$ is odd. The above relation is valid for complex $\omega$ in the UHP. Restricting to the imaginary axis $\omega=i\omega_{I}$ we obtain
\eq{
\chi(i\omega_I)=\frac{1}{\pi}\int_{0}^{\infty}\frac{\dd{z^2}}{z^2+\omega_I^2}\,\,\Im\chi(z)>0\,,
\label{land1}
}
which is strictly positive (otherwise it means that the imaginary part is always zero and therefore the function is everywhere zero) and monotonically decreasing. On the other hand, away from the imaginary axis we can calculate the imaginary part from \cref{fullim2}
\eq{
\Im\chi(\omega)=\frac{\Im\omega^2}{\pi}\int_{0}^{\infty}\frac{\dd{z^2}}{|z^2-\omega^2|^2}\,\,\Im\chi(z)\,,
\label{land2}
}
which means that $\Im\chi(\omega)$ has the same sign as $\Im\omega^2=2\omega_{I}\omega_R$ and in particular it is nonzero. We conclude that $\chi(\omega)$ cannot have any zeros in the UHP. 

When we have momentum dependence, as in the main text, one can use the parameterization $\tilde\chi(\omega)\equiv\chi(\omega,\vbq+\omega\vbxi)$. Then for any real $\vbq$ and $\vbxi$ with $\xi<1$ the function $\tilde\chi(\omega)$ is analytic in the UHP for complex $\omega$. Moreover, the positivity condition for $\chi$ implies the same also for $\tilde\chi$. However, the reality property is slightly different for $\tilde\chi$. Remember that for a generic four-momentum $p^\mu=(\omega,\vbk)$ we have $\chi(p)^*=\chi(-p^*)$ and, in particular, for $p^\mu=ip^\mu_{I}$ then $\chi(p)$ is real. But in terms of $\tilde\chi$, assuming rotational invariance, we have
\eq{
\tilde\chi(\omega)^*=\chi(\omega,(\vbq+\omega\vbxi)^2)^*=\chi(-\omega^*,(-\vbq-\omega^*\vbxi)^2)\neq\tilde\chi(-\omega^*)\,.
}
Therefore, the reality condition does not imply that $\tilde\chi(\omega)^*$ and $\tilde\chi(-\omega^*)$ are equal for general $\boldsymbol{q}$ and $\boldsymbol{\xi}$. However, if we restrict ourselves to the points for which $\vbq\cdot\vbxi=0$ then $\tilde\chi(\omega)^*=\tilde\chi(-\omega^*)$. With this condition the function $\tilde\chi(\omega)$ will be real on the imaginary axis. By using \cref{fullim}, we write
\eq{
\chi(\omega,\vbq+\omega\vbxi)=\frac{1}{\pi}\int_0^\infty\frac{\dd{z^2}}{z^2-(\omega+i\epsilon)^2}\,\,\Im\chi(z,\vbq+z\vbxi)\,,\qquad (\vbq\cdot\vbxi=0)\,,
} 
using $\Im\tilde{\chi}(-z)=-\Im\tilde{\chi}(z)$. Then the analog of \cref{land1} and \cref{land2} can be easily derived. We conclude that $\chi(\omega,\vbq+\omega\vbxi)$ cannot have any zeros in the UHP of complex $\omega$ when $\vbq\cdot\vbxi=0$.

One may wonder how restrictive this is. Is it possible to find appropriate $\vbq$ and $\vbxi$ such that $k^2=(\vbq+\omega\vbxi)^2$, with $\vbq\cdot\vbxi=0$, for any point in the analyticity region? If this is true then the function cannot have zeros in the whole region of analyticity. Unfortunately the answer is no. A little algebra shows that in order to relate any point $(\omega,\vbk)$ to a form $(\omega,\vbq+\omega\vbxi)$ with $\vbq\cdot\vbxi=0$, keeping $k^2$ fixed, we need
\eq{
	\xi^2=\frac{\vbk_R\cdot\vbk_I}{\omega_R\omega_I}\,,\qquad q^2=k_R^2-k_I^2-(\omega_R^2-\omega_I^2)\xi^2\,.
} 
From the above expression it is clear that the conditions $0<\xi^2<1$ and $q^2>0$ are not satisfied for many points in the analyticity region. Therefore, we cannot exclude the possibility of having zeros, at least from the arguments given above, for $\chi(\omega,\vbk)$ in the whole analyticity region.

Notice that the condition of positivity is crucial for the above argument. For instance, let us consider the following function
\eq{
\frac{\log(-(\omega+i\epsilon)^2+c_1^2k^2+\omega_0^2)-\log(\omega_1^2)}{-(\omega+i\epsilon)^2+c_2^2k^2+\omega_2^2}\,.
} 
This function goes to zero at infinity and it is analytic in the FLC provided that $0<c_i<1$ and $\omega_i^2>0$. This ensures that the poles of the denominator and the branch cut of the $\log$ function lie outside the FLC. However, we see that it has zeros at $-(\omega+i\epsilon)^2+c_1^2k^2+(\omega_0^2-\omega_1^2)=0$ which for $\omega_0^2-\omega_1^2<0$ lie in the FLC (see App.~\ref{dampwave} for details). However, if we calculate the imaginary part of this function we see that in order to have positivity we must have $c_1>c_2$ and $\omega_0^2-\omega_2^2>\omega_1^2>0$, i.e.~$\omega_0^2>\omega_1^2$. Therefore, the condition of positivity forbids zeros for the $\log$ function in the FLC, consistent with the above result.

%%%%%%%%%%%%%%%%%%%%%%%%%%%%%%%%%%%%%%%%%%%%%%%%%%%%
\section{Damped wave response}\label{dampwave}
%%%%%%%%%%%%%%%%%%%%%%%%%%%%%%%%%%%%%%%%%%%%%%%%%%%%
The simplest possible retarded Green's function is the one associated to the wave equation with damping, 
\eq{
	\ddot\psi+\gamma\dot\psi-c_s^2\nabla^2\psi+\omega_0^2\psi=F\,.
}
The Green's function in Fourier space is given by
\eq{
	\spl{
		G(\omega,k)=\frac{F}{\psi}&=\frac{1}{-\omega^2-i\gamma\omega+c_s^2k^2+\omega_0^2}\\
		&=\frac{1}{-(\omega+i\gamma/2)^2+c_s^2k^2+(\omega_0^2-\gamma^2/4)}\equiv\frac{1}{P^2+m^2}\,,
		\label{Gwv}
}}
where in the second line we have rearranged the terms to write it similarly to the Lorentz-invariant case. We have defined $P^\mu\equiv(\Omega,\boldsymbol{k})$ with $\Omega=\omega+i\gamma/2$, $\boldsymbol{k}=c_s\vbk$ and $m^2=\omega_0^2-\gamma^2/4$. We would like to find the conditions on the parameters $\gamma$, $c_s^2$ and $\omega_0^2$ (assuming they are real numbers) such that it corresponds to a physical Green's function, i.e.~with no singularities in the FLC of the imaginary part of $p^\mu=(\omega,\vbk)$. Singularities of \cref{Gwv} satisfy
\eq{
	P_1^2-P_2^2+m^2=0\,,\qquad P_1^\mu P_{2\mu}=0\,,
}
where we have decomposed $P^\mu=P^\mu_1+iP^\mu_2$ into real and imaginary parts. Next, we consider different possibilities for $P_1^\mu$. If it is spacelike then it can be written as $P_1^\mu=(0,\vbk_1)$ and therefore we have
\eq{
P_2^2=m^2+k_1^2\,,\qquad \vbk_1 \cdot \vbk_2=0\,.
}
If it is timelike we can write it as $P_1^\mu=(\Omega_1,0)$ and therefore the location of the singularities must satisfy
\eq{
P_2^2=m^2-\Omega_1^2\,,\qquad \Omega_2=0\,,
}
which, in particular, means that $P_2$ is necessarily spacelike. Finally, for null vectors we write $P_1^\mu=\Omega_1(1,\hat{{n}})$, where $\hat{n}^2=1$, and we have
\eq{
P_2^2=m^2\,,\qquad -\Omega_2+(\hat{{n}}\cdot\vbk_2)=0\,.
}
In the cases above we have assumed nonzero $P_1^\mu$; otherwise we would get $P_2^2=m^2$ without any further restriction. Let us for the moment assume that $c_s^2 \ge 0$ then in components $P^\mu_1=(\omega_R,c_s\vbk_R)$ and $P^\mu_2=(\omega_I+\gamma/2,c_s\vbk_I)$. 

\paragraph{Case 1: $m^2>0$.} This means $\omega_0^2>\gamma^2/4$. In all the cases discussed above we can see that if a solution exists (for some choice of $P_1^\mu$) then $P_2^\mu$ must be spacelike. Therefore, we must have
\eq{
	-c_sk_I<\omega_I+\frac{\gamma}{2}<c_sk_I\,.
	\label{spac}
}
Notice also that all $P_2^\mu$ that satisfy \eqref{spac} correspond to poles: it is indeed straightforward to see that for any $P_2^\mu$ one can find a solution for $P_1^\mu$. But then it means that we must require $\gamma>0$ and $c_s \le 1$ to avoid singularities in the FLC in the plane of $(\omega_I,\vbk_I)$.

\paragraph{Case 2: $m^2<0$.} This means that $\omega_0^2<\gamma^2/4$. Then solutions for $P_2^\mu$ can be spacelike, timelike or null. We have already discussed the spacelike solutions above, see Eq.~\eqref{spac}, so there is nothing new here. For timelike or null solutions we note that they always satisfy $0 \ge P_2^2\geq m^2$ and therefore we have
\eq{
	c_sk_I \le \abs{\omega_I+\frac{\gamma}{2}}\leq\sqrt{c_s^2k^2_I+\frac{\gamma^2}{4}-\omega_0^2}\,.
	\label{tim}
}
It is straightforward to check that all timelike $P_2^\mu$ that satisfy this correspond to poles.
But then it means that we must require $\gamma>0$, $\omega_0^2 \ge 0$ and $c_s \le 1$ to avoid singularities in the FLC in the plane of $(\omega_I,\vbk_I)$. See \cref{poleloc}. The case $m^2 =0$ does not give further constraints.

\paragraph{Case 3: $c_s^2<0$.} In this case we write $c_s=i\tilde{c}_s$ which means that in components $P^\mu_1=(\omega_R,-\tilde{c}_s\vbk_I)$ and $P^\mu_2=(\omega_I+\gamma/2,\tilde{c}_s\vbk_R)$. Then the location of poles will be the same as \cref{spac} and \cref{tim} with $\vbk_I$ replaced with $\vbk_R$. It means that the singularities always appear in the FLC in the plane of $(\omega_I,\vbk_I)$. Therefore, $c_s^2<0$ is not a valid possibility.

\vspace{.4cm}

In summary, we need $\gamma>0$, $0 \le c_s^2 \le 1$ and $\omega_0^2 \ge 0$ for a consistent Green's function of the form \cref{Gwv}. The locations of the poles are shown in \cref{poleloc}. 
\fg{
	\includegraphics[width=0.45\textwidth]{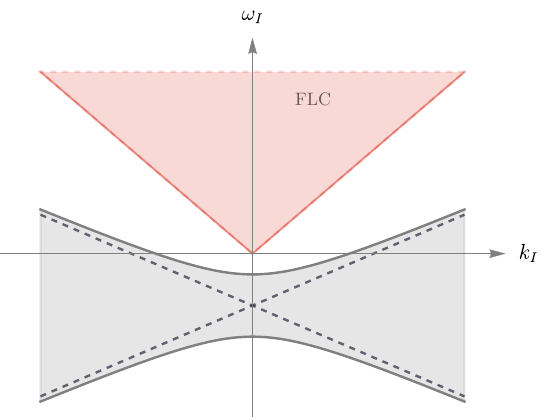}
	\caption{Singularities of $G(\omega,k)$ defined in \cref{Gwv} are located in the grey region. The solid gray lines correspond to the upper limit in \cref{tim} while the dashed lines are associated to \cref{spac} (or lower limit of \cref{tim}). For a physical Green's function singularities must be outside the FLC (pink shade).}
	\label{poleloc}
}
Before closing this section we write the imaginary part of \cref{Gwv} for later usage. If the damping parameter is finite we have
\eq{
\Im G(\omega,k)=\frac{\gamma\omega}{(-\omega^2+c_s^2k^2+\omega_0^2)^2+\gamma^2\omega^2}\,.
}
Notice that the positivity condition on the imaginary part implies $\gamma>0$. For infinitesimal damping we must replace $\gamma\to\epsilon$. Then the imaginary part becomes
\eq{
\Im G(\omega,k)=\pi\frac{|\omega|}{\omega}\delta(-\omega^2+c_s^2k^2+\omega_0^2)\,.
}

\footnotesize
%%%%%%%%%%%%%%%%%%%%%%%%%%%%%%%%%%%%%%%%%%%%%%%%%%%%%%%%
\bibliography{refs} 

\providecommand{\href}[2]{#2}\begingroup\raggedright\begin{thebibliography}{10}

\bibitem{Adams:2006sv}
A.~Adams, N.~Arkani-Hamed, S.~Dubovsky, A.~Nicolis and R.~Rattazzi,
  \emph{{Causality, analyticity and an IR obstruction to UV completion}},
  \href{https://doi.org/10.1088/1126-6708/2006/10/014}{\emph{JHEP} {\bfseries
  10} (2006) 014} [\href{https://arxiv.org/abs/hep-th/0602178}{{\ttfamily
  hep-th/0602178}}].

\bibitem{Bellazzini:2020cot}
B.~Bellazzini, J.~Elias~Mir\'o, R.~Rattazzi, M.~Riembau and F.~Riva,
  \emph{{Positive moments for scattering amplitudes}},
  \href{https://doi.org/10.1103/PhysRevD.104.036006}{\emph{Phys. Rev. D}
  {\bfseries 104} (2021) 036006}
  [\href{https://arxiv.org/abs/2011.00037}{{\ttfamily 2011.00037}}].

\bibitem{Caron-Huot:2020cmc}
S.~Caron-Huot and V.~Van~Duong, \emph{{Extremal Effective Field Theories}},
  \href{https://doi.org/10.1007/JHEP05(2021)280}{\emph{JHEP} {\bfseries 05}
  (2021) 280} [\href{https://arxiv.org/abs/2011.02957}{{\ttfamily
  2011.02957}}].

\bibitem{Tolley:2020gtv}
A.~J. Tolley, Z.-Y. Wang and S.-Y. Zhou, \emph{{New positivity bounds from full
  crossing symmetry}},
  \href{https://doi.org/10.1007/JHEP05(2021)255}{\emph{JHEP} {\bfseries 05}
  (2021) 255} [\href{https://arxiv.org/abs/2011.02400}{{\ttfamily
  2011.02400}}].

\bibitem{Arkani-Hamed:2020blm}
N.~Arkani-Hamed, T.-C. Huang and Y.-T. Huang, \emph{{The EFT-Hedron}},
  \href{https://doi.org/10.1007/JHEP05(2021)259}{\emph{JHEP} {\bfseries 05}
  (2021) 259} [\href{https://arxiv.org/abs/2012.15849}{{\ttfamily
  2012.15849}}].

\bibitem{Paulos:2016fap}
M.~F. Paulos, J.~Penedones, J.~Toledo, B.~C. van Rees and P.~Vieira, \emph{{The
  S-matrix bootstrap. Part I: QFT in AdS}},
  \href{https://doi.org/10.1007/JHEP11(2017)133}{\emph{JHEP} {\bfseries 11}
  (2017) 133} [\href{https://arxiv.org/abs/1607.06109}{{\ttfamily
  1607.06109}}].

\bibitem{EliasMiro:2022xaa}
J.~Elias~Miro, A.~Guerrieri and M.~A. Gumus, \emph{{Bridging positivity and
  S-matrix bootstrap bounds}},
  \href{https://doi.org/10.1007/JHEP05(2023)001}{\emph{JHEP} {\bfseries 05}
  (2023) 001} [\href{https://arxiv.org/abs/2210.01502}{{\ttfamily
  2210.01502}}].

\bibitem{EliasMiro:2023fqi}
J.~Elias~Miro, A.~Guerrieri and M.~A. Gumus, \emph{{Extremal Higgs couplings}},
   \href{https://arxiv.org/abs/2311.09283}{{\ttfamily 2311.09283}}.

\bibitem{Bellazzini:2023nqj}
B.~Bellazzini, G.~Isabella, S.~Ricossa and F.~Riva, \emph{{Massive gravity is
  not positive}},
  \href{https://doi.org/10.1103/PhysRevD.109.024051}{\emph{Phys. Rev. D}
  {\bfseries 109} (2024) 024051}
  [\href{https://arxiv.org/abs/2304.02550}{{\ttfamily 2304.02550}}].

\bibitem{Guerrieri:2022sod}
A.~Guerrieri, H.~Murali, J.~Penedones and P.~Vieira, \emph{{Where is M-theory
  in the space of scattering amplitudes?}},
  \href{https://doi.org/10.1007/JHEP06(2023)064}{\emph{JHEP} {\bfseries 06}
  (2023) 064} [\href{https://arxiv.org/abs/2212.00151}{{\ttfamily
  2212.00151}}].

\bibitem{Baumann:2015nta}
D.~Baumann, D.~Green, H.~Lee and R.~A. Porto, \emph{{Signs of Analyticity in
  Single-Field Inflation}},
  \href{https://doi.org/10.1103/PhysRevD.93.023523}{\emph{Phys. Rev. D}
  {\bfseries 93} (2016) 023523}
  [\href{https://arxiv.org/abs/1502.07304}{{\ttfamily 1502.07304}}].

\bibitem{Grall:2021xxm}
T.~Grall and S.~Melville, \emph{{Positivity bounds without boosts: New
  constraints on low energy effective field theories from the UV}},
  \href{https://doi.org/10.1103/PhysRevD.105.L121301}{\emph{Phys. Rev. D}
  {\bfseries 105} (2022) L121301}
  [\href{https://arxiv.org/abs/2102.05683}{{\ttfamily 2102.05683}}].

\bibitem{Freytsis:2022aho}
M.~Freytsis, S.~Kumar, G.~N. Remmen and N.~L. Rodd, \emph{{Multifield
  positivity bounds for inflation}},
  \href{https://doi.org/10.1007/JHEP09(2023)041}{\emph{JHEP} {\bfseries 09}
  (2023) 041} [\href{https://arxiv.org/abs/2210.10791}{{\ttfamily
  2210.10791}}].

\bibitem{Creminelli:2022onn}
P.~Creminelli, O.~Janssen and L.~Senatore, \emph{{Positivity bounds on
  effective field theories with spontaneously broken Lorentz invariance}},
  \href{https://doi.org/10.1007/JHEP09(2022)201}{\emph{JHEP} {\bfseries 09}
  (2022) 201} [\href{https://arxiv.org/abs/2207.14224}{{\ttfamily
  2207.14224}}].

\bibitem{Hui:2023pxc}
L.~Hui, I.~Kourkoulou, A.~Nicolis, A.~Podo and S.~Zhou, \emph{{$S$-matrix
  positivity without Lorentz invariance: a case study}},
  \href{https://arxiv.org/abs/2312.08440}{{\ttfamily 2312.08440}}.

\bibitem{Creminelli:2023kze}
P.~Creminelli, M.~Delladio, O.~Janssen, A.~Longo and L.~Senatore,
  \emph{{Non-analyticity of the $S$-matrix with spontaneously broken Lorentz
  invariance}},  \href{https://arxiv.org/abs/2312.08441}{{\ttfamily
  2312.08441}}.

\bibitem{Kronig1926}
R.~Kronig, \emph{{On the theory of the dispersion of X-rays}}, {\emph{J. Opt.
  Soc. Am.} {\bfseries 12} (1926) 547}.

\bibitem{Kramers1927}
H.~Kramers, \emph{{La diffusion de la lumi\`ere par les atomes}}, {\emph{Atti
  Cong. Intern. Fisici, (Transactions of Volta Centenary Congress) Como.}
  {\bfseries 2} (1927) 545}.

\bibitem{Keldysh1989TheDF}
L.~V. Keldysh, D.~A. Kirzhnitz and A.~A. Maradudin, \emph{The dielectric
  function of condensed systems},  1989,
  \href{https://api.semanticscholar.org/CorpusID:118667756}{https://api.semanticscholar.org/CorpusID:118667756}.

\bibitem{Leon}
M.~Leontovich, \emph{{Generalization of the Kramers-Kronig formulas to media
  with spatial dispersion}}, {\emph{J. Exptl. Theoret. Phys. (U.S.S.R.)}
  {\bfseries 40} (1961) 907}.

\bibitem{Chaikin_Lubensky_1995}
P.~M. Chaikin and T.~C. Lubensky, \emph{Principles of Condensed Matter
  Physics}. Cambridge University Press, 1995.

\bibitem{Jackson:1998nia}
J.~D. Jackson, \emph{{Classical Electrodynamics}}. Wiley, 1998.

\bibitem{Calzetta:2008iqa}
E.~A. Calzetta and B.-L.~B. Hu, \emph{{Nonequilibrium Quantum Field Theory}}.
  Oxford University Press, 2009,
  \href{https://doi.org/10.1017/9781009290036}{10.1017/9781009290036}.

\bibitem{Nieves:1988qz}
J.~F. Nieves and P.~B. Pal, \emph{{$P$ and {CP} Odd Terms in the Photon
  Selfenergy Within a Medium}},
  \href{https://doi.org/10.1103/PhysRevD.39.652}{\emph{Phys. Rev. D} {\bfseries
  39} (1989) 652}.

\bibitem{Carroll:1998zi}
S.~M. Carroll, \emph{{Quintessence and the rest of the world}},
  \href{https://doi.org/10.1103/PhysRevLett.81.3067}{\emph{Phys. Rev. Lett.}
  {\bfseries 81} (1998) 3067}
  [\href{https://arxiv.org/abs/astro-ph/9806099}{{\ttfamily
  astro-ph/9806099}}].

\bibitem{Nakai:2023zdr}
Y.~Nakai, R.~Namba, I.~Obata, Y.-C. Qiu and R.~Saito, \emph{{Can we explain
  cosmic birefringence without a new light field beyond Standard Model?}},
  \href{https://doi.org/10.1007/JHEP01(2024)057}{\emph{JHEP} {\bfseries 01}
  (2024) 057} [\href{https://arxiv.org/abs/2310.09152}{{\ttfamily
  2310.09152}}].

\bibitem{book:plasmapysics}
S.~Ichimaru, \emph{Basic Principles Of Plasma Physics: A Statistical Approach
  Mathematics}. CRC Press, 1973.

\bibitem{Weinberg:1995mt}
S.~Weinberg, \emph{{The Quantum theory of fields. Vol. 1: Foundations}}.
  Cambridge University Press, 6, 2005,
  \href{https://doi.org/10.1017/CBO9781139644167}{10.1017/CBO9781139644167}.

\bibitem{AnriARukhadze1961}
A.~A. Rukhadze and V.~P. Silin, \emph{Electrodynamics of media with spatial
  dispersion},
  \href{https://doi.org/10.1070/PU1961v004n03ABEH003357}{\emph{Soviet Physics
  Uspekhi} {\bfseries 4} (1961) 459}.

\bibitem{PhysRevE.78.026608}
V.~A. Markel, \emph{Can the imaginary part of permeability be negative?},
  \href{https://doi.org/10.1103/PhysRevE.78.026608}{\emph{Phys. Rev. E}
  {\bfseries 78} (2008) 026608}.

\bibitem{PhysRevE.70.048601}
R.~A. Depine and A.~Lakhtakia, \emph{Comment i on ``resonant and antiresonant
  frequency dependence of the effective parameters of metamaterials''},
  \href{https://doi.org/10.1103/PhysRevE.70.048601}{\emph{Phys. Rev. E}
  {\bfseries 70} (2004) 048601}.

\bibitem{PhysRevE.70.048602}
A.~L. Efros, \emph{Comment ii on ``resonant and antiresonant frequency
  dependence of the effective parameters of metamaterials''},
  \href{https://doi.org/10.1103/PhysRevE.70.048602}{\emph{Phys. Rev. E}
  {\bfseries 70} (2004) 048602}.

\bibitem{Dubovsky:2007ac}
S.~Dubovsky, A.~Nicolis, E.~Trincherini and G.~Villadoro, \emph{{Microcausality
  in curved space-time}},
  \href{https://doi.org/10.1103/PhysRevD.77.084016}{\emph{Phys. Rev. D}
  {\bfseries 77} (2008) 084016}
  [\href{https://arxiv.org/abs/0709.1483}{{\ttfamily 0709.1483}}].

\bibitem{Hartman:2015lfa}
T.~Hartman, S.~Jain and S.~Kundu, \emph{{Causality Constraints in Conformal
  Field Theory}}, \href{https://doi.org/10.1007/JHEP05(2016)099}{\emph{JHEP}
  {\bfseries 05} (2016) 099}
  [\href{https://arxiv.org/abs/1509.00014}{{\ttfamily 1509.00014}}].

\bibitem{ggKK}
D.~Melrose and R.~Stoneham, \emph{Generalised kramers-kronig formula for
  spatially dispersive media},
  \href{https://doi.org/10.1088/0305-4470/10/1/004}{\emph{Journal of Physics A:
  Mathematical and General} {\bfseries 10} (2001) L17}.

\bibitem{Froissart:1961ux}
M.~Froissart, \emph{{Asymptotic behavior and subtractions in the Mandelstam
  representation}}, \href{https://doi.org/10.1103/PhysRev.123.1053}{\emph{Phys.
  Rev.} {\bfseries 123} (1961) 1053}.

\bibitem{Martin:1962rt}
A.~Martin, \emph{{Unitarity and high-energy behavior of scattering
  amplitudes}}, \href{https://doi.org/10.1103/PhysRev.129.1432}{\emph{Phys.
  Rev.} {\bfseries 129} (1963) 1432}.

\bibitem{lan84}
L.~D. Landau and E.~M. Lifshitz, \emph{Electrodynamics of Continuous Media}.
  Pergamon, New York, 1984.

\bibitem{lindhard}
J.~Lindhard, \emph{On the properties of a gas of charged particles},
  {\emph{Kgl. Danske Videnskab. Selskab Mat.-fys. Medd.} {\bfseries Vol: 28,
  No. 8} (1954) }.

\bibitem{1986PThPS8643W}
S.~{Weinberg}, \emph{{Superconductivity for Particular Theorists}},
  \href{https://doi.org/10.1143/PTPS.86.43}{\emph{Progress of Theoretical
  Physics Supplement} {\bfseries 86} (1986) 43}.

\bibitem{schakel1998boulevard}
A.~M.~J. Schakel, \emph{Boulevard of broken symmetries},  1998.

\bibitem{Landry:2022nog}
M.~J. Landry and H.~Liu, \emph{{A systematic formulation of chiral anomalous
  magnetohydrodynamics}},  \href{https://arxiv.org/abs/2212.09757}{{\ttfamily
  2212.09757}}.

\bibitem{losyakov}
O.~Dolgov, D.~Kirzhnits and V.~Losyakov, \emph{{Admissible values of
  permittivity and magnetic permeability of matter}}, {\emph{J. Exp. Theor.
  Phys} {\bfseries 56} (1982) }.

\bibitem{Kirzhnits:1989}
D.~A. Kirzhnits, \emph{{General properties of electromagnetic response
  functions}},
  \href{https://doi.org/10.1016/B978-0-444-87366-8.50008-4}{\emph{Mod. Probl.
  Condens. Matter Sci.} {\bfseries 24} (1989) 41}.

\bibitem{Holstein:1990qy}
B.~R. Holstein, \emph{{Pion polarizability and chiral symmetry}},
  {\emph{Comments Nucl. Part. Phys.} {\bfseries 19} (1990) 221}.

\bibitem{Peskin:1995ev}
M.~E. Peskin and D.~V. Schroeder, \emph{{An Introduction to quantum field
  theory}}. Addison-Wesley, Reading, USA, 1995.

\bibitem{Raman:2021pkf}
P.~Raman and A.~Sinha, \emph{{QFT, EFT and GFT}},
  \href{https://doi.org/10.1007/JHEP12(2021)203}{\emph{JHEP} {\bfseries 12}
  (2021) 203} [\href{https://arxiv.org/abs/2107.06559}{{\ttfamily
  2107.06559}}].

\bibitem{Sommer:1970mr}
G.~Sommer, \emph{{Present state of rigorous analytic properties of scattering
  amplitudes}}, \href{https://doi.org/10.1002/prop.19700181102}{\emph{Fortsch.
  Phys.} {\bfseries 18} (1970) 577}.

\bibitem{Hartman:2017hhp}
T.~Hartman, S.~A. Hartnoll and R.~Mahajan, \emph{{Upper Bound on Diffusivity}},
  \href{https://doi.org/10.1103/PhysRevLett.119.141601}{\emph{Phys. Rev. Lett.}
  {\bfseries 119} (2017) 141601}
  [\href{https://arxiv.org/abs/1706.00019}{{\ttfamily 1706.00019}}].

\bibitem{Delacretaz:2021ufg}
L.~V. Delacretaz, A.~L. Fitzpatrick, E.~Katz and M.~T. Walters,
  \emph{{Thermalization and hydrodynamics of two-dimensional quantum field
  theories}},
  \href{https://doi.org/10.21468/SciPostPhys.12.4.119}{\emph{SciPost Phys.}
  {\bfseries 12} (2022) 119}
  [\href{https://arxiv.org/abs/2105.02229}{{\ttfamily 2105.02229}}].

\bibitem{Heller:2022ejw}
M.~P. Heller, A.~Serantes, M.~Spali\'nski and B.~Withers, \emph{{Rigorous
  Bounds on Transport from Causality}},
  \href{https://doi.org/10.1103/PhysRevLett.130.261601}{\emph{Phys. Rev. Lett.}
  {\bfseries 130} (2023) 261601}
  [\href{https://arxiv.org/abs/2212.07434}{{\ttfamily 2212.07434}}].

\bibitem{Heller:2023jtd}
M.~P. Heller, A.~Serantes, M.~Spali\'nski and B.~Withers, \emph{{The
  Hydrohedron: Bootstrapping Relativistic Hydrodynamics}},
  \href{https://arxiv.org/abs/2305.07703}{{\ttfamily 2305.07703}}.

\bibitem{Jackiw:1974cv}
R.~Jackiw, \emph{{Functional evaluation of the effective potential}},
  \href{https://doi.org/10.1103/PhysRevD.9.1686}{\emph{Phys. Rev. D} {\bfseries
  9} (1974) 1686}.

\bibitem{Abbott:1981ke}
L.~F. Abbott, \emph{{Introduction to the Background Field Method}}, {\emph{Acta
  Phys. Polon. B} {\bfseries 13} (1982) 33}.

\bibitem{Chou:1984es}
K.-c. Chou, Z.-b. Su, B.-l. Hao and L.~Yu, \emph{{Equilibrium and
  Nonequilibrium Formalisms Made Unified}},
  \href{https://doi.org/10.1016/0370-1573(85)90136-X}{\emph{Phys. Rept.}
  {\bfseries 118} (1985) 1}.

\bibitem{Wang:1998wg}
E.~Wang and U.~W. Heinz, \emph{{A Generalized fluctuation dissipation theorem
  for nonlinear response functions}},
  \href{https://doi.org/10.1103/PhysRevD.66.025008}{\emph{Phys. Rev. D}
  {\bfseries 66} (2002) 025008}
  [\href{https://arxiv.org/abs/hep-th/9809016}{{\ttfamily hep-th/9809016}}].

\bibitem{PhysRevB.22.3385}
G.-z. Zhou, Z.-b. Su, B.-l. Hao and L.~Yu, \emph{Closed time path green's
  functions and critical dynamics},
  \href{https://doi.org/10.1103/PhysRevB.22.3385}{\emph{Phys. Rev. B}
  {\bfseries 22} (1980) 3385}.

\bibitem{Weinberg:2005vy}
S.~Weinberg, \emph{{Quantum contributions to cosmological correlations}},
  \href{https://doi.org/10.1103/PhysRevD.72.043514}{\emph{Phys. Rev. D}
  {\bfseries 72} (2005) 043514}
  [\href{https://arxiv.org/abs/hep-th/0506236}{{\ttfamily hep-th/0506236}}].

\bibitem{kubo}
R.~Kubo, \emph{Statistical-mechanical theory of irreversible processes. i.
  general theory and simple applications to magnetic and conduction problems},
  \href{https://doi.org/10.1143/JPSJ.12.570}{\emph{Journal of the Physical
  Society of Japan} {\bfseries 12} (1957) 570}.

\bibitem{Hartnoll:2009sz}
S.~A. Hartnoll, \emph{{Lectures on holographic methods for condensed matter
  physics}}, \href{https://doi.org/10.1088/0264-9381/26/22/224002}{\emph{Class.
  Quant. Grav.} {\bfseries 26} (2009) 224002}
  [\href{https://arxiv.org/abs/0903.3246}{{\ttfamily 0903.3246}}].

\bibitem{Uhlmann_2016}
A.~Uhlmann, \emph{Anti- (conjugate) linearity},
  \href{https://doi.org/10.1007/s11433-015-5777-1}{\emph{Science China Physics,
  Mechanics \& Astronomy} {\bfseries 59} (2016) }.

\bibitem{Weldon:1982aq}
H.~A. Weldon, \emph{{Covariant Calculations at Finite Temperature: The
  Relativistic Plasma}},
  \href{https://doi.org/10.1103/PhysRevD.26.1394}{\emph{Phys. Rev. D}
  {\bfseries 26} (1982) 1394}.

\bibitem{Shuryak:1980tp}
E.~V. Shuryak, \emph{{Quantum Chromodynamics and the Theory of Superdense
  Matter}}, \href{https://doi.org/10.1016/0370-1573(80)90105-2}{\emph{Phys.
  Rept.} {\bfseries 61} (1980) 71}.

\bibitem{Nicolis:2023pye}
A.~Nicolis, A.~Podo and L.~Santoni, \emph{{The connection between nonzero
  density and spontaneous symmetry breaking for interacting scalars}},
  \href{https://doi.org/10.1007/JHEP09(2023)200}{\emph{JHEP} {\bfseries 09}
  (2023) 200} [\href{https://arxiv.org/abs/2305.08896}{{\ttfamily
  2305.08896}}].

\bibitem{Podo:2023ute}
A.~Podo and L.~Santoni, \emph{{Fermions at finite density in the path integral
  approach}}, \href{https://doi.org/10.1007/JHEP02(2024)182}{\emph{JHEP}
  {\bfseries 02} (2024) 182}
  [\href{https://arxiv.org/abs/2312.14753}{{\ttfamily 2312.14753}}].

\bibitem{2015imbpbookC}
P.~{Coleman}, \emph{{Introduction to Many-Body Physics}}. 2015.

\bibitem{PhysRevB.25.6310}
Z.~H. Levine and S.~G. Louie, \emph{New model dielectric function and
  exchange-correlation potential for semiconductors and insulators},
  \href{https://doi.org/10.1103/PhysRevB.25.6310}{\emph{Phys. Rev. B}
  {\bfseries 25} (1982) 6310}.

\bibitem{Landau:1980mil}
L.~D. Landau and E.~M. Lifshitz, \emph{{Statistical Physics, Part 1}}, vol.~5
  of \emph{Course of Theoretical Physics}. Butterworth-Heinemann, Oxford, 1980.

\end{thebibliography}\endgroup
\bibliographystyle{JHEP}
%%%%%%%%%%%%%%%%%%%%%%%%%%%%%%%%%%%%%%%%%%%%%%%%%%%%%%%%
\end{document}